\documentclass[pre,twocolumn,amsmath,amssymb,superscriptaddress]{revtex4-1}

\usepackage{mathtools,amscd}

\usepackage{url}
\usepackage{hyperref}
\hypersetup{
    colorlinks,
    citecolor=blue,
    filecolor=blue,
    linkcolor=blue,
    urlcolor=blue
}

\usepackage{tabularx}

\usepackage{mathrsfs,ragged2e}
\usepackage{verbatim}
\usepackage{graphicx}	
\usepackage{dcolumn}	
\usepackage{bm}		
\usepackage{bbm,upgreek}
\usepackage{mathtools}
\usepackage{physics}
\usepackage{dsfont}
\usepackage{cancel}
\usepackage{xcolor}
\usepackage{harpoon}
\usepackage{accents}
\usepackage[normalem]{ulem}

\usepackage{color}
\usepackage[cal=boondoxo]{mathalfa}
\usepackage{palatino}

\makeatletter
\def\maketitle{
\@author@finish
\title@column\titleblock@produce
\suppressfloats[t]}
\makeatother

\usepackage{graphicx}

\usepackage{dsfont}
\usepackage{cancel}
\usepackage{xcolor}
\usepackage{harpoon}
\usepackage{accents}

\definecolor{red}{RGB}{0, 0, 0}

\definecolor{orange}{RGB}{1, .7, 0}

\newcommand{\EC}{\textit{E.~coli}}

\newcommand*{\Scale}[2][4]{\scalebox{#1}{$#2$}}%
\newcommand*{\inliner}[1]{{\smash{\Scale[0.9]{#1}}}}

\DeclareRobustCommand{\SkipTocEntry}[5]{}
\newcommand{\TocSkip}{\addtocontents{toc}{\SkipTocEntry}}

\newcommand{\beginsupplement}{%
        \setcounter{table}{0}
        \renewcommand{\thetable}{S\arabic{table}}%
        \setcounter{figure}{0}
        \renewcommand{\thefigure}{S\arabic{figure}}%
        \renewcommand{\theHfigure}{S\arabic{figure}}
     }

\begin{document}

\newcommand*{\idea}[1]{\medskip \noindent \textbf{#1}}



\title{Noise robustness and metabolic load determine the principles of central dogma regulation}

\author{Teresa W. Lo}
\thanks{These two authors contributed equally}
\author{Han James Choi}
\thanks{These two authors contributed equally}
\author{Dean Huang}
\affiliation{Department of Physics, University of Washington, Seattle, Washington 98195, USA}
\author{Paul A. Wiggins}
\email{pwiggins@uw.edu}
\affiliation{Department of Physics, University of Washington, Seattle, Washington 98195, USA}
\affiliation{Department of Bioengineering, University of Washington, Seattle, Washington 98195, USA}
\affiliation{Department of Microbiology, University of Washington, Seattle, Washington 98195, USA}

\begin{abstract}
\textbf{Abstract:} The processes of gene expression are inherently stochastic, even for essential genes required for growth. How does the cell maximize fitness in light of noise? To answer this question, we build a mathematical model to explore the trade-off between metabolic load and growth robustness. The model predicts novel principles of central dogma regulation: Optimal protein expression levels for many genes are in vast overabundance. Essential genes are transcribed above a lower limit of one message per cell cycle. Gene expression is achieved by load balancing between transcription and translation. We \textcolor{red}{present evidence} that each of these novel regulatory principles is observed. These results reveal that robustness and metabolic load determine the global regulatory principles that govern \textcolor{orange}{gene expression} processes, and these principles have broad implications for cellular function.

\medskip
\noindent

\textbf{Short title:} Noise robustness shapes central dogma regulation

\medskip
\noindent

\textbf{One-sentence summary:} Fitness maximization predicts protein overabundance, a transcriptional floor, and the balancing of transcription and translation.

\end{abstract}



\maketitle

\noindent

\TocSkip\section*{Introduction}

What rationale determines the optimal transcription and translation level of a gene in the cell?  Protein expression levels optimize cell fitness \cite{Dekel:2005fu,Keren:2016pt}: Too low of an expression level of essential proteins slows growth by compromising the function of essential processes {\color{red}\cite{Peters:2016jf, 
Gallagher:2020sp}}, whereas the overexpression of proteins slows growth by increasing the metabolic load \cite{Lengeler1998}. This trade-off na\"ively predicts that the cell maximizes its fitness by a Goldilocks principle in which cells express just enough protein for function \cite{Belliveau:2021pj}; however, achieving growth robustness is nontrivial, since all processes at the cellular scale are stochastic, including gene expression \cite{Raser:2005we}. This biological noise leads to significant cell-to-cell variation in protein numbers, even for essential proteins that are required for growth \cite{Newman:2006nl,Taniguchi2010}. The optimal expression program must therefore  ensure robust expression of hundreds of distinct essential gene products. In this paper, we explore the consequences of growth robustness on the central dogma regulatory program.


%


\begin{figure*}
  \centering
    \includegraphics[width=0.95\textwidth]{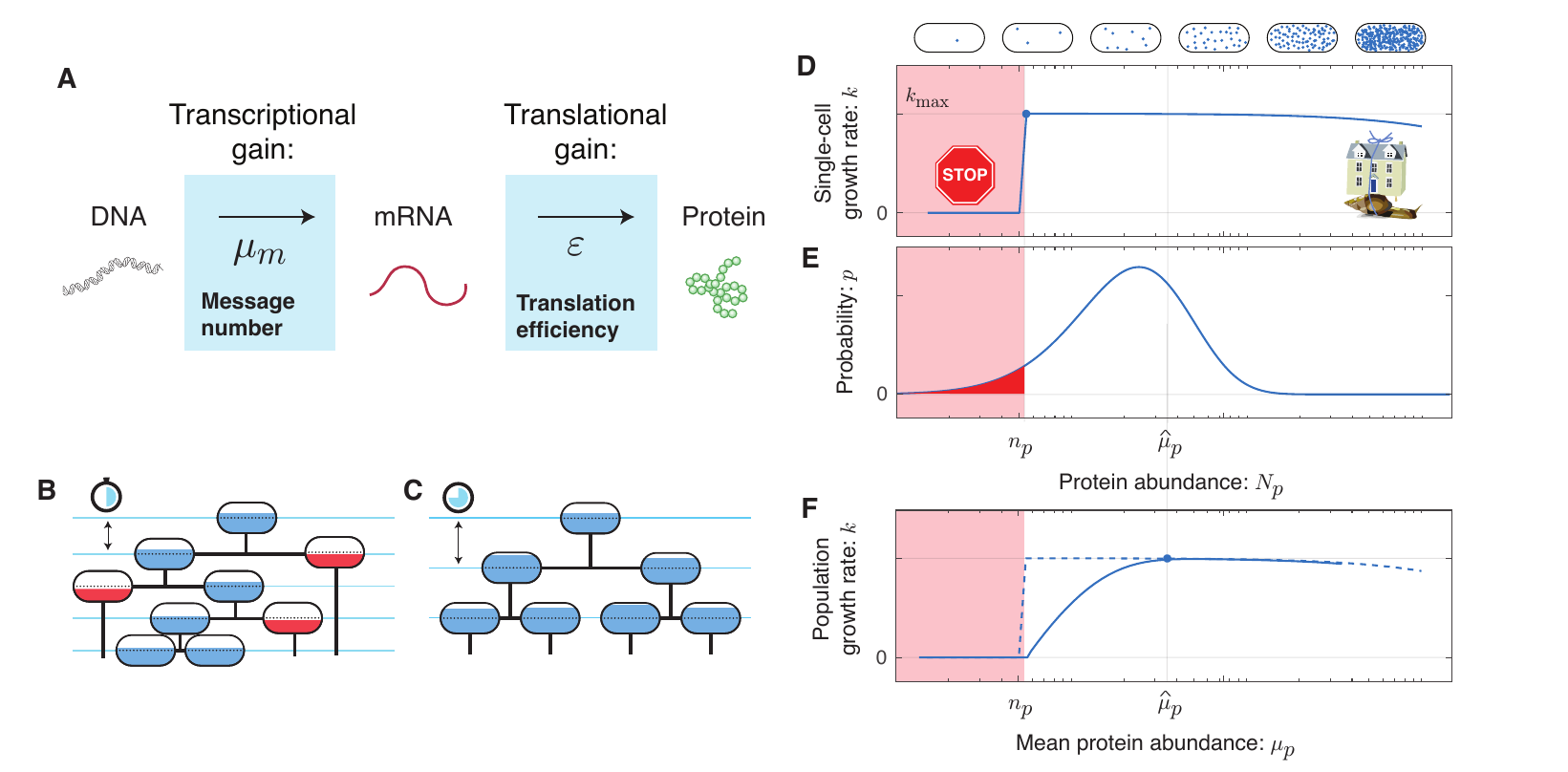}
      \caption{{\color{red}\textbf{The RLTO Model. Panel A: Gene expression is stochastic.} The central dogma describes a two-stage stochastic process where genes are first transcribed and then translated.  The transcription process transcribes an average of $\mu_m$ messages per cell cycle. The translation process translates an average of $\varepsilon$ proteins per message.  
            \textbf{Panel B \& C:} A schematic cell lineage tree is shown during exponential growth. \textcolor{red}{For a specific protein $i$, the cell fill represents the protein number $N_p$ relative to its threshold number $n_p$ required for cell growth. \textbf{Panel B:} Reducing the mean expression level reduces doubling time; however, stochasticity in expression results in  below-threshold cells (red fill) which grow slowly.  
            \textbf{Panel C:} Increasing protein expression increases the doubling time; however, all cells are above threshold (blue fill).}
            \textbf{ Panel D: The fitness landscape as a function of protein number.} 
        Growth arrests for protein number $N_p$ smaller than the threshold level $n_p$ (red) due to the failure of essential processes. High expression levels are penalized due to the metabolic cost of protein expression. This trade-off leads to a highly asymmetric fitness landscape: The relative metabolic cost of overabundance is small relative to the cost of growth arrest due to the large size of the total metabolic load $N_0$. 
               \textbf{Panel E: The gene expression process is stochastic.} There is significant cell-to-cell variation in protein abundance ($N_p$) around the mean level ($\mu_p$).  Even for mean expression levels significantly above the threshold level $n_p$, some cells fall below threshold (red). The distribution in protein number is modeled using a gamma distribution \cite{Taniguchi2010}, parameterized by message number $\mu_m$ and translation efficiency $\varepsilon$.
               \textbf{Panel F: The robustness load trade-off determines the optimal expression level.} The population growth rate depends on the distribution of the protein number. The asymmetry of the fitness landscape drives the optimal expression level far above the threshold level due to the high fitness cost of low protein abundance. 
             \label{fig:RLTOmodel}}}
\end{figure*}

\TocSkip\section*{Results}

\idea{Defining the RLTO  Model.} To study the consequences of growth robustness on \textcolor{orange}{gene expression} quantitatively, we propose and analyze a minimal model: the Robustness-Load Trade-Off (RLTO) Model. The model includes three critical components: (i) Protein levels are stochastic and the single-cell growth rate depends upon them, (ii) gene transcription and translation generate a metabolic load, and (iii) cell growth is dependent on a large number of essential genes. {\color{red} These model characteristics result in a highly-asymmetric fitness landscape.
  The optimization of expression on this asymmetric landscape predicts new phenomenology absent from previous models (\textit{e.g.}\ \cite{Hausser:2019fi}). }

\begin{figure}
  \centering
    \includegraphics[width=.5\textwidth]{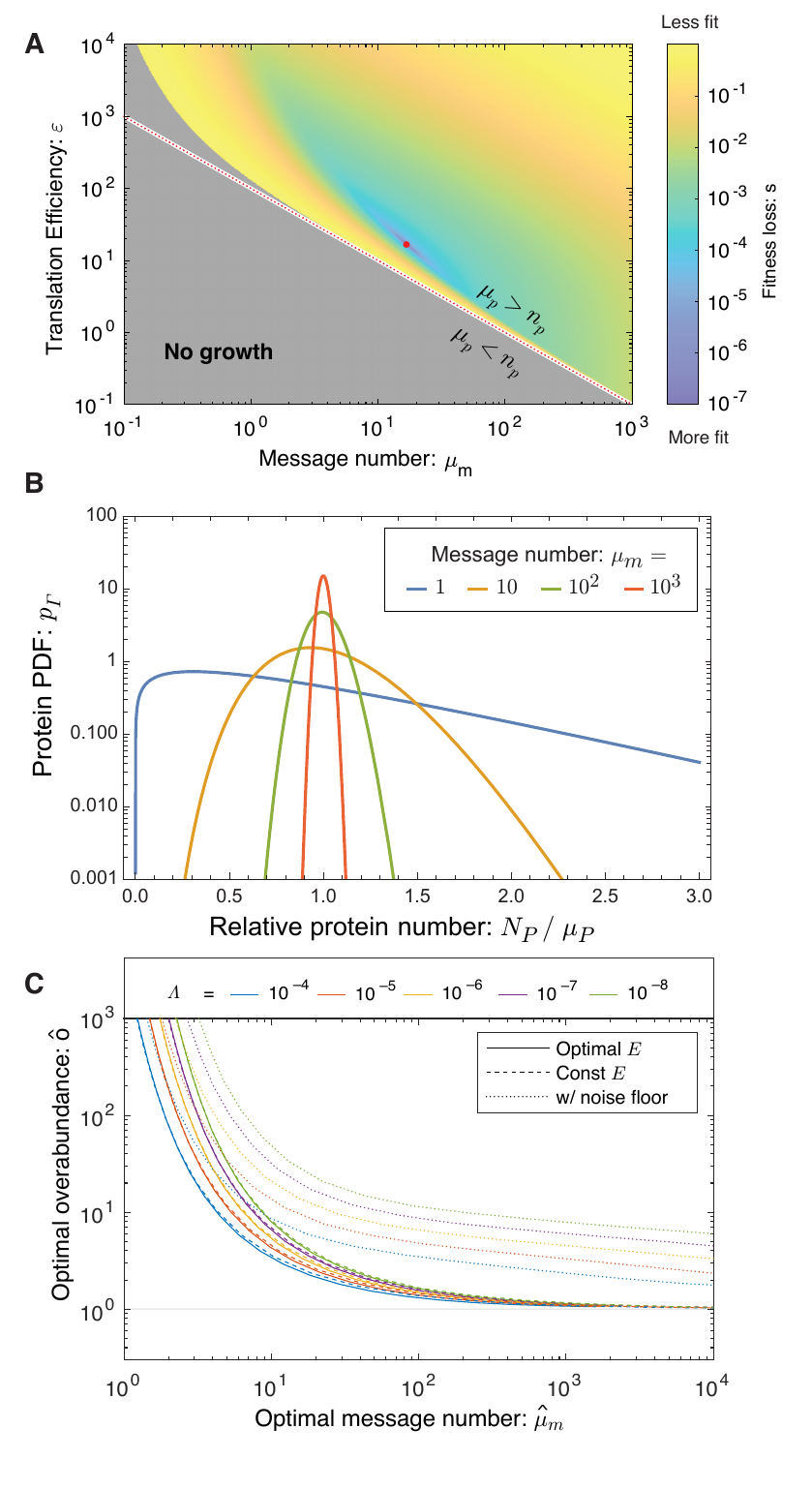}
    \caption{ 
     {\color{red} \textbf{The RLTO model predicts overabundance is optimal. Panel A: Fitness landscape determines optimal message number and translation efficiency.} The fitness loss ($s\equiv \ln k_{\rm max}/k$)  is shown as a function of message number ($\mu_m$) and translation efficiency ($\varepsilon$). The red dotted curve represents programs where the mean protein number is equal to the threshold ($\mu_p = n_p$) and the red dot represents the optimal regulatory program ($\hat{\mu}_m$, $\hat{\varepsilon}$).
    \textbf{Panel B: Gene-expression noise.} Due to the stochasticity of gene expression processes \textcolor{red}{at equilibrium}, the protein number $N_p$ is gamma-distributed \textcolor{red}{\cite{Taniguchi2010}}. 
      For high-expression genes, expression has low noise and the protein number is tightly distributed around its mean; however, for low-expression genes, expression is noisy and the distribution is extremely wide. 
      \textbf{Panel C: Overabundance is optimal for all genes.} For high-expression genes, low overabundance is optimal ($\mu_p \approx n_p$); however, for low-expression genes, vast overabundance is optimal ($\mu_p\gg n_p$).  From a quantitative perspective, overabundance depends on the relative load $\Lambda$; however, the qualitative dependence is invariant to over an orders-of-magnitude range of values.  \label{fig:overep}}}
\end{figure}

{\color{red} The protein number $N_p$ expressed from gene $i$ is the product of two sequential stochastic processes: transcription and translation \cite{Crick:1970vy}, leading to cell-to-cell variation in protein number, which we will refer to as \textit{noise}. In our analysis, we will model gene expression using the \textcolor{orange}{canonical steady-state noise} model \cite{Paulsson:2000xi}. (See Fig.~\ref{fig:RLTOmodel}A.) In this model, the numbers of proteins $N_p$ for gene $i$ are predicted to be gamma-distributed \cite{Friedman:2006oh}, in close agreement with observation \cite{Taniguchi2010}. The distribution is described by two gene-specific statistical parameters: \textit{message number} ($\mu_m$), defined as the mean number of messages transcribed per cell cycle for gene $i$, and  the \textit{translation efficiency} ($\varepsilon$), the mean number of proteins translated from each message transcribed for gene $i$. The mean protein abundance is their product: $\mu_p = \mu_m\varepsilon$. These  parameters can be expressed in terms of ratios of the rates of the underlying \textcolor{orange}{gene expression} processes, as described in Supplementary Material Sec.~\ref{sec:noise_intro}.

How should the effect of essential protein expression on growth rate be modeled in the context of the RLTO model? Much recent work has focused on cellular resource allocation to functional sectors (\textit{e.g.}~\cite{Scott:2010ec}). In this approach, an optimization is performed by \textcolor{orange}{the \textit{coordinated} modulation of} the abundance of all proteins in a particular sector, leading to a trade-off between functional capacities of the cell. However, in the RLTO model, the optimization is fundamentally different: We consider the \textcolor{orange}{\textit{uncoordinated} modulation of the abundance} of  protein species~$i$ due to noise. For these incoherent changes, we generically expect proteins to exhibit rate-limited kinetics: Increases in the protein number $N_p$ above a threshold level $n_p$ has minimal effect on the rate since other chemical species (proteins, metabolites, \textit{etc.}) are rate limiting \cite{Steinfeld1999}. However, if the protein number $N_p$ falls below the threshold $n_p$, then protein species~$i$ becomes rate limiting and leads to a significant slowdown of the growth rate. In the RLTO model, we coarse-grain the details of this growth slowdown as growth arrest. (See Fig.~\ref{fig:RLTOmodel}.) There is already some precedent for the use of this type of threshold (\textit{e.g.}~\cite{Charlebois2011-bc}), but we will demonstrate that the detailed form of the fitness landscape is not important. (See Supplementary Material Sec.~\ref{expmean}.) Although sufficiently detailed knowledge of the relevant molecular and cellular biology could be used to predict the protein thresholds $n_p$, we will treat these as gene-specific unknown parameters.

As shown in Materials and Methods, the relative cellular fitness with respect to the expression of gene $i$ can be computed by combining the fitness losses associated with robustness (Eq.~\ref{eqn:robust}) and metabolic load (Eq.~\ref{eq:metaload}):
\begin{equation}
\textstyle \frac{\Delta k}{k_0} =  \textstyle - (\Lambda+\frac{\varepsilon}{N_0})\mu_m - \frac{1}{\ln 2}\gamma(\frac{\mu_m}{\ln 2},\frac{n_p}{\varepsilon \ln 2}), \label{eqn:growthrate2}
\end{equation}
where the first term represents fitness loss due to metabolic load of transcription and translation while the second term represents loss due to the arrest of essential processes. \textcolor{orange}{$\gamma$ is the regularized incomplete gamma function and the central distribution function (CDF) of the gamma distribution. (See Supplementary Material Sec.~\ref{sec:noise_intro} and ~\ref{Sec:RLTOderivation}.)}  In summary, the model depends only on a single global parameter: the relative metabolic load $\Lambda$ and three gene-specific parameters: the threshold  number $n_p$, the message number $\mu_m$ and the relative translation efficiency $\varepsilon/N_0$. We propose that the cell is regulated to maximize the growth rate with respect to transcription (message number) and translation (translation efficiency). The fitness landscape predicted by the RLTO model for representative parameters is shown in Fig.~\ref{fig:overep}A.

}


\idea{RLTO  predicts protein overabundance.} The optimal regulatory program ($\mu_m$ and $\varepsilon$ values) can be predicted analytically. They depend on only a single global parameter, the relative load $\Lambda$, and the gene-specific threshold number $n_p$. Since the threshold number is not directly observable experimentally,  we will instead predict the optimal overabundance $o$, defined as the ratio of the mean protein number to the threshold number: 
\begin{equation}
o \equiv \textstyle \mu_p/n_p.
\end{equation}
As shown in Fig.~\ref{fig:overep}C, the RLTO model generically predicts that the optimal protein fraction is overabundant ($o>1$); however, the overabundance is not uniform for all proteins. 
For highly-transcribed genes ($\mu_m\gg 1$) like ribosomal genes, the overabundance is predicted to be quite small ($o\approx 1$); however,
for message numbers approaching unity, the overabundance is predicted to be extremely high $(o\gg 1)$.    
At a quantitative level, the relation between optimal overabundance and message number  depends on the relative load ($\Lambda$), but its phenomenology is qualitatively unchanged over orders of magnitude variation in  $\Lambda$. 

\idea{Understanding the rationale for overabundance.} To explore both the robustness of the protein overabundance prediction and to understand its mathematical rationale, we explored a collection of more complex models numerically. (Supplementary Material Sec.~\ref{expmean}.) The key mathematical feature that drives overabundance is not the assumption of growth arrest, but rather the strong asymmetry of the fitness landscape: the high cost of protein \textit{underabundance} and the low cost of protein \textit{overabundance}. (See Fig.~\ref{fig:RLTOmodel}EF.) \textcolor{orange}{The population growth rate (Panel~\ref{fig:RLTOmodel}F) can be understood qualitatively as the convolution of the single-cell growth rate (Panel~\ref{fig:RLTOmodel}D) with the probability density function (PDF) of the protein abundance (Panel~\ref{fig:RLTOmodel}E).} In the RLTO model, this asymmetry is parameterized by the relative load ($\Lambda$), defined as the relative metabolic cost of transcribing an additional message. Since we estimate that $\Lambda<10^{-5}$, this cost is very low relative to the total metabolic cost of the cell, therefore we expect this asymmetry, and the prediction of the RLTO model, to be robust.

\begin{figure*}
  \centering
    \includegraphics[width=.95\textwidth]{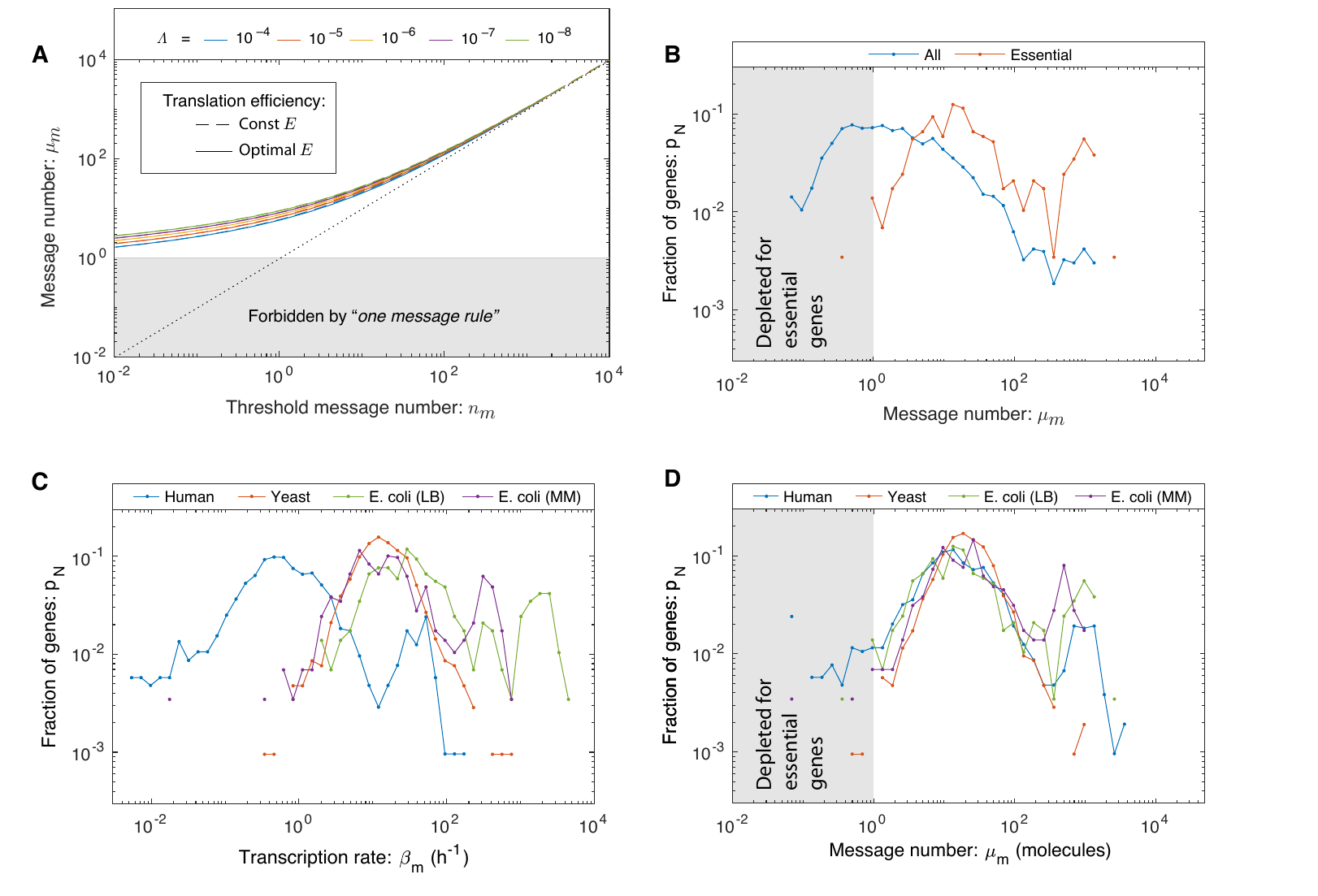}
      \caption{  {\color{red} \textbf{A lower threshold for transcription: The one message rule. Panel A: RLTO predicts the one-message rule.}} For high-expression genes, overabundance is low and  the message number $\mu_m$ is predicted to be comparable to the threshold level $n_m$ (dotted line); however, for low-expression genes there is a lower threshold ($\mu_m\ge 1$) below which expression is too noisy for robust growth.
 The threshold is weakly dependent on relative load $\Lambda$. 
 {\color{red} \textbf{Panel B: A one-message threshold is observed in \textit{E.~coli} for essential genes.} A histogram shows the distribution of gene message numbers for all genes (blue) versus essential genes (orange). As predicted by the RLTO model, virtually all essential genes are expressed above the one-message-per-cell-cycle threshold.
  \textbf{Panel C: The distribution of transcription rates for essential genes.} No alignment is observed between the distributions of transcription rates in three evolutionarily-divergent organisms. For instance, the per gene transcription rate is significantly lower in human cells relative to \textit{E.~coli}.
   \textbf{Panel D: The distribution of message numbers for essential genes in three evolutionarily-divergent organisms.} The alignment of distributions of message number per gene between human, yeast, and \textit{E.~coli} (under two distinct growth conditions) reveals a nontrivial commonality between central dogma regulatory programs. We propose that the rationale for this alignment is the one-message rule that predicts that all essential genes must be expressed above one message per cell cycle. Both yeast and \textit{E.~coli} come very close to satisfying this proposed threshold; however, a greater proportion of genes in human break the one-message threshold. We speculate that this is due in part to the \textit{ad hoc} nature of the essential-gene classification in the context of complex multicellular organisms. }
  \label{fig:onemessagerule}}
\end{figure*}

\idea{Overabundance is observed in a range of experiments.} The RLTO Model predicts that all essential proteins are overabundant. In general, the RLTO model predicts that protein numbers have very significant robustness (\textit{i.e.}~buffering) to protein depletion.  Although this result is potentially surprising, it is in fact consistent with many studies. For instance, Belliveau \textit{et al.}~have recently analyzed the abundance of a wide range of metabolic and other essential biological processes, and conclude that protein abundance appears to be in significant excess of what is required for function \cite{Belliveau:2021pj}. 
Likewise, CRISPRi approaches have facilitated the characterization of essential protein depletion. The qualitative results from these experiments are consistent with overabundance: Large-magnitude protein depletion is typically required to generate strong phenotypes \cite{Peters:2016jf,Silvis:2021oz,Donati:2021kq}.  In particular, Peters \textit{et al.}\ engineered a complete collection of CRISPRi essential-gene depletion constructs in \textit{Bacillus subtilis}. Importantly, when \textit{dcas9} is constitutively expressed, these constructs deplete essential proteins about three-fold below their endogenous expression levels \cite{Peters:2016jf}; however, roughly 80\% grew without measurable fitness loss in log-phase growth despite the depletion. When grouped by functional category, only ribosomal proteins were found to have statistically significant reductions in fitness \cite{Peters:2016jf}.  
As shown in Fig.~\ref{fig:overep}C, the RLTO model predicts that all but the highest expression proteins are expected to show minimal fitness reductions in response to a three-fold depletion of essential enzymes. The optimality of protein overabundance explains the  paradox of protein expression levels being simultaneously optimal \cite{Dekel:2005fu} and in excess of what is required for function \cite{Belliveau:2021pj,Peters:2016jf,Gallagher:2020sp,Donati:2021kq}. Although this qualitative picture of essential protein overabundance is clear, there has yet to be a quantitative and detailed measurement of protein overabundance, and in particular, an analysis of the relationship between protein overabundance and message number. 

\idea{RLTO predicts a one-message transcription threshold.} 
The RLTO model predicts protein overabundance, but is there a clear transcriptional signature? To analyze this question, we define the message threshold $n_m \equiv \mu_m/o$. (This parameterization is convenient since it is independent of the translation efficiency.) We can then analyze the relation between optimal message number and threshold message number, as shown in Fig.~\ref{fig:onemessagerule}A.  The model predicts that even for genes that have extremely small threshold message  numbers (\textit{e.g.}~$n_m=10^{-2}$), the optimal message number stays above one message transcribed per cell cycle.  Qualitatively, expressing messages below this level is simply too noisy even for proteins needed at the lowest expression levels. {\color{red}(See the blue curve in Fig.~\ref{fig:overep}B corresponding to the protein number distribution of $\mu_m=1$.)} The model therefore predicts a lower floor on transcription for essential genes of one message per cell cycle.

{\color{red}
\idea{A lower threshold is observed for message number.} 
To identify a putative transcriptional floor, we first analyzed the transcriptome in \textit{Escherichia coli}. We hypothesize that cells must express essential genes above the one-message threshold for robust growth.
The distinction between essential and nonessential genes is critical in this context, since nonessential genes can be inducibly expressed. For instance, in \textit{E.~coli}, the \textit{lac} operon is repressed in the absence of lactose and therefore need not satisfy the one-message threshold. The transcriptional threshold is only hypothesized to apply to genes whose products are required to maintain cell fitness under the measured conditions.

We generated histograms for \textit{E.~coli} growing rapidly on rich media for these two classes of genes.  (See Supplementary Material Sec.~\ref{sec:desripton}.)  The message numbers for \textit{nonessential} genes are widely distributed, with a significant fraction of genes falling below the one-message threshold; however, only one \textit{essential} gene is expressed below the one-message threshold (0.3\% of essential genes). (See Fig.~\ref{fig:onemessagerule}B.)  The threshold is not sharp, but rather a smooth depletion relative to a median of 18 messages per cell cycle. This observation is consistent with the predictions of the RLTO model. 

To further test this prediction, we then analyzed \textit{E.~coli} transcription under slow-growth conditions. Since these cells are less transcriptionally active, we hypothesized that this analysis would constitute a more stringent test of the one-message rule. (See Supplementary Material Sec.~\ref{sec:desripton}.) To our surprise, although the transcription rate is indeed reduced in slow growth, the essential gene message numbers  still satisfy the one-message rule (with a two gene exception, 0.7\%), again consistent with the predictions of the RLTO model. (See Fig.~\ref{fig:onemessagerule}D.)

Next, we analyzed eukaryotic transcriptomes in \textit{Saccharomyces cerevisiae} (yeast) and \textit{Homo sapiens} (human). (See Supplementary Material Sec.~\ref{sec:desripton}.) For yeast, there is a well-defined notion of essential genes \cite{Leeuwen:2020eh}. As predicted, yeast essential genes obey the one-message threshold (with two exceptions, 0.2\%). (See Fig.~\ref{fig:onemessagerule}D.) The interpretation is less clear-cut in human cells: An essential gene classification has been generated in the context of proliferation in cell culture \cite{Wang:2015lb}. In order to try to capture a generic picture, we average the human transcriptome of cell types. We find that the vast majority of essential genes obey the  one-message rule; however, there are significantly more genes that break the rule (81 genes, 8\%) than in the other organisms. 

\idea{Message number distribution is conserved.} To what extent is this human data consistent with the RLTO model? For human cells, our test of the one-message rule is too simplistic in two  respects: (i) We ignore the significant transcriptional differences associated with distinct cell types and (ii) the essential gene classification itself is defined by the ability of mutants or knockdowns to proliferate in cell culture; in marked contrast to the \textit{in vivo} context where cell proliferation is tightly regulated \cite{Wang:2015lb}. Due to these subtleties, we  decided to take a complementary approach: We considered the distribution of three different transcriptional statistics for each gene: transcription rate, cellular message number, defined as the average number of messages instantaneously, and message number ($\mu_m$), defined as the number of messages transcribed in a cell cycle. (See Supplementary Material Sec.~\ref{sec:desripton}.) The RLTO model predicts a one-message threshold with respect to message number, but not the other two statistics. We therefore predict that the message number distributions in each organism (\textit{E.~coli}, yeast, and human) should align for low expression genes with respect to message number, but not for the other two transcriptional statistics. Consistent with the predictions of the RLTO model, there is a striking alignment of message number for essential genes between all three model organisms and growth conditions for message number.  (See Fig.~\ref{fig:onemessagerule}D.) This alignment is non-trivial:  It is not observed with respect to other transcriptional statistics (Fig.~\ref{fig:onemessagerule}C and Supplementary Material Fig.~\ref{fig:onemessagerule}). 

\textcolor{orange}{What is the significance of the similarity in the distributions of message number between organisms? Another strategy for satisfying the one message rule would be for transcription to be increased. For instance, mammalian cells have about 1000 times the number of  proteins relative to bacterial cells. (See Supplemental Tab.~\ref{tab1refs}.) One might therefore naively predict that the message number should be increased 1000-fold as well. This is not observed. In fact, the message number distributions of all the model organisms analyzed abut the one message threshold. The proximity to the threshold  suggests that organisms do as little transcription as possible while satisfying the one message rule. This appears to be a conserved transcriptional regulatory strategy from \textit{E.~coli} to human. }
}

\begin{figure*}
  \centering
    \includegraphics[width=0.95\textwidth]{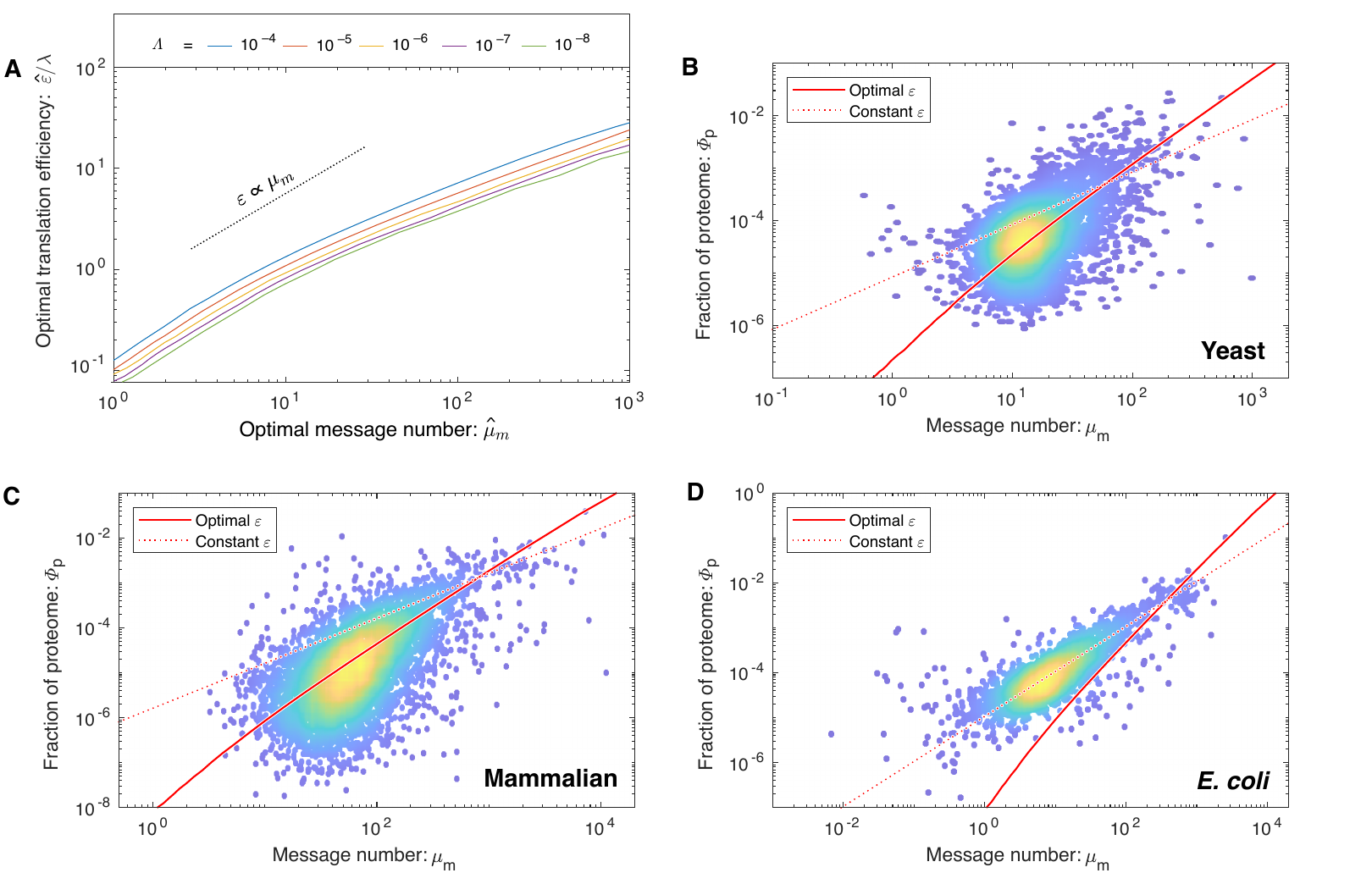}
      \caption{  
      {\color{red} \textbf{ How are transcription and translation balanced? Panel A: The RLTO model predicts load balancing.} The ratio of the optimal translation efficiency ($\hat{\varepsilon}$) to the message cost ($\lambda$) is roughly independent of the relative  load ($\Lambda$). The  translation efficiency $\varepsilon$ is predicted to be roughly proportional message number $\mu_m$. 
       \textbf{Panel B: RLTO predicts the protein-message-abundance relation in yeast.} The observed proteome fraction is compared to two models: the RLTO optimal model (solid red line) and constant-translation-efficiency model (dotted red line). Both models make parameter-free predictions.  The RLTO optimum predicts the global trend. (Data from Ref.~\cite{Ghaemmaghami:2003jj}.)
\textbf{Panel C: Mammalian proteome fraction.}  The RLTO prediction (solid) is superior to the constant-translation-efficiency prediction (dashed).   \textbf{Panel D: \textit{E.~coli} proteome fraction.} In contrast, the constant-translation-efficiency prediction (dashed) is superior to RLTO prediction (solid).      }
      \label{fig:proteom}} %
\end{figure*}

{\color{red} \idea{Translation efficiency is predicted to increase with transcription.}  What does the RLTO model predict about how the cell should balance the gene expression process between transcription and translation? Minimizing transcription (at fixed protein abundance) reduces the metabolic load; however, it decreases robustness. Growth rate maximization balances these two costs. Quantitatively, the maximization of the growth rate (Eq.~\ref{eqn:growthrate2}) with respect to the translation efficiency can  be performed analytically, predicting the optimal translation efficiency,  shown in  Fig.~\ref{fig:proteom}A. We provide an exact expression in the Supplementary Material Sec.~\ref{sec:opti}; however, an approximate expression for the translation efficiency is more clearly interpretable:}
\begin{equation}
\hat{\varepsilon} \approx 0.1 \lambda \hat{\mu}_m. \label{eqnsym}
\end{equation}
The optimal translation efficiency has two important qualitative features for central dogma regulation. The first prediction is that as the message cost ($\lambda$) rises, the optimal translation  efficiency ($\hat{\varepsilon}$) increases in proportion while the message number decreases. We present evidence for this prediction in the Supplementary Material Sec.~\ref{sec:phos}.

The second prediction is that the optimal translation efficiency is also approximately proportional to message number ($\hat{\varepsilon} \propto \mu_m$). Therefore, the RLTO model predicts that low expression levels should be achieved with low levels of transcription and translation, whereas high-expression genes are achieved with high levels of both. We call this relation between optimal transcription and translation the \textit{load balancing principle}. The most direct test of load balancing is measuring the protein-message abundance relation. Due to load balancing, the RLTO model predicts protein number (and proteome fraction) to scale like: 
\begin{equation}
\hat{\mu}_{p} \propto \hat{\mu}_{m}^2, \label{eqn:prop}
\end{equation} 
whereas  a constant-translation-efficiency model has linear scaling ($\mu_{p} \propto \mu_{m}$). Computing proteome fraction, rather than protein number, results in a parameter-free prediction. (See Supplemental Sec.~\ref{sec:ppf}.)  

\begin{figure}
  \centering
    \includegraphics[width=0.45\textwidth]{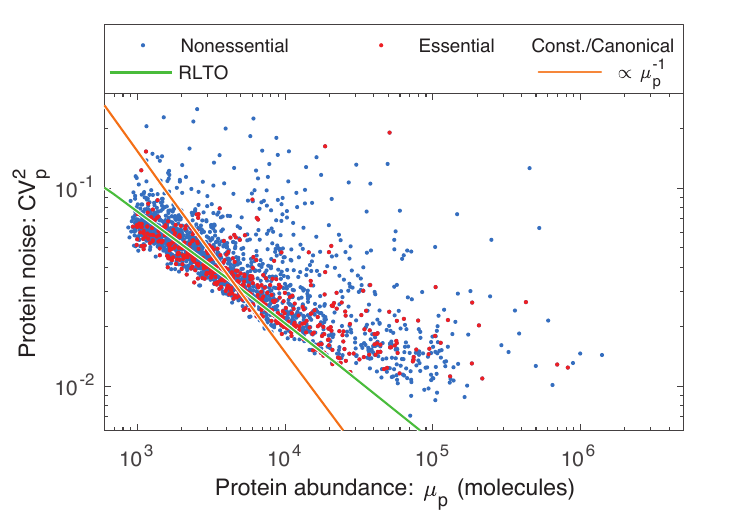}
      \caption{  {\color{red}
         \textbf{RLTO predicts the magnitude of noise in yeast.} The observed gene expression noise is yeast is shown for essential and nonessential genes. Two protein-message abundance models are compared to the data: The RLTO model (green) versus the constant-translation-efficiency (canonical model, orange). The RLTO model predicts both the magnitude of the noise, as well as its scaling with protein abundance. The reduced slope of the RLTO  model  is the consequence of load balancing, which reduces the noise for the noisiest, low-expression genes. (Data from Ref.~\cite{Newman:2006nl}.)
     \label{fig:noisefig}}} %
\end{figure}

{\color{red} \idea{Load balancing is observed in eukaryotic cells.}}  To test the RLTO predictions, we compare observed proteome measurements in three evolutionarily divergent species, \textit{E.~coli} \cite{Balakrishnan:2022ai}, yeast \cite{Ghaemmaghami:2003jj} and mammalian cells \cite{Schwanhausser:2011ec}, to two models: the RLTO  and the constant-translation-efficiency models. The results of the parameter-free predictions are shown in {\color{red} Fig.~\ref{fig:proteom}BCD} for each organism. The RLTO model clearly captures the global trend in the proteome-fraction message-number relation in eukaryotic cells {\color{red} and a direct fit to a power law with an unknown exponent is consistent with  Eq.~\ref{eqn:prop} (Supplementary Material Sec.~\ref{sec:empmodel0}).}

In \textit{E.~coli}, the constant-translation-efficiency model better describes the data. Why does this organism appear not to load balance? In the supplementary material, we demonstrate that the observed translation efficiency is consistent with the RLTO model, augmented by a ribosome-per-message limit. Hausser 
\textit{et al.}\ have proposed just such a limit, based on the ribosome footprint on mRNA molecules \cite{Hausser:2019fi}. 
(See Supplementary Material Sec.~\ref{sec:translationallimit}.) Although this augmented model is consistent with central dogma regulation in \textit{E.~coli}, it is not a complete rationale. This proposed translation-rate limit could be circumvented by increasing the lifetime of \textit{E.~coli} messages, which would increase the translation efficiency. A more in-depth analysis specific to \textit{E.~coli} is needed to understand why the observed message lifetime is so short. 

\idea{RLTO model predicts observed noise in yeast.} Although the protein fraction measurements support the RLTO predictions for the translation efficiency in eukaryotic cells, these measurements do not provide a compelling rationale for why load balancing maximizes the growth rate. To understand its rationale, we explore its implications for noise. 

{\color{red}In a typical biological context, $\mu_m \ll \varepsilon$ and as a result, noise production is dominated by the transcription step of the gene expression process \cite{Paulsson:2000xi,Friedman:2006oh}.  (A table of central dogma parameters for each model organism appears in the Supplementary Material Tab.~\ref{tab1refs}.)
Quantitatively, the \textcolor{orange}{canonical steady-state noise model}  predicts that the noise should be inversely related to the message number \cite{Paulsson:2000xi,Friedman:2006oh}:
\begin{equation}
{\rm CV}_p^2 = \textstyle \frac{\ln 2}{\mu_m},
\end{equation}}
however, it is the relation between mean protein abundance $\mu_p$ and noise ($\inliner{{\rm CV}_p^2}$) which is typically reported \cite{Taniguchi2010,Newman:2006nl}. \textcolor{red}{Based on the scaling of the optimal translation efficiency with the message number in eukaryotic cells (Eq.~\ref{eqnsym}), we find the protein number to scale with message number (Eq.~\ref{eqn:prop}), which} predicts that noise should scale with protein abundance $\inliner{{\rm CV}_p^2 \propto \mu_p^{-1/2}}$ in yeast \textcolor{red}{(see Supplementary Material Sec. \ref{sec:non-canon})}; however, due to the observed absence of translation-efficiency scaling in bacteria, the noise should scale as $\inliner{{\rm CV}_p^2 \propto \mu_p^{-1}}$ in bacteria, as observed \cite{Taniguchi2010}. Does the yeast noise show the predicted scaling? The parameter-free RLTO noise prediction closely matches the observed noise in both magnitude and scaling, as shown in Fig.~\ref{fig:noisefig}.

\idea{Reducing noise is the rationale for load balancing.} This noise analysis also provides a conceptual insight into the rationale for load balancing. The load balanced (RLTO-green) and constant-translation-efficiency (orange) predictions for the noise are shown in Fig.~\ref{fig:noisefig}. Load balancing results in decreased noise for low-expression, noisy genes over what is achieved with constant translation efficiency. This decreased noise is predicted to increase growth robustness. In principle, the noise could be reduced further by tipping the balance even more towards transcription; however, the RLTO model predicts that this approach is too metabolically costly, and the optimal strategy is that observed for noise scaling in yeast.

\begin{figure}
  \centering
   \includegraphics[width=0.48\textwidth]{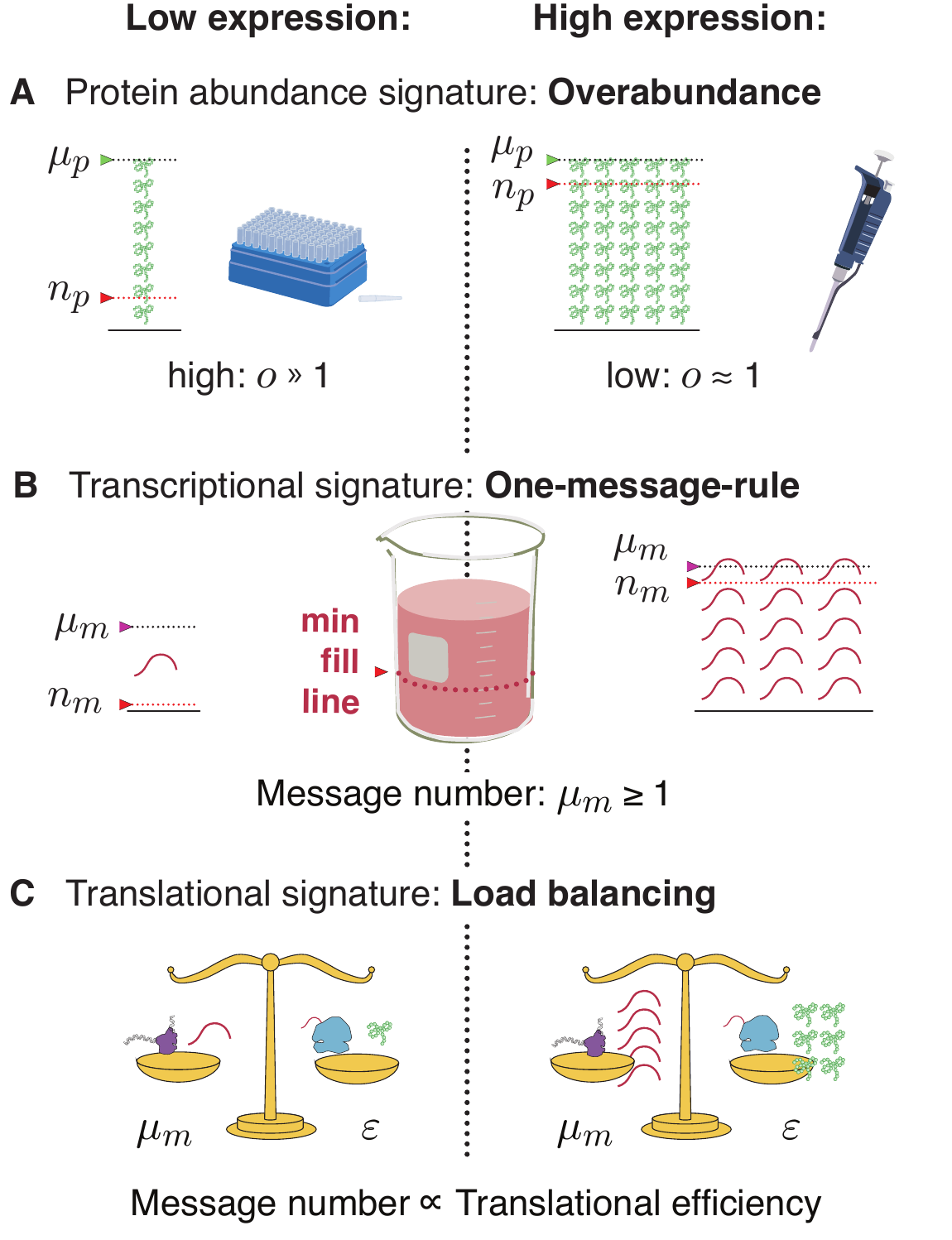}
      \caption{ \textbf{Central dogma regulatory principles. Panel A: Overabundance.}  Low-expression essential genes are expressed with high overabundance; whereas, high-expression essential genes are expressed with low overabundance. {\color{red} Lab supply analogy:  Low-cost items that are used stochastically (\textit{e.g.}\ pipette tips) are purchased in great excess, while the higher cost items that are less stochastic (\textit{e.g.}\ pipette)  are purchased  as needed.}   
     \textbf{Panel B: One-message rule.} Robust expression of essential genes requires them to be transcribed above a threshold of one message per cell cycle. \textbf{Panel C: Load balancing.} In eukaryotic cells, optimal fitness is achieved by balancing transcription and translation: The optimal message number is proportional to the optimal translation efficiency. High (low) expression levels are achieved by high (low) levels of transcription followed by high (low) levels of translation per message. 
      \label{fig:principles} }
\end{figure}

\TocSkip\section*{Discussion}

{\color{red}
\idea{What are the biological implications of noise?} Many important proposals have been made, including bet-hedging strategies, the necessity of feedback in gene regulatory networks, \textit{etc.}\ \cite{Raser:2005we}.  Our model suggests that  overcoming cell-to-cell variation  may fundamentally reshape the metabolic budget: Typically, proteins constitute 50-60\% of the dry mass of the cell \cite{Lengeler1998} and therefore overabundance could increase the overall protein budget by a significant factor. Why does the cell tolerate this significant increase in metabolic load above what would be predicted by a resource allocation analysis (\textit{e.g.}~\cite{Dourado2021})? This \textcolor{orange}{strategy} dramatically reduces the consequence of stochastic expression of proteins on the rate of single-cell proliferation.


A second source of stochasticity, environmental fluctuations, has been proposed as a rationale for overabundance \cite{Mori:2017al}, especially in the context of metabolic genes \cite{Lambert:2014tp}. In short, cells express protein to hedge against starvation \cite{Mori:2017al} or changes in the carbon source, \textit{etc.}\ \cite{Lambert:2014tp}. 
How does this hypothesis compare to our growth robustness hypothesis? There are some similarities between these environmental-fluctuation models and the RLTO model: In both models, it is a fluctuations-based mechanism that drives overabundance; however, there are important distinctions between the model predictions. In the environmental fluctuation model, there is a trade-off between log-phase fitness and the rapidity of adaptation \cite{Mori:2017al}; whereas in the RLTO model, overabundance corresponds to the log-phase optimum. Organisms experiencing prolonged periods of balanced growth would therefore be expected to reduce overabundance. 
Furthermore, the environmental fluctuation model most naturally explains overabundance for proteins related to metabolic processes, whereas the RLTO model predicts overabundance generically, dependent only on message number, which appears to be much more consistent with experiments exploring essential-protein depletion \cite{Gallagher:2020sp}.
}

\idea{Implications for nonessential genes.} In our analysis, we have focused on essential genes in order to motivate the growth-threshold in the RLTO model. To what extent do nonessential genes share the same optimization? In support of the proposal that RLTO optima describe nonessential genes is the success of the model in predicting the translation efficiency for all genes, not just essential genes. (See Fig.~\ref{fig:proteom} and Fig.~\ref{fig:noisefig}.) 
Furthermore, the definition of a gene as \textit{essential} depends on context: For instance, in the context of \textit{E.~coli} growth on lactose, the gene \textit{lacZ} is essential, although it is nonessential on other carbon sources \cite{MONOD:1952hr}. Under growth conditions where the \textit{lacZ} gene is essential, we predict that LacZ should be overabundant, consistent with observation \cite{Lambert:2014tp}. Finally, our modeling suggested that RLTO model phenomenology is the results of asymmetry of the cost of under versus overabundance. For nonessential genes whose activity significantly increases fitness, we still expect fitness asymmetry due to the low relative metabolic cost of increased expression.  
We therefore expect all gene products, most especially those with low expression, to be overabundant, under conditions where their activity increases fitness.

\idea{Implications of overabundance for inhibitors.} The generic nature of overabundance, especially for low-expression proteins, has important potential implications for the targeting of these proteins with small-molecule inhibitors (\textit{e.g.}\ drugs). 
For the highest expression proteins, like the constituents of the ribosome, relatively small decreases in the active fraction (\textit{e.g.}~a three-fold reduction) are expected to lead to growth arrest \cite{Peters:2016jf}. This may help explain why inhibitors targeting translation make such effective antimicrobial drugs.
However, we predict that the lowest expression proteins require a much  higher fraction of the protein to be inactivated, with the lowest-expression proteins expected to need more than a 100-fold depletion. This predicted robustness  makes these  proteins  much less attractive drug targets 
\cite{Bosch:2021gd}.

\idea{The principles that govern central dogma regulation.}  We propose that robustness to noise fundamentally shapes the central dogma regulatory program for all genes and predicts
 a number of key regulatory principles. (See Fig.~\ref{fig:principles}.) 
For high-expression genes, load balancing implies that gene expression consists of both high-amplification translation and transcription.
The resulting expression level has low overabundance relative to the threshold required for function. In contrast, for essential low-expression genes, a three-fold strategy is implemented: (i) overabundance raises the mean protein levels far above the threshold required for function, (ii) load balancing, and (iii) the one-message rule \textcolor{orange}{ensures} that message number is sufficiently large to lower the noise of  inherently-noisy, low-expression genes. 
We anticipate that these regulatory principles, in particular protein overabundance, will have \textcolor{red}{important implications}, not only for our understanding of central dogma regulation specifically, but for understanding the rationale for protein expression level and function in many biological processes.

{\color{red}
\TocSkip\section*{Materials and Methods}

\idea{RLTO model.} The effect of stochastic cell arrest can be implemented analytically as follows:  The probability of growth is the probability that all essential proteins are above threshold, $P_+$.
The population growth rate $k$ is \cite{Huang:2022fu}:
\begin{equation}
\textstyle\frac{k}{k_0} =  1 +\textstyle\frac{1}{\ln 2 }\ln P_+, \label{eqn:robust}
\end{equation} 
for a population of cells subject to stochastic arrest with probability $1-P_+$ per cell cycle where $k_0$ is the growth rate of the non-arrested cells. For each gene $i$, the \textcolor{orange}{canonical steady-state noise model}  predicts the protein number CDF in terms of message number $\mu_m$ and translation efficiency $\varepsilon$ \cite{Paulsson:2000xi}.
Assuming the below-threshold probability is small, the probability that the cell is below threshold for gene $i$ is:
\begin{equation}
\ln P_{+,i} = -\gamma( \textstyle\frac{\mu_{m}}{\ln 2},\textstyle\frac{n_{p}}{\varepsilon \ln 2}), 
\end{equation}
where $\gamma$ is the regularized incomplete gamma function and the CDF of the gamma distribution. (See Supplementary Material Sec.~\ref{Sec:RLTOderivation}.)

While protein underabundance slows cell growth by the arrest of essential processes, protein overabundance slows growth by increasing the metabolic load. To implement the metabolic-load contribution to cell fitness, we use a minimal model that realizes the metabolic cost of both transcription and translation that is analogous to those previously used in the context of resource allocation (\textit{e.g.}~\cite{Scott:2010ec}). 
The metabolic load of transcription and translation of gene $i$ is:  
\begin{equation}
\textstyle\frac{k}{k_0} =  1-  \textstyle \frac{\lambda+\varepsilon}{N_0} \mu_m, \label{eq:metaload}
\end{equation}
where $k_0$ is the growth rate in the absence of the metabolic load of gene $i$, $N_0$ is the total cellular metabolic load, and $\lambda$ is the metabolic message cost. (See the Supplementary Material Sec.~\ref{Sec:RLTOderivation} for a detailed development of the model.) The $\lambda$-term represents the metabolic cost of transcription and the $\varepsilon$-term represents the metabolic cost of translation of gene $i$. We define the relative load as $\Lambda \equiv \lambda/N_0$ as the ratio of the metabolic load of a single message to the total metabolic cost of the cell. In \textit{E.~coli}, we estimate that  $\Lambda$ is roughly $10^{-5}$ and it is smaller still for eukaryotic cells. 
 
\idea{Data analysis.} We provide a detailed description of the data analysis for the one message rule, loading balancing, and the noise analysis in the Supplementary Material.





}

\idea{Acknowledgments.} The authors would like to thank B.~Traxler, A.~Nourmohammad, J.~Mougous, K.~Cutler, M.~Cosentino-Lagomarsino, S.~van Teeffelen, C.~Manoil, L.~Gallagher, J.~Bailey and S.~Murray. 

\noindent
\textbf{Funding:}
\noindent
National Institutes of Health grant R01-GM128191

\noindent
\textbf{Author contributions:}
\noindent
T.W.L., H.J.C., D.H. and P.A.W.~conceived the research, performed the analysis, and  wrote the paper. 

\noindent
\textbf{Competing interests:}
\noindent
The authors declare no competing interests.

\idea{Data availability.} All data needed to evaluate the conclusions in the paper are present in the paper and/or the Supplementary Materials.

\idea{Supplementary Materials}

\noindent
Supplementary Text

\noindent
\textcolor{red}{Figs. S1 to S12}

\noindent
\textcolor{red}{Tables S1 to S3}

\noindent
\textcolor{red}{Data Tables S1 to S11}


\bibliographystyle{Versions/Science}
\TocSkip\bibliography{message2}

\onecolumngrid

\newpage 


\setcounter{page}{1}
\beginsupplement{

\title{Supplementary Material: Noise robustness and metabolic load determine the principles of central dogma regulation}
\maketitle
\onecolumngrid

\appendix 

\tableofcontents

\bigskip
\bigskip
\bigskip

\twocolumngrid


\section{Detailed development of the RLTO model}

In this section, we provide a detailed development of the RLTO model. First, \textcolor{red}{we describe the stochastic kinetic model for the central dogma, which introduces key quantities for the RLTO model (Sec.~\ref{sec:noise_intro}). Next,} we provide a derivation of the growth rate as a function of the model parameters (Sec.~\ref{Sec:RLTOderivation}) as well as other methods (Secs.~\ref{sec:opti}, \ref{sec:ppf}, \ref{sec:incproload}, and \ref{sec:estimateofLambda}). For each of the results discussed in the main paper, we provide more detailed analyses, which include both supplemental results (Secs.~\ref{sec:loadbals}, \ref{sec:TransEff_s}, \ref{sec:phos}, \ref{sec:euklbs}, and \ref{sec:protcost_s}) that support the story described in the main paper, as well as supplemental discussions (Secs.~\ref{sec:discland_s}, \ref{sec:rationale}, \ref{sec:bacteria}, \ref{sec:depletion}, \ref{sec:Hausser}, and \ref{sec:translationallimit}). 

{\color{red}
\subsection{Methods: Detailed description of the noise model} \label{sec:noise_intro}

\subsubsection{Stochastic kinetic model for the central dogma.}  

The \textcolor{red}{\textcolor{orange}{canonical steady-state noise model}} for the central dogma describes multiple steps in the gene expression process \cite{Paulsson:2000xi,Friedman:2006oh,Taniguchi2010}:
 Transcription generates mRNA messages.  These messages are then translated to synthesize the protein gene products \cite{Crick:1970vy}.  Both mRNA and protein are subject to degradation and dilution \cite{Hargrove:1989vo}. 
 At the single cell level, each of these processes are stochastic. We will model these processes with the stochastic kinetic scheme   \cite{Crick:1970vy}:
\begin{equation}
  \begin{CD}
   {\rm DNA} @>\beta_m>> {\rm mRNA} & @>\beta_p>>  {\rm Protein} \\
     & & @V\gamma_mVV & @V\gamma_pVV\\
     & & \varnothing && &  \; \varnothing, \label{stochmodel}
  \end{CD}
\end{equation}
where $\beta_m$ is the transcription rate (s$^{-1}$), $\beta_p$ is the translation rate (s$^{-1}$), $\gamma_m$ is the message  degradation rate (s$^{-1}$), and $\gamma_p$ is the  protein effective degradation rate (s$^{-1}$).  The message lifetime is $T_m\equiv \gamma_m^{-1}$. For most proteins in the context of rapid growth, dilution is the dominant mechanism of protein depletion and therefore $\gamma_p$ is approximately the growth rate  \cite{KOCH:1955oa, Martin-Perez:2017jx, Taniguchi2010}: $\gamma_p = T^{-1}\ln 2$,  where $T$ is the doubling time. 

\label{sectelegraph}

\subsubsection{Statistical model for  protein abundance.} 

\label{sec:statnoiseS}
To study the stochastic dynamics of gene expression, we used a stochastic Gillespie simulation \cite{Gillespie1992,Gillespie1977}. (See Sec.~\ref{SecGS}.) In particular, we were interested in the explicit relation between the kinetic parameters $(\beta_m, \gamma_m, \beta_p, \gamma_p)$ and experimental observables.

Consistent with previous reports \cite{Paulsson:2000xi,Friedman:2006oh}, we find that the distribution of protein number per cell (at cell birth) was described by a gamma distribution:
$N_p \sim \Gamma(a,\theta)$, where $N_p$ is the protein number at cell birth and $\Gamma$ is the gamma distribution which is parameterized by a scale parameter $\theta$ and a shape parameter $a$. 
(See Sec.~\ref{sec:gammaconv}.) We refer to this distribution as the \textit{\textcolor{orange}{canonical steady-state noise model}};
The relation between the four kinetic parameters and these two statistical parameters has already been reported, and have clear biological interpretations \cite{Friedman:2006oh}: The scale parameter: 
\begin{eqnarray} 
\theta = \varepsilon \ln 2, \label{eqn:eff_rates} 
\end{eqnarray}
is proportional to the translation efficiency:
\begin{eqnarray}
\varepsilon \equiv \textstyle \frac{\beta_p}{\gamma_m}, \label{eqn:TE}
\end{eqnarray} 
where  $\beta_p$ is the translation rate and $\gamma_m$ is the message degradation rate.   $\varepsilon$ is understood as the mean number of proteins translated from each message transcribed. 
The shape parameter $a$ can also be expressed in terms of the kinetic parameters \cite{Friedman:2006oh}: 
\begin{equation}
a = \textstyle\frac{\beta_m}{\gamma_p }; \label{eqn:shape2}
\end{equation}
however, we will find it more convenient to express the scale parameter in terms of the cell-cycle message number:
\begin{eqnarray}
    \mu_{m} \equiv \beta_m T =  a \ln 2,
\end{eqnarray}
which can be interpreted as the mean  number of messages transcribed per cell cycle. Forthwith, we will abbreviate this quantity \textit{message number} in the interest of brevity.

\begin{figure}
    \centering
    \includegraphics[width=0.45\textwidth]{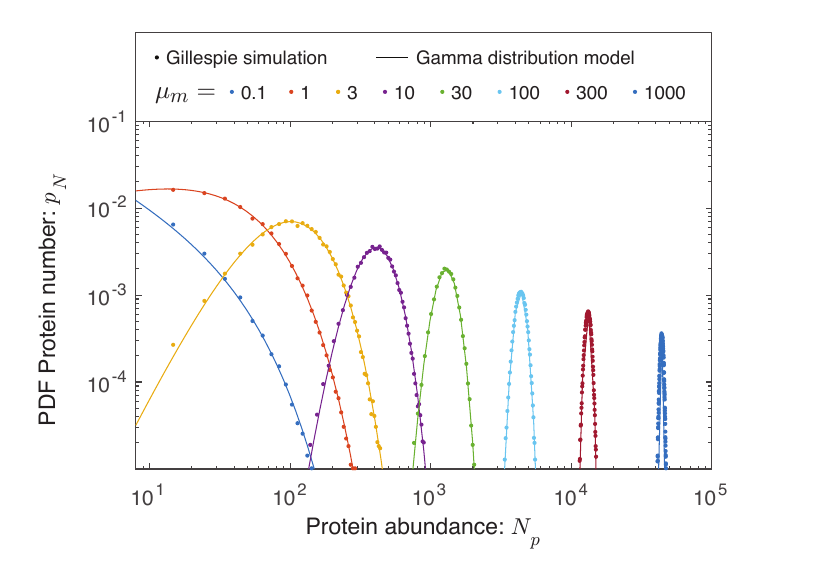}
    \caption{\textbf{The protein abundance is approximately gamma distributed.} Protein abundance was modeled for eight different transcription rates using a Gillespie simulation, including the stochastic partitioning of the proteins between daughter cells at cell division. The range in abundance matches the observed range of expression levels in the cell. 
      We observed that the simulated protein abundances were well fit by gamma distributions.  }
    \label{fig:gillespie}
\end{figure}

\subsubsection{Gillespie simulation of stochastic kinetic scheme}
\label{SecGS}


Protein distributions based on the kinetic scheme defined in Sec. \ref{sectelegraph}  were simulated with a Gillespie algorithm, \textcolor{red}{with specific parameter values for \textit{E.~coli}}. Assuming the lifetime of the cell cycle ($T = 30$ min) \cite{Bernstein:2002rp}, mRNA lifetime ($T_m = 2.5$ min)  \cite{Chen:2015wt}, and translation rate ($\beta_p \approx 500$ hr$^{-1}$), the protein distributions for several mean expression levels were numerically generated for exponential growth with 100,000 stochastic cell divisions, with protein partitioned at division following the binomial distribution. 

The gamma distributions for each mean message number with scale and shape parameters determined by the corresponding translation efficiency and message number  ($\theta = \varepsilon \ln 2 $, $a = \frac{\mu_m}{\ln 2} $) as used for the Gillespie simulation were also plotted with the protein distributions. We observe an excellent match between these Gillespie simulations and the \textcolor{orange}{canonical steady-state} statistical noise model (\textit{i.e.}\ gamma function)  as shown in Fig.~\ref{fig:gillespie}.

\subsubsection{Gamma function and distribution conventions}
\label{sec:gammaconv}
There are a number of  conflicting conventions for the gamma function and distribution arguments. We will use those defined on Wikipedia and the CRC Encyclopedia of Mathematics \cite{Weisstein2009}. The gamma distributed random variable $X$ will be written:
\begin{equation}
    X\sim \Gamma( a,\theta),
\end{equation}
where $a$ is the shape parameter and $\theta$ is the scale parameter. The PDF of the distribution is:
\begin{equation}
p_\Gamma(x|a,\theta) \equiv \textstyle \frac{x^{a-1}}{\theta^{a}\Gamma(a)} e^{-x/\theta},
\end{equation}
where $\Gamma(a)$ is the gamma function. The CDF is therefore:
\begin{eqnarray}
P_\Gamma( x|a,\theta) &=& \int_0^x {\rm d}x'\ p_\Gamma(x'|a,\theta),\\ 
&=& P_\Gamma( \textstyle \frac{x}{\theta}|a,1),\\
&=& \int_0^{x/\theta} {\rm d}x'\ \textstyle \frac{x^{a-1}}{\Gamma(a)} e^{-x},\\
&=& \gamma( a, x/\theta),
\end{eqnarray}
where $\gamma$ is the regularized incomplete gamma function. } 

\subsection{Methods: The derivation of the RLTO growth rate}

\label{Sec:RLTOderivation}

{\color{red}

\subsubsection{The metabolic load of protein and the resource allocation model}

To model the effect of the metabolic load on cell growth, we will expand on a model used by Hwa and co-workers \cite{Scott:2010ec}. For conciseness, we will call the original model the \textit{\textcolor{orange}{resource allocation} model}. 

Consider a cell where the total number of proteins in the cell is $N$. The synthesis of these proteins requires two sets of processes: (i) the metabolic processes responsible for synthesizing the precursors (\textit{i.e.}~amino acids, \textit{etc}) and (ii) the translation process. Proteins involved in the metabolic processes be referred  to as the $P$ sector and number $N_P$. The proteins involved in the translational process will be referred to as the $R$ sector and number $N_R$. In addition to the P and R sectors, there is a third  Q sector with protein number $N_Q$. The total protein number per cell is therefore:
\begin{equation}
N = N_R+N_P+N_Q.
\end{equation}
The proteome fractions are defined $\Phi_X \equiv N_X/N$ and have the normalization condition:
\begin{equation}
1 = \Phi_R+\Phi_P+\Phi_Q. \label{eqn:norm}
\end{equation}
The key assumption in the \textcolor{orange}{resource allocation model} is that the abundances of the R and P sectors can change in size to accommodate changes in the nutrient quality and translation load associated with a particular growth condition \cite{Scott:2010ec}. In contrast, the Q sector has a fixed proteome fraction, $\Phi_Q$, irrespective of growth conditions.  In the \textcolor{orange}{resource allocation model}, the size of these adjustable R and P sectors are chosen to optimize the growth rate $k$.

The condition for balanced growth requires that the overall protein output of the translation process match the growth rate: 
\begin{equation}
kN = k_R N_{*R},  \label{eqn:hwarate} 
\end{equation}
where $k_R$ is the effective translation rate per protein and $N_{*R}$ is the number of productive R sector proteins, which is subset of the total number $N_R$:
\begin{equation}
N_R = N_{*R}+N_{0R},   
\end{equation}
and $N_{0R}$ represents unproductive R sector protein.
We can rewrite Eq.~\ref{eqn:hwarate} in terms of the proteome fraction:
\begin{equation}
k = k_R (\Phi_R-\Phi_{0R}). \label{eqn:fracr}  
\end{equation}
In the expression above, we will assume that the parameters $k_R$ and $\Phi_{0R}$ are fixed,  but the total fraction $\Phi_R$ is chosen to optimize the growth rate $k$. 

For any productive sectors $i$, we will write analogous equations to Eq.~\ref{eqn:fracr} linking sector fraction size $\Phi_i$ to function:
\begin{equation}
k = k_i (\Phi_i-\Phi_{0i}), \label{eqn:general} 
\end{equation}
where, as before, $\Phi_{0i}$ represents a fixed-size fraction of unproductive protein.

To determine the unknown optimum growth rate and sector sizes, Eq.~\ref{eqn:general} can be re-written:
\begin{equation}
\Phi_{*i} = \Phi_i-\Phi_{0i} = \textstyle \frac{k}{k_i},  \label{eqn:phiki}
\end{equation}
and then summed over all sectors (excluding Q):
\begin{equation}
\Phi_* = k \sum_{i\ne Q} k_i^{-1}, \label{growthrate}
\end{equation}
where we  define the total productive fraction of the proteome:
\begin{equation}
\Phi_{*} \equiv \sum_{i\ne Q} \Phi_{*i} = 1-\Phi_0,    
\end{equation}
and $\Phi_0$ represents the total fraction of unproductive protein:
\begin{equation}
\Phi_0 = \Phi_Q + \sum_{i\ne Q} \Phi_{0i},
\end{equation}
including the entire Q sector. 
 Since $\Phi_Q$ and the $\Phi_{0i}$ are all assumed to be fixed, Eq.~\ref{growthrate} determines the growth rate. (In \textit{E.~coli}, Hwa and coworkers estimate that $\Phi_{*} \approx 0.55$.)

To understand the meaning of Eq.~\ref{growthrate}, we first define an ideal growth rate as 
\begin{equation}
k_{\rm ideal}^{-1} = \sum_{i\ne Q} k_i^{-1},
\end{equation}
which would be the growth rate in the absence of unproductive protein; however, due to the presence of the unproductive protein, the growth rate is proportional to the productive fraction:
\begin{equation}
k = k_{\rm ideal} \Phi_{*}.
\end{equation}
The optimal protein fractions can be determined using Eq.~\ref{eqn:phiki}:
\begin{equation}
\Phi_i = \textstyle \frac{k_{\rm ideal}}{k_i} \Phi_{*} +  \Phi_{0i}.
\end{equation}

How does the growth rate change when the unproductive protein fraction is changed by $\delta \Phi$? The productive fraction is reduced:
\begin{equation}
\Phi_*\rightarrow \Phi_*' = \Phi_*-\delta \Phi.
\end{equation}
The ratio of the new growth rate $k'$ to the original is therefore:
\begin{equation}
\textstyle \frac{k'}{k} = 1-\textstyle \frac{\delta \Phi}{\Phi_*.} \label{eqn:almost}
\end{equation}
Note that our generalized \textcolor{orange}{resource allocation model} is written for arbitrary number of functional sectors $i\ne Q$ and the key determinant of the change in growth rate is the fraction of productive protein $\Phi_*$. This equation will be used to model the fitness cost of the metabolic load.

\subsubsection{The metabolic load of mRNA}

What is the cost of transcription? It is perhaps useful to first consider the estimates of biosynthetic cost of macromolecules in the cell in descending order in \textit{E.~coli} \cite{phillips2013physical}: 

\medskip

\begin{center}
    \begin{tabularx}{0.4 \textwidth}{X|r}
    \textbf{Macromolecule} & \textbf{Biosynthetic cost} \\
    & ($10^9$ ATP) \\ 
    \hline
Protein & 4.5 \\ 
Phospholidid & 3.2  \\
RNA &  1.6 \\
Lipopolysacchride & 3.8 \\
DNA & 0.35 \\
Peptidoglycan & 0.17 \\
Glycogen & 0.03
    \end{tabularx}
\end{center}

So clearly the cost of RNA  is itself not insignificant. Although a significant fraction of the RNA is rRNA rather than mRNA, the mRNA itself in \textit{E.~coli} undergoes multiple rounds of transcription due to its short lifetime, increasing its cost to what is required to synthesize the molecules observed in the \textit{E.~coli} cell at any time $t$. 
Furthermore,  transcription is dependent on protein enzymes, which themselves must be synthesized. We therefore conclude from this estimate that the cost of transcription is likely a significant determinant of the metabolic load.

Experimentally, Kafri and coworkers have measured the fitness cost of transcription and translation independently using the DAmP (Decreased Abundance by mRNA Perturbation) system
in \textit{Saccharomyces cerevisiae} \cite{Kafri:2016fu}. As expected, they report that the metabolic cost of transcription is comparable to translation and that the reduction in growth rate is linear in transcription, in close analogy to Eq.~\ref{eqn:almost}.

\subsubsection{Metabolic load in the RLTO model}

To produce a minimal model to study the trade-off between robustness and metabolic load, we must consider both the metabolic cost of transcription and translation. We will write that the metabolic load (in protein equivalents) associated with gene $i$ is:
\begin{equation}
\delta N_i = \lambda \mu_{m,i}+ \mu_{p,i},
\end{equation}  
where $\lambda$ is the message cost, the  metabolic load associated with an mRNA molecule relative to a single protein molecule of the gene product.   $\mu_{m,i}$ is the mean number of messages transcribed per cell cycle (mRNA molecules per cell cycle) for gene $i$.  $\mu_{p,i}$ is the mean number of protein translated per cell cycle for gene $i$.
We will describe the mean protein number in terms of the translation efficiency $\varepsilon_i$, the number of proteins translated per message: \begin{equation}
\mu_{p,i} = \varepsilon_i \mu_{m,i}.   \label{eqn:protnumS}
\end{equation}

How does the cell growth rate change due to the metabolic load associated with the expression of gene $i$? The change in the metabolic load is:
\begin{equation}
\delta \Phi_i = \textstyle \frac{\delta N_i}{N},    
\end{equation}
where $N$ represents the total metabolic load of all components of the cell, in units of protein equivalents. Using Eq.~\ref{eqn:almost}, the resulting change in growth rate is:
\begin{equation}
\textstyle \frac{k}{k_0} = 1-\textstyle \frac{(\lambda + \varepsilon_i)\mu_{m,i}}{\Phi_* N}, \label{eqn:almost3}
\end{equation}
where $k_0$ is the growth rate in the absence of the metabolic load of gene $i$.

In our analysis, the exact size of the total metabolic load $N$ will not be important. In the interest of simplicity we will therefore adsorb the productive fraction $\Phi_*$ into an effective 
total metabolic load: 
\begin{equation}
N_0 \equiv \Phi_* N,
\end{equation}
and write a concise relation between the load from gene $i$ and the growth rate:
\begin{equation}
\textstyle \frac{k}{k_0} = 1-\textstyle \frac{(\lambda + \varepsilon_i)\mu_{m,i}}{N_0}. \label{eqn:almost2}
\end{equation}
This equation has an intuitive interpretation: growth slows in proportion to the relative added metabolic load.
Since $\Phi_*$ is order unity, we will ignore the distinction between the $N$ and  $N_0$ quantities hence forth.

Although the global parameters $N_0$ and  $\lambda$ provide an intuitive representation of the model, the relative growth rate depends on fewer parameters. Let $k$ and $k_0$ be the growth rates in the presence and absence of the metabolic load of gene $i$. The relative growth rate is:  
\begin{equation}
\textstyle\frac{k}{k_0} =  1 -  (\Lambda+E_i) \mu_{m,i}, 
\end{equation}
where we have introduced two new reduced parameters: the relative load, defined as $\Lambda \equiv \lambda/N_0$, represents the ratio of the metabolic load of a single message to the total load and the relative translation efficiency, defined $E_i \equiv \varepsilon_i/N_0$, which is the ratio of the number of proteins translated per message to the total metabolic load $N_0$.  (Note that due to the high multiplicity $N_0 \gg (\lambda+\varepsilon_i) \mu_{m,i}$, we can ignore the distinction between $N_0$ and $N_0'$ in the denominator.) If we neglect the difference between the total metabolic load and the number of proteins, the proteome fraction for gene $i$ is: 
\begin{equation}
\Phi_i = E_i\mu_{m,i}.  \label{eqn:protomeFracS}   
\end{equation}
Both reduced parameters, $\Lambda$ and $E_i$ are extremely small. In \textit{E.~coli}, we estimate that both $\Lambda$ and $E_i$ are roughly $10^{-5}$ and they are smaller still for eukaryotic cells. (See Sec.~\ref{sec:estimateofLambda}.)

\label{sec:derivationgr}
\subsubsection{Growth rate with stochastic arrest}
For completeness, we provide a derivation of the growth rate with stochastic cell-cycle arrest that we have previously described \cite{Huang:2022fu}. Starting from the exponential mean expression for the population growth rate \cite{Huang:2022fu}:
\begin{equation}
   k = \textstyle \frac{\ln 2}{ \overline{T}}, \label{growthraterr}
\end{equation}
where $k$  is the population growth rate and 
\begin{equation}
 \overline{T} \equiv -\frac{1}{k} \ln \mathbb{E}_T \exp( -kT ), \label{expmeaneqn}
\end{equation}
is the exponential mean, where $T$ is the stochastic cell cycle duration and $\mathbb{E}$ is the expectation operator \cite{Huang2022b,Huang:2022fu}. \textcolor{orange}{We take a coarse-grained model which considers changes in growth rate due to fluctuations in protein number to be negligible. Note that Eq.~\ref{expmeaneqn} is equivalent to the \textit{Euler-Lotka} equation \cite{Levien2021,Powell1956}.}  

Let $P_+$ be the probability of growth. When the cells are growing, the cell cycle duration $\tau$ is determined by the metabolic load predictions (Eq.~\ref{eqn:almost2}). The probability mass function is therefore:
\begin{equation}
p_T(t) = \begin{cases} P_+, & t = \tau \\
(1-P_+), & t \rightarrow \infty 
\end{cases}.
\end{equation}
Evaluating the expectation in Eq.~\ref{expmeaneqn} gives:
\begin{equation}
\overline{T}  = -k\ln P_++\tau.     
\end{equation}
Using Eq.~\ref{growthraterr}, we can solve for the growth rate $k$:
\begin{equation}
k = \tau^{-1} \ln (2P_+).    \label{eqn:dergrowth} 
\end{equation}
As expected, the growth rate goes down as the probability of growth $P_+$ decreases, stopping completely at $P_+=\frac{1}{2}$.
We can then compute the ratio of the growth with ($k$) and without arrest ($k_0$):
\begin{equation}
\textstyle\frac{k}{k_0} = 1+\textstyle\frac{1}{\ln 2}\ln P_+.    \label{eqn:dergrowth2} 
\end{equation}
where $k_0$ is computed by evaluating Eq.~\ref{eqn:dergrowth} at $P_+=1.$

\begin{figure*}
  \centering
    \includegraphics[width=0.95\textwidth]{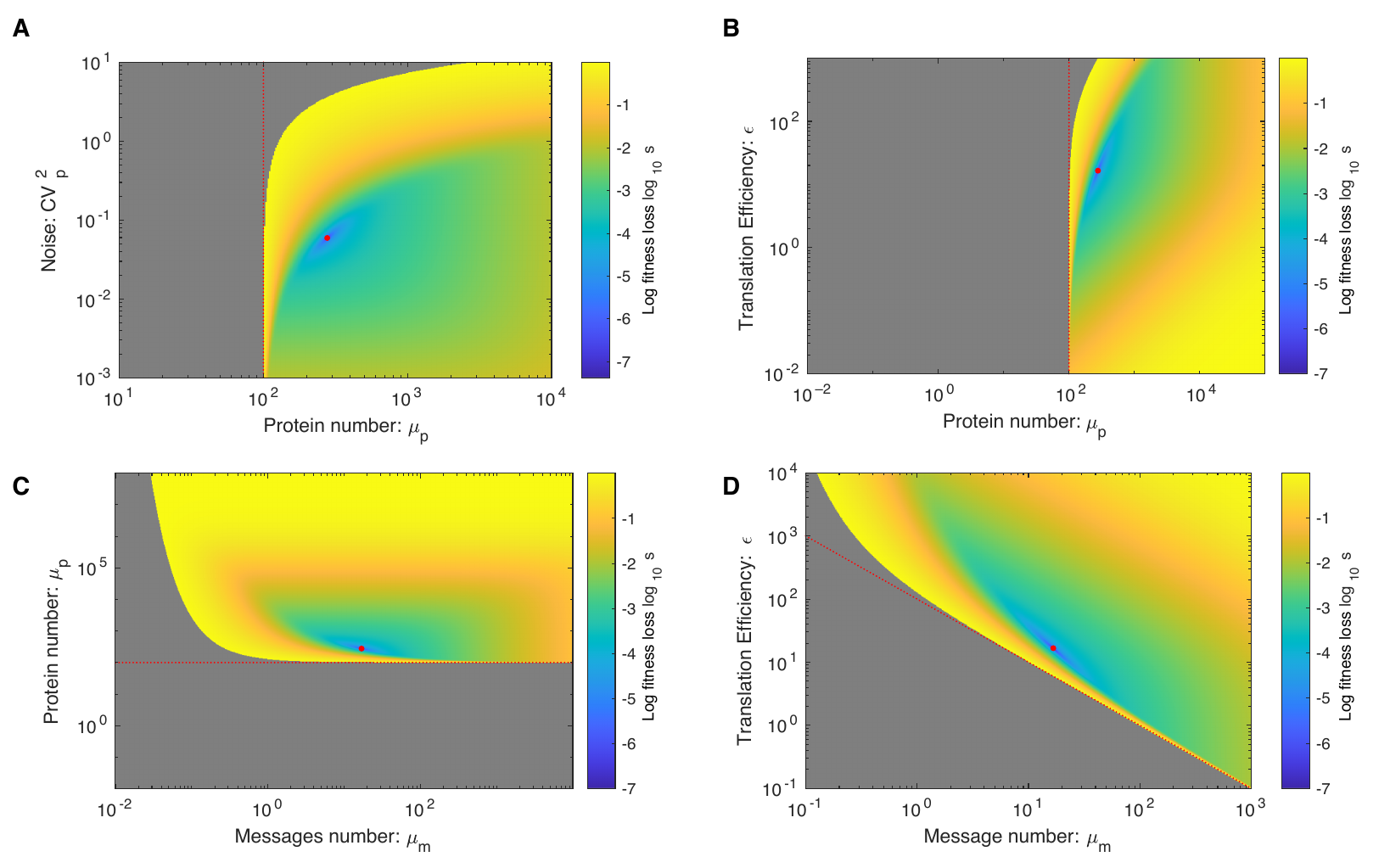}
      \caption{ \textbf{Four perspectives on the fitness landscape.} In each landscape, the red circle represents the fitness optimum. The red dotted line represents the mean protein number equal to the protein threshold $n_p=10^2$.  Here, fitness is quantified by the log growth rate: $s = \ln k_{\rm max}/k$.  \textbf{Panel A: Mean protein number versus noise.} \textbf{Panel B: Mean protein number versus translation efficiency.} \textbf{Panel C: Mean message  versus protein numbers.} \textbf{Panel D: Message number versus translation efficiency.} 
       \label{fig:over_combo1}}
\end{figure*}

\subsubsection{RLTO  growth rate}

In the RLTO model, we will assume the probability of growth is the probability that all essential protein numbers are above threshold.  We will further assume that each protein number is independent, and therefore: 
\begin{equation}
P_+ = \prod_{i\in {\cal E}} {\rm Pr}\{ N_{p,i}>n_{p,i}\},
\end{equation}
where ${\cal E}$ is the set of essential genes. Clearly, this assumption of independence fails in the context of polycistronic messages. We will discuss the significance of this feature of bacterial cells  elsewhere, but we will ignore it in the current context.

As we will discuss, the probability of arrest of any protein $i$ to be above threshold is extremely small. It is therefore convenient to work in terms of the CDFs which are very close to zero:
\begin{eqnarray}
\ln P_+ &=& \sum_{i\in {\cal E}} \ln \left(1-{\rm Pr}\{ N_{p,i}<n_{p,i}\}\right), \label{eqn:full}\\
&\approx& -\sum_{i\in {\cal E}} {\rm Pr}\{ N_{p,i}<n_{p,i}\},\\
&=& -\sum_{i\in {\cal E}} \textstyle \gamma( \frac{\mu_{m,i}}{\ln 2},\frac{n_{p,i}}{\varepsilon_i\ln 2}), \label{eqn:gammader}
\end{eqnarray}
where $\gamma$ is the regularized incomplete gamma function and the CDF of the gamma distribution (see ~\ref{sec:statnoiseS} and \ref{sec:gammaconv}). 
}

\subsubsection{When the ln approximation is avoided...}

The approximation discussed in the previous section is extremely well justified at the optimal central dogma parameters; however, there are a set of figures where we cannot use it. In the fitness landscape figures (Fig.~\ref{fig:over_combo1}), we compute the fitness not just at the optimal values but far from them. {\color{red} When  cell arrest has a large effect on the growth rate, we cannot approximate the natural log with a series expansion, and we must use the full expression in Eq.~\ref{eqn:full}.}

{\color{red}

\subsubsection{Single-gene equation}
By summing the fitness losses from the metabolic load and cell arrest (Eqs.~\ref{eqn:almost2}, \ref{eqn:dergrowth2},  and \ref{eqn:gammader}), we can write an expression for the growth rate including contributions from essential gene $i$:
\begin{eqnarray}
\textstyle \frac{k}{k_0} &=& 1-\textstyle\frac{1}{N_0} (\lambda+ \varepsilon_i) \mu_{m,i}   + ...\nonumber \\
&\ &  -\textstyle \frac{1}{\ln 2} \gamma( \frac{\mu_{m,i}}{\ln 2},\frac{n_{p,i}}{\varepsilon\ln 2}), \label{eq:paw1}
\end{eqnarray}
where the second term on the RHS represents the fitness loss due to the metabolic load and the third term represents the fitness loss due to stochastic cell arrest due to protein $i$ falling below threshold.
The fitness landscape for different gene expression parameters is shown from four different perspectives in Fig.~\ref{fig:over_combo1}. From this point forward, we will drop the subscript $i$ for the sake of brevity unless otherwise noted.
}

\subsubsection{Summary of RLTO parameter values for figures}

The parameter values for the RLTO model used for each figure in the paper are shown in Tab.~\ref{tab:my_label}. 

{\color{red}
\begin{table*}[]
    \centering
\resizebox{\textwidth}{!}{    \begin{tabular}{ |c|c|c|c|c|c|c|}
    \hline
                         Figure  &        Relative load: & Protein cost:         & Protein threshold: & Message number:              & Translation efficiency:  & Noise floor  \\
                          number &           $\Lambda$      &    $\lambda$                &  $n_p$             & $\mu_m$                      &  $\varepsilon$       & $C_0$     \\
                     and panel:  &         (No units)   &   (molecules$^{-1}$)  &  (molecules)       & (molecules/cell cycle)       &  (No units)    & (No units)            \\
 \hline
         \ref{fig:overep}A  &           $10^{-5}$     &     $10$             &     $10^2$         &     Range                    &     Range                & 0 \\
        \ref{fig:overep}C        &     Range            &        NA             &     Range          &     Local optimum              &    Local optimum         & 0        \\
\hline
         \ref{fig:proteom}A      &  Weak dependence ($10^{-5}$)  &        NA       &     Range          &     Local optimum              &    Local optimum         & 0        \\
         \ref{fig:proteom}B      &  Weak dependence ($10^{-5}$)  &        NA       &     Range          &     Local optimum              &    Local optimum         & 0        \\
         \ref{fig:proteom}C      &  Weak dependence ($10^{-5}$)  &        NA       &     Range          &     Local optimum              &    Local optimum          & 0       \\
         \ref{fig:proteom}D      &  Weak dependence ($10^{-5}$)  &        NA       &     Range          &     Local optimum              &    Local optimum            & 0     \\
 \hline
         \ref{fig:noisefig}      &  Weak dependence ($10^{-5}$)  &        NA       &     Range          &     Local optimum              &    Local optimum            & 0     \\
 \hline

         \ref{fig:over_combo1}A  &           $10^{-5}$     &     $10$             &     $10^2$         &     Range                    &     Range                & 0 \\
         \ref{fig:over_combo1}B  &           $10^{-5}$     &     $10$             &     $10^2$         &     Range                    &     Range                & 0 \\
         \ref{fig:over_combo1}C  &           $10^{-5}$     &     $10$             &     $10^2$         &     Range                    &     Range                & 0 \\
         \ref{fig:over_combo1}D  &           $10^{-5}$     &     $10$             &     $10^2$         &     Range                    &     Range                & 0 \\
\hline
         \ref{fig:over_combo2}A &       Range      &     $10^{2}$         &    Range           &     Local optimum              &    Local optimum                 & 0 \\
         \ref{fig:over_combo2}B &       Range      &     $10^{2}$         &    Range           &     Local optimum              &    Local optimum                  & 0  \\
         \ref{fig:over_combo2}C &       Range      &     $10^{2}$         &    Range           &     Local optimum              &    Local optimum                 & 0   \\
         \ref{fig:over_combo2}D &       Range      &     $10^{2}$         &    Range           &     Local optimum              &    Local optimum                 & 0 \\
\hline
         \ref{fig:te_combo2}A &         Range      &     $10^{2}$         &     Range          &    Local optimum                &   Local optimum               & 0 \\
         \ref{fig:te_combo2}B &         Range      &     $10^{2}$         &     Range           &   Local optimum                &   Local optimum              & 0 \\
         \ref{fig:te_combo2}C &         Range      &     $10^{2}$         &     Range           &    Local optimum               &   Local optimum               & 0  \\
         \ref{fig:te_combo2}D &         Range      &     $10^{2}$         &     Range           &    Local optimum               &    Local optimum             & 0  \\
\hline
         \ref{fig:overepS}       &            $10^{-5}$        &        NA             &     Range             &     Local optimum              &    30         & 0.1        \\
\hline
         \ref{fig:overepB}A  &     Range            &        NA             &     Range          &     Local optimum              &    Local optimum         & 0        \\
 \hline
         \ref{fig:loadrat2}      &  Weak dependence ($10^{-5}$)  &        NA       &     Range          &     Local optimum              &    Local optimum         & 0        \\
         \hline
\ref{SIfig:onemessagerule}A        &     Range            &        NA             &     Range          &     Local optimum              &    Local optimum         & 0        \\

 \hline

         \ref{fig:noisefigS}B      &  Weak dependence ($10^{-5}$)  &        NA       &     Range          &     Local optimum              &    Local optimum         & 0        \\
\hline
    \end{tabular}}
    \caption{\textbf{RLTO model parameters by figure.} \textit{Range} appears if a range of parameters is used.  \textit{NA} appears if the parameter value is irrelevant. \textit{Local optimum} appears if the parameter values in optimized to maximize the growth rate. 
    }
    \label{tab:my_label}
\end{table*}}


\subsection{Discussion: The fitness landscape of a trade-off. }

\label{sec:discland_s}

 The fitness landscape predicted by the RLTO model for representative parameters is shown in Fig.~\ref{fig:over_combo1}. The figure displays a number of important model phenomena: There is no growth for mean protein number $\mu_p$  below the threshold number $n_p$, and for high noise, $\mu_p$ must be in significant excess of $n_p$. 
Rapid growth can be achieved by the two mechanisms: (i) high expression levels ($\mu_p$) are required for high noise amplitude ($\inliner{\rm{CV}_p^2}$) or (ii) lower expression levels coupled with lower noise.
This trade-off leads to a ridge-like feature of nearly optimal models.
The optimal fitness corresponds to a balance between increasing the mean protein number ($\mu_p$) and decreasing noise ($\inliner{{\rm CV}_p^2}$). This optimal central dogma program strategy leads to significant overabundance. 

\begin{figure*}
  \centering
    \includegraphics[width=0.95\textwidth]{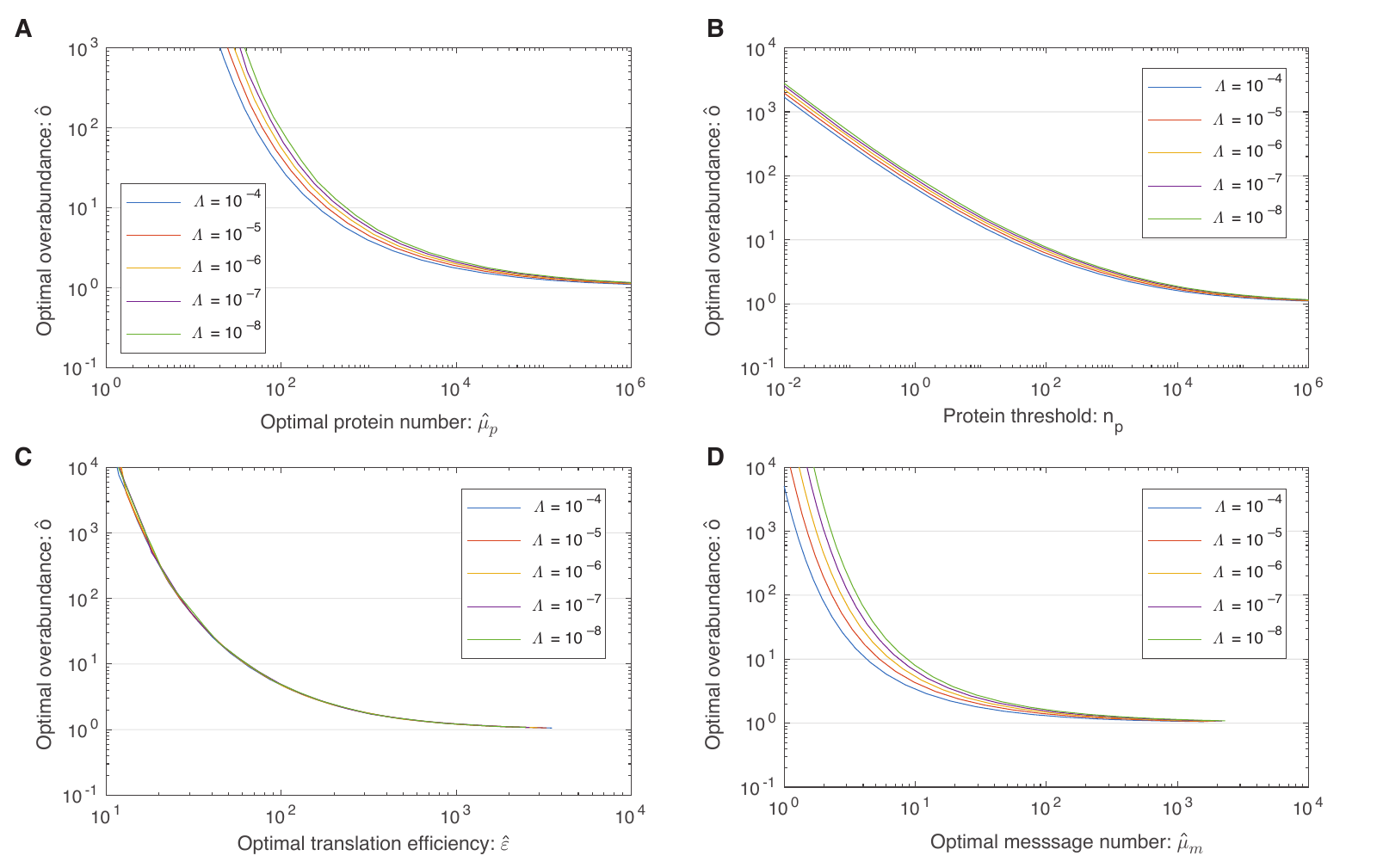}
      \caption{ \textbf{Four perspectives on the dependence of optimal overabundance on relative load $\Lambda$.} All these calculations are performed for protein cost $\lambda = 100$ in order to give real numbers in molecules per cell. 
      \textbf{Panel A: Overabundance as a function of protein number.} Overabundance decreases as protein number increases. These calculations are $\lambda$ \textit{dependent}. \textbf{Panel B: Overabundance as a function of the protein threshold.} Overabundance decreases as protein threshold increases. These calculations are $\lambda$ \textit{dependent}.  \textbf{Panel C: Overabundance as a function of translation efficiency.} Overabundance decreases as translation efficiency increases. These calculations are $\lambda$ \textit{dependent}.  \textbf{Panel D: Overabundance as a function of message number.} Overabundance decreases as message number increases. These calculations are $\lambda$ \textit{independent}. 
       \label{fig:over_combo2}}
\end{figure*}

\begin{figure*}
  \centering
    \includegraphics[width=0.95\textwidth]{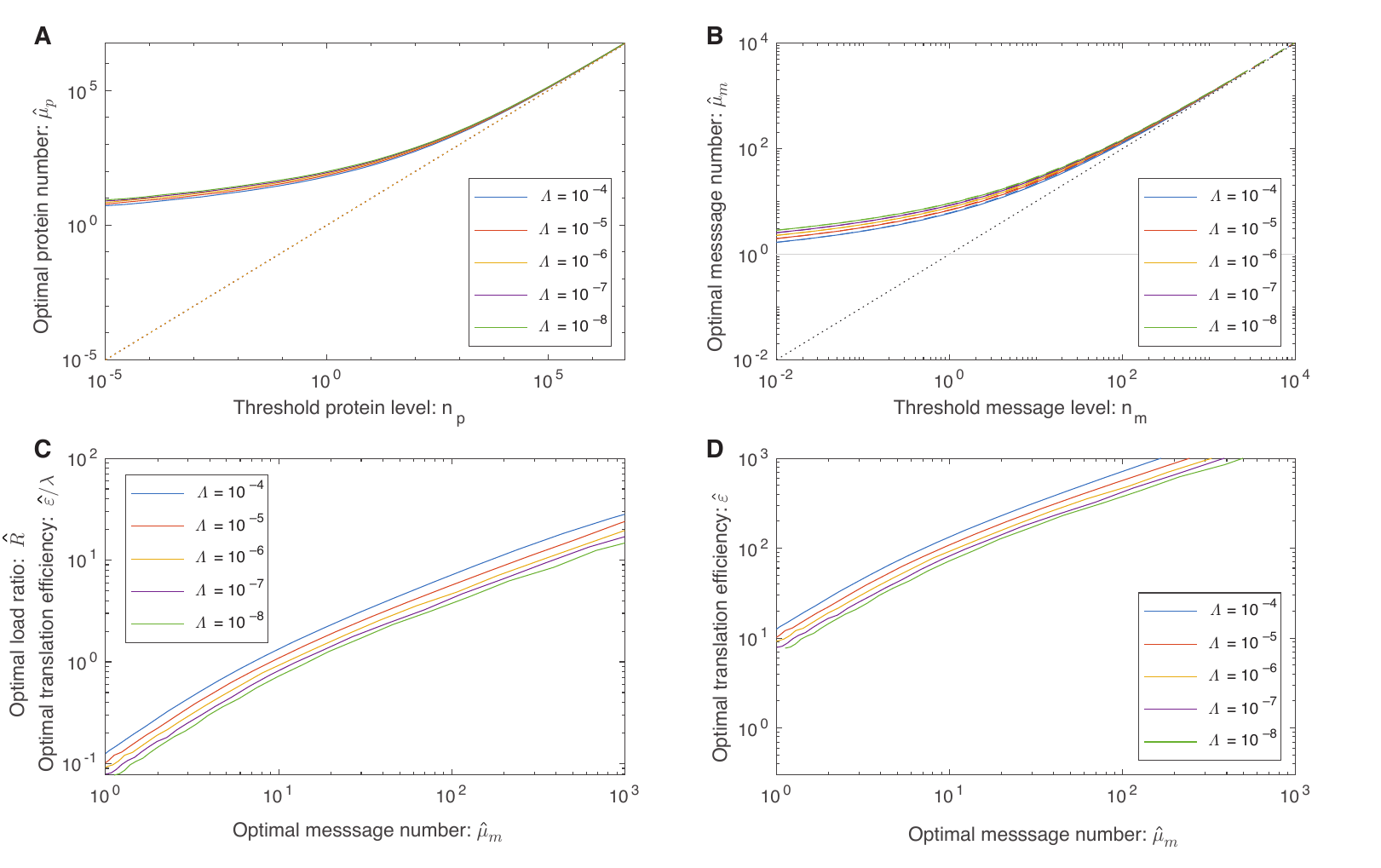}
      \caption{ \textbf{Four perspectives on load balancing.} All these calculations are performed for protein cost $\lambda = 100$ in order to give real numbers in molecules per cell. \textbf{ Panel A: Protein number versus protein threshold.} At high expression levels, the protein number tracks the protein threshold; however, the one message rule forces the protein number to threshold for low expression levels. These calculations are $\lambda$ \textit{dependent}. \textbf{Panel B: Message number versus message threshold.}  At high expression levels, the message number tracks the message threshold; however, the one message rule forces the message number to a threshold close to $\mu_m = 1$ for low expression levels. These calculations are $\lambda$ \textit{independent}. \textbf{ Panel C \& D: Message number versus translation efficiency.}  The optimal translation efficiency grows almost linearly with the optimal message number. The scaled translation efficiency ($\hat{\varepsilon}/\lambda$) is independent of $\lambda$ while the translation efficiency ($\hat{\varepsilon}$) is dependent on $\lambda$.  The ratio $\varepsilon/\lambda$ has a second interpretation: the load ratio $R$. $R$ is defined as the metabolic cost of translation over transcription of the gene.
       \label{fig:te_combo2}}
\end{figure*}

\subsection{Methods: Central dogma optimization}
\label{sec:opti}

\subsubsection{Optimization of transcription and translation (eukaryotes) }

{\color{red}
The relative growth rate is: 
\begin{equation}
\textstyle \frac{\Delta k}{k_0} =  \textstyle - (\Lambda+\frac{\varepsilon}{N_0})\mu_m - \frac{1}{\ln 2}\gamma(\frac{\mu_m}{\ln 2},\frac{n_p}{\varepsilon \ln 2}), \label{SIeqn:growthrate2}
\end{equation}
where $\gamma$ is the regularized incomplete gamma function, which is the CDF of the gamma distribution and represents the probability of arrest due to gene $i$. (Note that this equation is identical to Eq.~\ref{eq:paw1} but with the gene subscript $i$ implicit.)
We set the partial derivative of the growth rate with respect to message number equal to zero: }
\begin{equation}
0 = -\textstyle\frac{\lambda+\hat{\varepsilon}}{N_0} - \frac{1}{(\ln 2)^2} \gamma_{,1}( \frac{\hat{\mu}_m}{\ln 2},\frac{n_p}{\hat{\varepsilon} \ln 2}), \label{eqn:der1}
\end{equation}
where we use the canonical comma notation to show which argument of $\gamma$ has been differentiated. Next we differentiate with respect to the translation efficiency to generate a second optimization condition:
\begin{equation}
0 = -\textstyle\frac{\hat{\mu}_m}{N_0} + \frac{n_p}{\hat{\varepsilon}^2(\ln 2)^2} \gamma_{,2}( \frac{\hat{\mu}_m}{\ln 2},\frac{n_p}{\hat{\varepsilon}  \ln 2}). \label{eqn:der2}
\end{equation}
We will work in the large multiplicity limit where the overall metabolic load is much smaller than the metabolic load associated with any single gene: $N_0 \gg (\lambda+\hat{\varepsilon})\hat{\mu}_m$. 
Next, we eliminate the threshold $n_p$ in favor of the optimal overabundance:
\begin{equation}
\hat{o} \equiv \textstyle\frac{\hat{\mu}_p}{n_p} = \frac{\hat{\varepsilon} \hat{\mu}_m}{n_p},
\end{equation}
in both Eqs.~\ref{eqn:der1} and \ref{eqn:der2}. Eq.~\ref{eqn:der2} can now be solved for the optimal translation efficiency:
\begin{equation}
\hat{\varepsilon}  = \textstyle \frac{N_0}{\hat{o}(\ln 2)^2} \gamma_{,2}( \frac{\hat{\mu}_m}{\ln 2},\frac{\hat{\mu}_m}{\hat{o} \ln 2}). \label{eqn:der4}    
\end{equation}
If we reinterpret $\gamma$ as the CDF of the gamma distribution, we can rewrite this equation in terms of the gamma distribution PDF:
\begin{equation}
\hat{\varepsilon}  = \textstyle \frac{N_0}{\ln 2} p_\Gamma(\mu_m|  \frac{\hat{\mu}_m}{\ln 2},\hat{o} \ln 2), \label{eqn:der3}    
\end{equation}
which will be the optimization equation for the translation efficiency. 

To derive the optimization condition for the message number $\mu_m$, we substitute Eq.~\ref{eqn:der4} into Eq.~\ref{eqn:der1}:
\begin{equation}
\textstyle\frac{\lambda \ln 2}{N_0}  = -\frac{1}{\hat{o}\ln 2} \gamma_{,2}( \frac{\hat{\mu}_m}{\ln 2},\frac{\hat{\mu}_m}{\hat{o}\ln 2}) -  \frac{1}{\ln 2}\gamma_{,1}( \frac{\hat{\mu}_m}{\ln 2},\frac{\hat{\mu}_m}{\hat{o}\ln 2}).
\end{equation}
The two terms on the RHS can now be collected as the single partial derivative of message number $\mu_m$: 
\begin{equation}
\textstyle \Lambda \ln 2  = -\partial_{\hat{\mu}_m} \gamma( \frac{\hat{\mu}_m}{\ln 2},\frac{\hat{\mu}_m}{\hat{o}\ln 2}),\label{eqn:optimder}
\end{equation}
where the relative load is $\Lambda \equiv \lambda/N_0$. 

The two optimization equations are summarized below:
\begin{eqnarray}
 \Lambda \ln 2  &=& \textstyle -\partial_{\hat{\mu}_m} \gamma( \frac{\hat{\mu}_m}{\ln 2},\frac{\hat{\mu}_m}{\hat{o}\ln 2}), \label{eqn:oeqn}\\
\textstyle\frac{\hat{E}}{\Lambda}  &=& \textstyle\frac{\hat{\varepsilon}}{\lambda}  = \textstyle \frac{1}{\Lambda \ln 2} p_\Gamma(\hat{\mu}_m|  \frac{\hat{\mu}_m}{\ln 2},\hat{o} \ln 2). \label{eqn:der3B}    
\end{eqnarray}
The optimal overabundance is shown for a range of relative loads in Fig.~\ref{fig:over_combo2}. The optimal translation efficiency and scaled translation efficiency are  shown for a range of relative loads in Fig.~\ref{fig:te_combo2}.

\begin{figure}
  \centering
    \includegraphics[width=.475\textwidth]{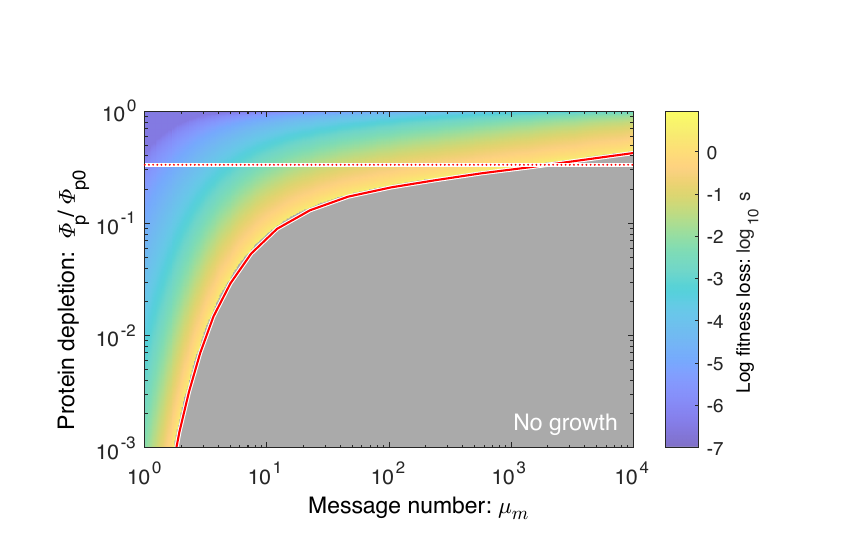}
      \caption{ 
      \textbf{Optimal expression levels are buffered.} The predicted fitness loss as a function of protein depletion level and message number for bacterial cells (including the noise floor). Due to the overabundance phenomenon, all proteins are buffered against depletion, but low-expression genes are particularly robust due to higher overabundance. The solid red line represents $1/o$, and predicts the range of depletion values for which cell growth is predicted. The dotted red line represents a three-fold depletion. 
       \label{fig:overepS}}
\end{figure}



\subsubsection{Optimization of  message number only}

\label{sec:messageonly}
Consider the special case of optimizing the message number only at fixed translation efficiency. Eq.~\ref{eqn:der1} is  the condition; however, in this case it makes sense to adsorb both the message and protein metabolic load into a single metabolic load. The optimum message number satisfies the equation:
\begin{equation}
\textstyle\frac{(\lambda + \varepsilon) \ln 2}{N_0} =  - [\partial_{\hat{\mu}_m}\gamma( \hat{\mu}_m,\hat{n}_m)]_{\hat{n}_m = \frac{\hat{\mu}_m}{\hat{o}}}.
\end{equation}
We define a modified relative  load:
\begin{equation}
\Lambda' \equiv \textstyle\frac{(\lambda + \varepsilon)}{N_0}, \label{eqn:modload}
\end{equation}
and substitute this into the optimum message number equation:
\begin{equation}
\Lambda' \ln 2 =  - [\partial_{\hat{\mu}_m}\gamma( \hat{\mu}_m,\hat{n}_m)]_{\hat{n}_m = \frac{\hat{\mu}_m}{\hat{o}}}.
\end{equation}
which is clearly closely related to Eq.~\ref{eqn:optimder}. 

We compare this modified expression to the original for optimum overabundance as a function of message number in Fig.~\ref{fig:overep}C and demonstrate that the two make nearly identical predictions.

\subsubsection{ Inclusion of the noise floor}

\label{sec:noisefloor}

In bacterial cells, the noise is dominated by the noise floor for high expression genes. Including the noise floor, the coefficient of variation squared is (\cite{Taniguchi2010} and Sec.~\ref{sec:aonS}):
\begin{equation}
{\rm CV}_p^2 = \textstyle \tilde{a}(\mu_m)^{-1} = \frac{\ln 2}{\mu_m}+C_0,    \label{eqn:noiseC0}
\end{equation}
where $C_0 = 0.1$ for bacterial cells \cite{Taniguchi2010}. In spite of the addition of noise from the noise floor, the observed distribution of protein number is still well described by the gamma distribution \cite{Taniguchi2010}; however, we need to modify the statistical parameters to account for the noise floor. (See the definition of the \textit{statistical noise model} in Sec.~\ref{sec:statnoiseS}.) The modified gamma parameters are:
\begin{eqnarray}
a &=& \tilde{a}, \\
\theta &=& \varepsilon   \frac{\mu_m}{\tilde{a}}, 
\end{eqnarray}
chosen such that the noise is determined by Eq.~\ref{eqn:noiseC0} but the protein number remains:
\begin{equation}
\mu_p = \varepsilon \mu_m,
\end{equation}
the product of the message number and translation efficiency.

The qualitative effect of the noise floor is to increase the noise, especially for low-copy messages.  Above $\mu_m = 7$ messages, the noise is dominated by the noise floor. Increases in transcription above this point have little effect on reducing the noise. As a consequence, the overabundance stays high, even for high copy messages. We compare this modified expression to the original for optimum overabundance as a function of message number in Fig.~\ref{fig:overep}C and demonstrate that bacterial cells are predicted to have much higher overabundance at high expression levels.

\subsection{Discussion: Understanding the rationale for overabundance}

\label{sec:rationale}

Essential protein overabundance is the signature prediction of the RLTO model. Its mathematical rationale is the highly-asymmetric fitness landscape.    
To understand why we expect this rationale to be generic, consider the form of the optimization condition for message number: 
\begin{equation}
\Lambda \ln 2  = \textstyle-\frac{\partial}{\partial {\hat{\mu}_m}} \gamma(\frac{\hat{\mu}_m}{\ln 2},\frac{\hat{\mu}_m}{\hat{o}\ln 2}). \label{equilcond}
\end{equation}
The growth rate is maximized when the probability of  slow-growth (\textit{e.g.}~arrest) is roughly equal to the relative load of adding one more message. Since the cell makes roughly $10^5$ messages per cell cycle, the relative load is extremely small and therefore the probability of slow growth must be as well. Making this probability very small requires vast overabundance for the inherently-noisy, low-expression proteins. 

The reason we expect the RLTO model protein abundance predictions to be robust is that we generically expect the fitness cost of overabundance to be small due to the high multiplicity (\textit{i.e.}\ total number of genes); whereas, the fitness cost of arrest of essential processes is very high. 


\subsection{Discussion: RLTO predicts larger overabundance in bacteria.}

\label{sec:bacteria}

There are two distinctive features of bacterial cells that could affect the model predictions: (i) the translation efficiency is constant \cite{Balakrishnan:2022ai} and bacterial gene expression is subject to a large-magnitude noise floor that increases the noise for high-expression genes \cite{Taniguchi2010}.
The optimization of message number at fixed translation efficiency does result in a slightly modified optimization condition for the message number (Sec.~\ref{sec:messageonly}); however, the predicted overabundance is only subtly perturbed (Fig.~\ref{fig:overep}C). In contrast, the noise floor increases the predicted overabundance, especially for high-expression proteins. As a result, the RLTO model predicts that the vast majority of bacterial proteins are expressed in significant overabundance. (See Fig.~\ref{fig:overep}C.)



\subsection{Discussion: RLTO predicts proteins are buffered to depletion.}

\label{sec:depletion}

A principle motivation for our analysis is the observation that many protein levels appear to be buffered.
To explore the prediction of the RLTO model for protein depletion, we first computed the optimal message numbers and translation efficiencies for a range of protein thresholds. To model the effect of protein depletion, 
 we computed the change in growth rate as function of protein depletion (equivalent to a reduction of the translation efficiency relative to the optimum.) The growth rate is shown in Fig.~\ref{fig:overepS} for the RLTO model with parameters representative of a bacterial cell. (See Sec.~\ref{sec:noisefloor}.) 

In general, the RLTO model predicts that protein numbers have very significant robustness (\textit{i.e.}~buffering) to protein depletion. This is especially true for low expression proteins that are predicted to have the largest overabundance. For these genes, even a ten-fold  depletion leads to very subtle reductions in the growth rate. For a three-fold reduction in the growth rate, only the very highest-expression genes (\textit{e.g.}~ribosomal genes) are expected to lead to qualitative phenotypes.

\subsection{Results: Detailed development of load balancing}

\label{sec:loadbals}

\subsubsection{Prediction of the optimal load ratio.}

 The two-stage amplification of the central dogma implies that the  expression and noise levels can be controlled independently by  the balance of transcription to translation. 
How does the cell achieve high and low gene expression optimally, and how does this strategy depend on the message cost?

To understand the optimization, we 
first define the load ratio $R$ for a gene as the metabolic cost of translation relative to transcription:
\begin{equation}
R \equiv \textstyle \frac{\mu_p}{ \lambda \mu_m} = \frac{\varepsilon}{\lambda }.   
\end{equation}
In Sec.~\ref{sec:opti}, we show that the optimal load ratio is:
\begin{equation}
\hat{R} \equiv \textstyle \frac{1}{\Lambda \ln 2}\ p_{\Gamma}(\hat{\mu}_m|\frac{\hat{\mu}_m}{\ln 2},\frac{1}{\hat{o} \ln 2}),  \label{eqnsym2}
\end{equation}
where $p_{\Gamma}$ is the PDF of the gamma distribution. The optimal load ratio is shown in Fig.~\ref{fig:te_combo2}C.

The dependence of the optimal load ratio $\inliner{\hat{R}}$ on $\Lambda$ is extremely weak, but it is strongly dependent on message number.
As a result, for low transcription genes ($\mu_m\ll 10$), the metabolic load is predicted to be dominated by transcription; whereas, for highly transcribed genes ($\mu_m\gg 10$), the metabolic load is dominated by translation. These predictions are robust since they are independent of the relative load $\Lambda$.   


\subsubsection{Measurements of the load ratio}

\label{sec:loadratio}

Unfortunately, there is somewhat limited data to which to compare the model. The best source we found was Kafri \textit{et al.}\ \cite{Kafri:2016ye} who analyzed the differences in fitness between transcription and transcriptional-and-translation of a fluorescent protein driven by the pTDH3 promoter in yeast. This promoter is one of the strongest in yeast. Based on the RLTO model, we would predict this promoter to have a very high translation efficiency and therefore a large load ratio; however, the translation efficiency is much lower than one would predict based on a global analysis and likewise its load ratio is roughly unity, which based on the smaller than expected translation efficiency is broadly consistent with our expectations. A satisfactory test of this prediction will require larger-scale measurements that probe more representative genes.

\subsection{Results: Translation efficiency is predicted to increase with transcription.}

\label{sec:TransEff_s}
Now that we have defined the optimal load ratio (Eq. \ref{eqnsym2}), the equation for optimal translation efficiency can be written concisely:
\begin{equation}
\hat{\varepsilon} = \lambda \hat{R}\ \ \ \ \ \text{or}\ \ \ \ \ \hat{E} = \Lambda \hat{R}, \label{SIeqnsym}
\end{equation}
where $\hat{R}$ depends weakly on the relative load $\Lambda$.
The RLTO model predicts that optimal partitioning of amplification between transcription (gain $\mu_m$) and translation (gain $\varepsilon$) has two important qualitative features: (i) As the message cost ($\lambda$) rises, the optimal translation efficiency increases in proportion. (ii) The optimal translation efficiency is also approximately proportional to message number ($\hat{\varepsilon} \propto \mu_m$). (See Fig.~\ref{fig:proteom}A.) Therefore, the RLTO model predicts that low expression levels should be achieved with low levels of transcription and translation, whereas high expression genes are achieved with high levels of both. We call this relation between optimal transcription and translation \textit{load balancing}.


\subsection{Results: RLTO predicts that message number  responds to message cost}

We will first focus on analyzing the implications of the message cost dependence in Eq.~\ref{SIeqnsym}. 
At a fixed load ratio,  Eq.~\ref{SIeqnsym} clearly implies that the translation efficiency increases as the message cost $\lambda$ increases; however, the message number (and load ratio) also respond to compensate to changes in $\lambda$.
To probe the dependence on message cost in an experimentally relevant context, consider optimal message numbers in a reference condition  (relative load $\Lambda_0$) relative to a second perturbed condition (relative load $\Lambda$).  The predicted relation between the optimal messages numbers is shown in Fig.~\ref{fig:overepB}A. 
The resulting relation between the optimal message numbers is roughly linear on a log-log plot, predicting the approximate power-law relation: 
\begin{equation}
\hat{\mu}_m(\Lambda)\propto \hat{\mu}_m(\Lambda_0)^\alpha, \label{eq:powerlaw}
\end{equation}
describing a non-trivial global change in the regulatory program.

\begin{figure*}
  \centering
    \includegraphics[width=.95\textwidth]{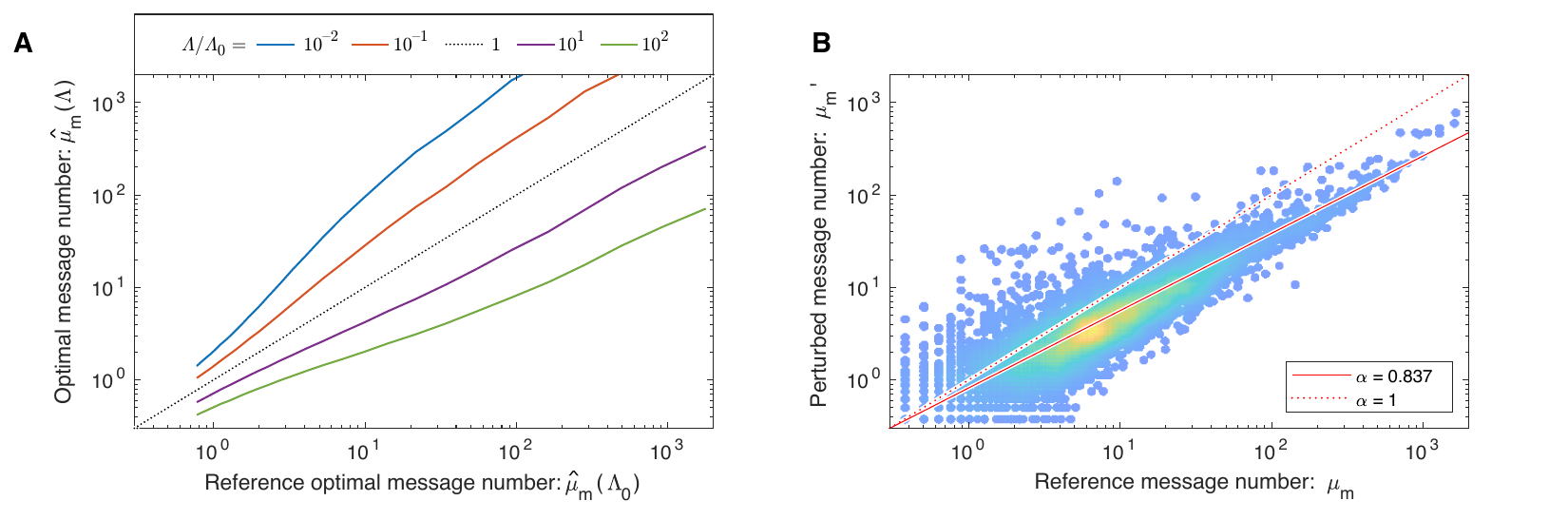}
      \caption{{\color{red} \textbf{The message cost affects transciption genome wide. Panel A: Message number decreases with increased relative load $\Lambda$.}} The optimal message number responds to changes in the message cost. 
      The RLTO model predicts an approximate power-law relation (linear on a log-log plot) between message numbers.    
      \textbf{Panel B: A power-law relation is observed.} To test whether central dogma regulation would adapt dynamically as predicted, we analyzed the relation between the yeast transcriptome under reference conditions and phosphate depletion (perturbed), which increases the message cost \cite{Kafri:2016ye}. (Data from Ref.~\cite{Metzl-Raz:2020ff}.) As predicted by the RLTO model, a global change in regulation is observed, which generates  a power-law relation with scaling exponent $\alpha=0.837\pm 0.01$. The observed exponent is smaller than one, as predicted by an increased relative load $\Lambda$. 
       \label{fig:overepB}}
\end{figure*}

\subsection{Results: RLTO predicts the yeast global regulatory response}
\label{sec:phos}

To test the RLTO predictions, 
we compared the relative message numbers for yeast growing under phosphate depletion, which increases the message cost \cite{Kafri:2016ye}, to a reference condition \cite{Metzl-Raz:2020ff}. As predicted, the relative transcriptome data was well described by a power law (Eq.~\ref{eq:powerlaw}) and the observed slope was smaller than one: $\hat{\alpha} = 0.837\pm 0.001$, as predicted by the increased message cost.  See Fig.~\ref{fig:overepB}B. 

The observation of this large-scale regulatory change has an important implication: 
This response supports a nontrivial hypothesis that the RLTO model not only can predict how the cell is optimized in an evolutionary sense, but can predict global regulatory responses as well.

Note that Metzl-Raz \textit{et al.} \cite{Metzl-Raz:2020ff} also explored conditions that increased the cost of protein; however, here the predictions of the model are ambiguous. The complication arises due to the observed decrease in size of the cells in the experimental condition, which decreases the total metabolic load $N_0$. As discussed above, the relative load  $\Lambda = \lambda/N_0$ is the key determinant in Eq.~\ref{eq:powerlaw}; however, even as the relative cost of transcription $\lambda$ decreases in the experimental condition, the total metabolic load $N_0$ also decreases, making no clear prediction about how the relative load $\Lambda$ changes.

\subsection{Methods: Prediction of the proteome fraction}

\label{sec:ppf}

\subsubsection{Parameter-free prediction of proteome fraction.} We now turn our focus to an analysis of the implications of \textit{load balancing}: the message number dependence of the optimal translation efficiency (Eq.~\ref{SIeqnsym}). The most direct  test of this prediction is measuring the relation between proteome fraction and message number. The RLTO model predicts proteome fraction (Eq.~\ref{eqn:protomeFracS}):
\begin{equation}
\hat{\Phi}_{p} = \hat{E} \hat{\mu}_{m} \propto \hat{\mu}_{m}^2, 
\end{equation}
where $\mu_{m}$ is the observed message number and the optimal relative translation efficiency is predicted by Eq.~\ref{SIeqnsym}. The proportionality is only approximate but gives important intuition for how protein number depends on message number in the RLTO model, in contrast to a constant-translation-efficiency model: $\Phi_{p} \propto \mu_{m}$. 
To compare these predictions to protein abundance measurements, we will renormalize the protein fraction to be defined relative to total protein number rather than $N_0$. This renormalization eliminates the $\Lambda$ dependence to result in  a parameter-free prediction of the proteome fraction. 

\subsubsection{ RLTO: proteome fraction}

Starting from Eq.~\ref{eq:protnumpred}, clearly:
\begin{equation}
\hat{\mu}_p \propto \textstyle \hat{\mu}_m\ p_\Gamma(\hat{\mu}_m|  \frac{\hat{\mu}_m}{\ln 2},\hat{o} \ln 2),  
\end{equation}
which can be used to predict the proteome fraction {\color{red} (where we have restored the explicit gene $i$ subscript)}:
\begin{equation}
\hat{\Phi}_{p,i} \equiv \textstyle \frac{\hat{\mu}_{p,i}}{\sum_j \hat{\mu}_{p,j}},  
\end{equation}
where the second subscript is the gene index. To predict the proteome fraction, we computed the proportionality constant $C$:
\begin{equation}
C  \equiv \textstyle \sum_i \left[ \mu_{m,i} p_\Gamma(\mu_m|  \frac{\mu_{m,i}}{\ln 2},o \ln 2)|_{o=\hat{o}(\mu_{m,i})}\right],
\end{equation}
where the message numbers $\mu_{mi}$ for gene $i$ are the experimentally observed message numbers, the implicit $o_i$ values are predicted by the RLTO model (Eq.~\ref{eqn:optimder}) for message number $\mu_{mi}$ and the sum index $i$ runs over all genes.
The predicted optimal proteome fraction is:
\begin{equation}
    \hat{\Phi}_{p,i} = C^{-1}\left[\mu_{m,i} p_\Gamma(\mu_{m,i}|  \frac{\mu_{m,i}}{\ln 2},o \ln 2)|_{o=\hat{o}(\mu_{m,i})}\right],
\end{equation}
which generates the predicted solid curves shown in Fig.~\ref{fig:proteom}BCD.

\subsubsection{ Constant-translation-efficiency model: proteome fraction}

For the constant translation efficiency model, we define the normalization:
\begin{equation}
C'  \equiv \sum_i \mu_{m,i},
\end{equation}
and the predicted proteome fraction is:
\begin{equation}
    \Phi_{p,i}' = {C'}^{-1}\mu_{m,i},
\end{equation}
which generates the predicted dotted curves shown in Fig.~\ref{fig:proteom}BCD.

\subsubsection{ Sources of experimental data for proteome fraction analysis}

For \textit{E.~coli} data, the protein abundance data was generated by mass spec measurements and the message abundance data was from ~\cite{Balakrishnan:2022ai}. For the yeast data, the protein abundance data is measured by mass spec and message abundances are determined by  ~\cite{Ghaemmaghami:2003jj}. For the mammalian data, we used mouse data. The protein abundance data is measured by mass-spec and message abundances are determined by  ~\cite{Schwanhausser:2011ec}. 

We estimated the message number $\mu_m$ as described in Sec.~\ref{sec:CCMN}. For the mouse data, the study provided message lifetimes, the cell cycle duration and abundances in molecules per cell \cite{Schwanhausser:2011ec}.
For the \textit{E.~coli} and yeast, the total number of proteins, messages \textit{etc}, cell cycle duration and message lifetimes for each organism and their sources are described in Tab.~\ref{tab1refs}.

\subsection{Results: Load balancing is observed in eukaryotic cells}

\label{sec:euklbs}

A non-trivial prediction of the RLTO Model is that translation efficiency and message number should be roughly proportional. 
Qualitatively, this strategy allows expression levels to be increased while distributing the added metabolic load between transcription, which reduces noise, and translation, which does not affect the noise. 
We predict the optimal translation efficiency versus message number which matches the observations in eukaryotic cells (Fig.~\ref{fig:proteom}BC). However, in \textit{E.~coli}, the translation efficiency and message number are \textit{not}  strongly correlated (Fig.~\ref{fig:proteom}D). Why does this organism appear not to load balance? We demonstrate that the observed translation efficiency is consistent with the RLTO model, augmented by a ribosome-per-message limit. (See Sec.~\ref{sec:incproload}.) Hausser 
\textit{et al.}\ have proposed just such a limit, based on the ribosome footprint of mRNA molecules \cite{Hausser:2019fi}. 
(See Sec.~\ref{sec:translationallimit}.) Although this augmented model is consistent with central dogma regulation in \textit{E.~coli}, it is not a complete rationale. This proposed translation-rate limit could be circumvented by increasing the lifetime of \textit{E.~coli} messages which would increase the translation efficiency. Why the message lifetime is as short as observed will require a more detailed \textit{E.~coli}-specific analysis.

\subsection{Discussion: Relation between load balancing and previous results}

\label{sec:Hausser}

Hausser \textit{et al.}\ have previously performed a more limited analysis of the trade-off between metabolic load and gene-expression noise \cite{Hausser:2019fi}. In this section, we will provide some more context into the differences between the two approaches. 

Hausser \textit{et al.}\ assume a symmetric (not an asymmetric) fitness landscape and consider only the metabolic cost of transcription (but not translation). Their model depends on two (not one) gene-specific parameters: an optimal protein number and a sensitivity, which defines the curvature of the fitness \cite{Hausser:2019fi}.  
The authors maximize fitness with respect to the transcription rate (but not the translation rate) and the condition they derive depends on the two (not one) unknown, gene-specific parameters. 
As a result, this condition is not predictive of global regulatory trends without  non-trivial, gene-specific measurements or  assumptions about the unknown sensitivity.  


The Hausser model assumes that the growth rate has the form:
\begin{equation}
k = k_0- \textstyle\frac{1}{2}|k''|(N_p-\mu_p)^2 - k_0\Lambda \mu_m,
\end{equation}
where $k$ is the growth rate, and we have rewritten the form of the fitness to better match our own definitions. Here $k''$ is the second derivative of the growth rate at the optimal protein number $\mu_p$ and $N_p$ is the stochastic protein number. If we take the expectation with respect to the protein number, we get:
\begin{equation}
k = k_0- \textstyle\frac{1}{2}|k''|\sigma_p^2 - k_0\Lambda \mu_m,
\end{equation}
and substituting the noise model for the variance of the protein number gives:
\begin{equation}
k = k_0- \textstyle\frac{1}{2}|k''|\mu_p^2(\frac{\ln 2}{\mu_m}+C_0) - k_0\Lambda \mu_m,
\end{equation}
where $C_0$ is the noise floor, and we have assumed the mean protein number is optimal ($\mu_p$). If we maximize the growth rate with respect to $\mu_m$, we get the following condition on the optimal message number:
\begin{equation}
\hat{\mu}_m^2 = \textstyle\frac{1}{2}\frac{|k''|}{k_0}\mu_p^2\frac{\ln 2}{\Lambda},
\end{equation}
which depends on the unknown curvature $k''$. To make global predictions about how transcription and translation are related, some added assumptions are necessary to describe how $k''$ scales with protein abundance. 
 
To illustrate how this expression does not make explicit global predictions, let's consider a number of plausible possibilities. First, we will assume that $k''$ is independent of $\mu_p$ and on average all proteins are equally sensitive to changes in protein number. In this case, we find:
\begin{eqnarray}
\mu_p &\propto& \hat{\mu}_m \sqrt{\Lambda} , \\
\hat{\varepsilon} &\propto&  \sqrt{\Lambda},
\end{eqnarray}
implying a constant translation efficiency which is inversely proportional to the square root of the relative load. 

Alternatively, we can assume that $k''\propto \mu_p^{-2}$ and, on average, the cell is equally sensitive to changes in the relative number of proteins (\textit{i.e.}\ $\Delta p/\mu_p$), regardless of expression level. In this case,
\begin{eqnarray}
\hat{\mu}_m &\propto& 1/\sqrt{\Lambda}, \\
\hat{\varepsilon} &\propto& \mu_p \sqrt{\Lambda},
\end{eqnarray}
implying a constant message number, irrespective of expression level, and a translation efficiency that is proportional to expression level. 

Finally, we will assume that $k''\propto \mu_p^{-1}$, which is the intermediate case. Here:
\begin{eqnarray}
\mu_p &\propto& \hat{\mu}_m^2 \Lambda, \\ \label{eqn:rescale}
\hat{\varepsilon} &\propto&  \hat{\mu}_m\Lambda,
\end{eqnarray}
implying that translation efficiency should increase with message number, analogous to our prediction. It appears that Hausser \textit{et al.} implicitly also favor this model, since they define their sensitivity parameter to include a power of protein number $\mu_p$. They justify this assumption by arguing that since $\sigma_p^2 \propto \mu_p$, it makes sense to define the \textit{sensitivity to noise} to include a factor of $\mu_p$ \cite{Hausser:2019fi}. At best, this is somewhat fuzzy logic since, as we demonstrate in the paper,  Eq.~\ref{eqn:rescale} implies that the protein variance does not scale $\sigma^2_p \propto \mu_p$!

The authors also propose a lower limit on the translation-transcription ratio; however, their limit is dependent on the noise floor, which only affects genes with the highest transcription rates in eukaryotic cells. The implementation of a more appropriate estimate of the noise, relevant for the vast majority of genes, does not lead to the same limit.

\begin{figure}
  \centering
    \includegraphics[width=0.47\textwidth]{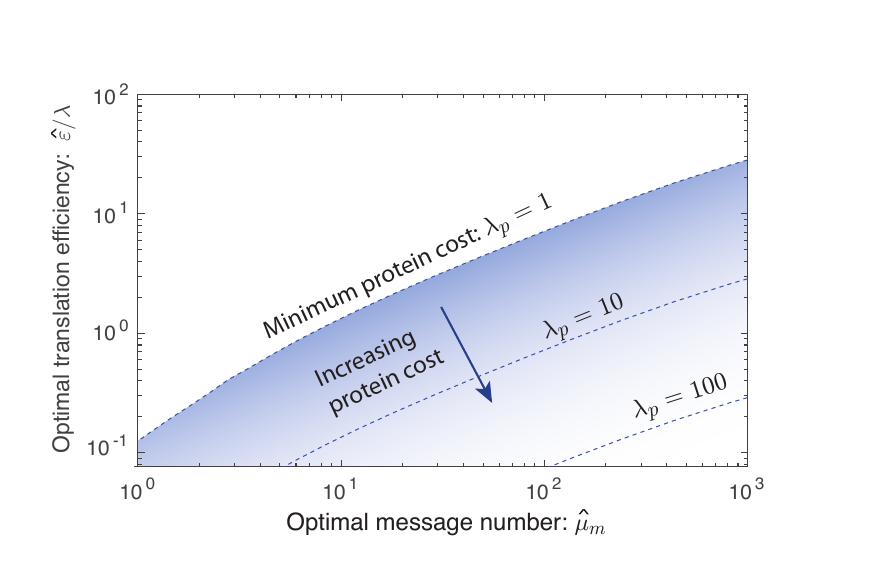}
      \caption{  \textbf{Increased protein cost decreases optimal translation efficiency.} A protein cost of $\lambda_p = 1$ corresponds to the metabolic cost of protein synthesis only, and is the minimum protein cost. For larger protein costs, the optimal translation efficiency is lower. As a result, the $\lambda_p = 1$ curve represents an upper bound of the optimal translation efficiency. 
       \label{fig:loadrat2}}
\end{figure}

\subsection{Methods: Analysis of translational limits \& gene-specific load analysis}

\label{sec:incproload}


To explore the consequences of a protein-specific load, we can modify the metabolic load term in the growth rate equation (Eq.~\ref{SIeqn:growthrate2}):
\begin{equation}
E \mu_m \rightarrow \lambda_p E \mu_m, 
\end{equation}
which includes an additional parameter: the protein cost $\lambda_p$, which is $1$ if the fitness cost is equal to the metabolic load and greater than one if the cost is higher. We will also treat the metabolic 
load per message $\lambda$ as a gene-specific parameter in this section only. The optimization can be repeated for this augmented model. 

To analyze the effect of increased protein load, we modify Eq.~\ref{SIeqn:growthrate2}:
\begin{equation}
\textstyle \ln \frac{k}{k_0} =  \textstyle - (\Lambda+\lambda_p E)\mu_m - \frac{1}{\ln 2}\gamma(\frac{\mu_m}{\ln 2},\frac{n_p}{\varepsilon \ln 2}), \label{eqn:growthrate3}
\end{equation}
to contain the supplemental load factor $\lambda_p$ which is unity if the only protein load is metabolic and $\lambda_p>1$ if there is additional load (\textit{e.g.}\ toxicity). 
The optimization conditions (Eqs.~\ref{eqn:der1} and \ref{eqn:der2}) become:
\begin{eqnarray}
0 &=& -\textstyle\frac{\lambda+\lambda_p\hat{\varepsilon}}{N_0} - \frac{1}{(\ln 2)^2} \gamma_{,1}( \frac{\hat{\mu}_m}{\ln 2},\frac{n_p}{\hat{\varepsilon} \ln 2}), \\
0 &=& -\textstyle\frac{\lambda_p\hat{\mu}_m}{N_0} + \frac{n_p}{\hat{\varepsilon}^2(\ln 2)^2} \gamma_{,2}( \frac{\hat{\mu}_m}{\ln 2},\frac{n_p}{\hat{\varepsilon}  \ln 2}). 
\end{eqnarray}
Using the same algebraic approach as before, we can derive the same optimal overabundance and load equations:
\begin{eqnarray}
\textstyle \Lambda \ln 2  &=& \textstyle -\partial_{\hat{\mu}_m} \gamma( \frac{\hat{\mu}_m}{\ln 2},\frac{\hat{\mu}_m}{\hat{o}\ln 2}),\\
\hat{R} &=&   \textstyle \frac{1}{\Lambda \ln 2} p_\Gamma(\hat{\mu}_m|  \frac{\hat{\mu}_m}{\ln 2},\hat{o} \ln 2);
\end{eqnarray} 
however, the relation between the load and the translation efficiency now has an extra factor: $\lambda_p$:
\begin{equation}
R = \frac{\lambda_p\varepsilon \mu_m}{\lambda \mu_m} =  \frac{\lambda_p\varepsilon}{\lambda},
\end{equation}
representing the modified total load ratio.

\subsection{Results: Increased protein-specific cost reduces the optimal translation efficiency.} 

\label{sec:protcost_s}

The relation between the overabundance and message number is unchanged (Eq.~\ref{equilcond}). This result can be rationalized in the following way: The optimal overabundance is determined by the noise which is determined by message number only. This relation is unaffected by the added parameter $\lambda_p$. However, the optimal translation efficiency is affected:
\begin{equation}
 \hat{\varepsilon} = \textstyle \frac{\lambda}{\lambda_p} \hat{R}, \label{eqn:opti2}
\end{equation}
where $\hat{R}$ is the optimal load ratio, defined by Eq.~\ref{SIeqnsym}. The optimal curves are shown in Fig.~\ref{fig:loadrat2}.

How do these added considerations affect the  RLTO  predictions? First, we consider message and protein length. What are the optimal translation efficiencies for two proteins, one ten times the length of the other, at fixed protein number? In this case, we will assume that both the transcriptional cost ($\lambda$) as well as the translational cost ($\lambda_p$) increase tenfold. These increases cancel, resulting in the same optimal translation efficiency since it is only the relative cost of transcription to translation that is determinative of the translation efficiency. 

Now consider a tenfold protein-specific increase in protein cost at fixed message cost and fixed protein number. The message number and translation efficiency would change by compensatory factors of $10$:
\begin{eqnarray}
 \hat{\mu}_m &\rightarrow& 10 \cdot \hat{\mu}_m,   \\
 \hat{\varepsilon}_m &\rightarrow& \textstyle \frac{1}{10} \cdot \hat{\varepsilon}_m,   
\end{eqnarray}
to maintain the protein number. 

Returning to our original motivation, we can understand how genes with a higher protein-to-message cost migrate downwards and rightwards off the optimal $\lambda_p = 1$ curve, predicting a cloud versus a narrow strip in proteome fraction measurements shown in Fig.~\ref{fig:proteom}. If the relative load $\Lambda$ were directly measured, we would expect the predicted optimal translation efficiency curve for $\lambda_p = 1$ to lie at the top edge of the observed data cloud rather than the bisecting it. This bisection is the consequence of fitting an effective relative load parameter to the abundance data in the unaugmented RLTO model.

\subsection{Discussion: Translation limits in \textit{E.~coli}}
\label{sec:translationallimit}
A critical assumption in the RLTO model to this point has been that the optimal central dogma parameters are realizable in the cell; however, translation can be limited by a number of different mechanisms. The superior performance of the constant- over the optimal-translation-efficiency model in \textit{E.~coli} (Fig.~\ref{fig:proteom}D) suggests that this assumption may not be satisfied for bacteria. How do translation limits affect the model phenomenology?

When considering possible  limits on translation, there are two natural mechanisms: (i) ribosome-number limit, where the number of ribosomes in the cell limits translation
and (ii) a ribosome-per-message limit, where the number of ribosomes per message is limiting.
Assuming the ribosome-number-limit mechanism, the original unconstrained optimization problem can be recast as a constrained optimization problem where the protein cost $\lambda_p$ is reinterpreted as a Lagrange multiplier to constrain the number of proteins translated (\textit{e.g.}\ \cite{garfken67:math}). In spite of this reformulation, we would still predict the same functional form for the coupling between the optimal translation efficiency and message number. (\textit{I.e.}\ it is still optimal to have a higher translation efficiency for highly-expressed genes even if the total number of proteins is fixed.) Therefore, the ribosome-number-limit mechanism  cannot be the rationale for the constant translation efficiency observed in \textit{E.~coli}.

Assuming the ribosome-per-message-limit mechanism, we limit the translation efficiency to a restricted range of values. If the unconstrained optimum lies above this range, the optimum is at the maximum limiting value. If the unconstrained optima for all genes lie above the realizable range, the model predicts a translation efficiency uncoupled from the message number, as observed.  These predictions are consistent with the observed central dogma regulatory program in \textit{E.~coli}. In added support of this hypothesis, Hausser \textit{et al.}\ have argued that \textit{E.~coli} translates close to just such a ribosome-per-message limit as a consequence of the finite ribosome complex footprint on a message \cite{Hausser:2019fi}.

\subsection{Methods: Estimate of the message cost and metabolic load}
\label{sec:estimateofLambda}

We can estimate the message cost $\lambda$ from the known total protein number for yeast and mammalian cells. (For \textit{E.~coli} this estimate is not possible since the protein cost in not determinative of the translation efficiency.) 

The optimal translation efficiency for gene $i$ is (Eq.~\ref{eqn:der3}):
\begin{equation}
\hat{\varepsilon}  = \textstyle \frac{\lambda}{\Lambda \ln 2} p_\Gamma(\hat{\mu}_{m}|  \frac{\hat{\mu}_{m}}{\ln 2},\hat{o} \ln 2),    
\end{equation}
and therefore the optimal protein number for gene $i$ is: 
\begin{equation}
\hat{\mu}_{p} = \hat{\mu}_{m}\hat{\varepsilon}  = \textstyle \frac{\lambda }{\Lambda \ln 2} \hat{\mu}_{m} p_\Gamma(\hat{\mu}_{m}|  \frac{\hat{\mu}_{m}}{\ln 2},\hat{o} \ln 2). \label{eq:protnumpred}
\end{equation}
We define the normalization constant $A$ {\color{red} (where we restore the explicit gene $i$ subscript)}:
\begin{equation}
A =  \textstyle  \sum_i \hat{\mu}_{m,i}\cdot \frac{1}{\Lambda \ln 2}  p_\Gamma(\hat{\mu}_{m,i}|  \frac{\hat{\mu}_{m,i}}{\ln 2},\hat{o}_i \ln 2),   \label{eqn:forA}
\end{equation}
where we have restored the explicit gene $i$ index running over all genes. Now, by summing Eq.~\ref{eq:protnumpred}, over all genes, we derive an expression for the total protein number $N_p^{\rm tot}$ in terms of the message cost $\lambda$ and the normalization constant $A$:
\begin{equation}
N_p^{\rm tot} = \textstyle \lambda A.
\end{equation}
Solving for the protein cost results in the estimate:
\begin{equation}
\hat{\lambda} = \textstyle \frac{N_p^{\rm tot}} {A}.\label{eqn:Lamhat}
\end{equation}
This message cost estimate $\hat{\lambda}$ can then be plugged into the metabolic load definition:
\begin{equation}
    N_0 \equiv L_0 +   \sum_{i}  (\lambda+ \varepsilon_i) \mu_{m,i}, \label{eq:load}
\end{equation}
to estimate its size: 
\begin{equation}
\hat{N}_0 \equiv L_0 +   \hat{\lambda} N_m^{\rm tot}+ N_p^{\rm tot} \label{eqn:Nhat}, 
\end{equation}
where we have ignored the non-protein and non-message contributions to the load ($L_0=0$). 

\subsubsection{Detailed protocol}
We first estimate the message numbers, as described in Sec.~\ref{sec:CCMN}, from  data. For each gene $i$, we set the optimal message number equal to the observed message number and then compute the optimal overabundance from the message number using Eq.~\ref{eqn:oeqn}. (Since the result is independent of the assumed $\Lambda$ value, we set an arbitrary initial value of $\Lambda = 10^{-5}$.) We then use these single gene optimal message number and overabundances to compute  $A$ using Eq.~\ref{eqn:forA}. In Eqs.~\ref{eqn:Lamhat} and \ref{eqn:Nhat}, we use the $N^{\rm tot}_p$ from Tab.~\ref{tab1refs}. $N_m^{\rm tot}$ is computed by summing the estimated message numbers.

\subsubsection{Estimate the message cost and metabolic load in yeast}

In yeast, the estimates are:
\begin{eqnarray}
A &=&  4.8\times 10^5,\\
\hat{\lambda} &=& 1.0\times 10^{2},\\
\hat{N}_0 &=& 6.2\times 10^7, \\
\hat{\Lambda} &=& 1.6\times 10^{-6},
\end{eqnarray}
where the data sources are described in detail in Sec.~\ref{sec:yeastdatasource}.

\subsubsection{Estimate the message cost and metabolic load in human cells}

In human cells, the estimates are:
\begin{eqnarray}
A &=&  4.3\times 10^6,\\
\hat{\lambda} &=& 7.1\times 10^{2},\\
\hat{N}_0 &=& 2.4\times 10^9, \\
\hat{\Lambda} &=& 2.9\times 10^{-7},
\end{eqnarray}
where the data sources are described in detail in Sec.~\ref{sec:humandatasource}.

\subsubsection{Estimate of the modified relative load in bacterial cells}

In bacterial cells, we will assume a constant translation efficiency model. We therefore use the modified relative load formula (Eq.~\ref{eqn:modload}) to estimate $\Lambda'$. We will assume that the load is dominated by proteins and messages:  
\begin{eqnarray}
N_0  &=& \sum_i (\lambda + \varepsilon) {\mu_{m,i}} = (\lambda + \varepsilon) N_m,
\end{eqnarray}
where $N_m$ is the total number of messages. We can then solve this equation for  $\Lambda'$:
\begin{equation}
\hat{\Lambda}'  = \textstyle\frac{\lambda + \varepsilon}{N_0} = \frac{1}{N_m} \approx 10^{-5},
\end{equation}
based on the total message number estimate for \textit{E.~coli}. (See Tab.~\ref{tab1refs}.)

\begin{figure}
  \centering
    \includegraphics[width=0.48\textwidth]{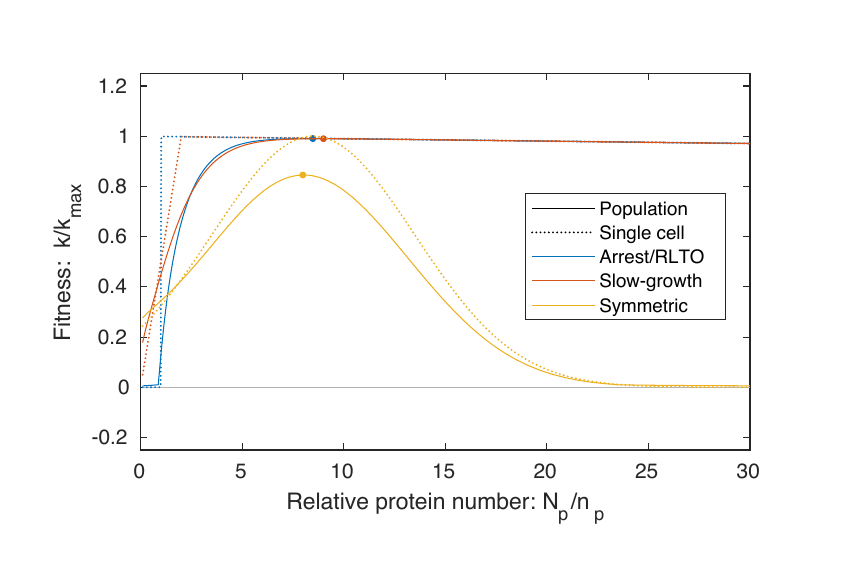}
      \caption{ \textbf{Exploring the mathematical mechanism of overabundance.} Single-cell and population growth rate are compared for three different models: arrest (RLTO), slow-growth, and symmetric models. In the arrest model (RLTO), the growth rate goes to zero below threshold protein level $n_p$. In the slow-growth model, the growth rate transitions continuously to zero as the $N_p$ is depleted below $n_p$.   In both the arrest and slow-growth models, there is a small negative slope above the threshold corresponding to the metabolic load. In the symmetric model, the fitness cost is symmetric about the optimum. Both the threshold-like and slow-growth models are optimized at mean expression levels $\mu_p$ far exceeding the threshold level $n_p$. This is a consequence of the highly-asymmetric dependence of the fitness on protein number $N_p$. This leads to the phenomenon of protein overabundance. In contrast, the symmetric model is optimized in close proximity to its single-cell optimum.
       \label{fig:threemod}}
\end{figure}

\section{Model robustness \& exploring alternatives to RLTO}
 \label{expmean}
 
In this section, we investigate the phenomenology of three different single-cell growth rate functions to determine what model features result in overabundance. We consider an \textit{arrest model} (the RLTO model), a \textit{slow-growth model}, and a \textit{symmetric model}.

\subsection{ Methods: Defining alternative models} 
In each case, we will assume that the protein number is described by a gamma distribution:
\begin{equation}
N_p \sim\Gamma(\textstyle \frac{\mu_m}{\ln 2}, \varepsilon \ln 2).
\end{equation}
We will assume the cell-cycle duration $T$ is determined by this stochastic protein number $N_p$ and then compute the population growth rate using Eq.~\ref{expmeaneqn} for a range of different message numbers $\mu_m$. In each case, $\tau_0 = 1/N_0$, $N_0 = 10^5$, $\varepsilon = 30$, $n_p= \varepsilon \ln 2$.  The mean expression level is $\mu_p=\mu_m \varepsilon$.

\subsubsection{Model 1: Arrest (RLTO) model}
The arrest (RLTO) model has cell cycle duration:  
\begin{equation}
T = \tau_0 \begin{cases} \infty, & N_p < n_p \\
   N_0 + N_p, &N_p > n_p \end{cases},   
\end{equation}
where protein expression below threshold $n_p$ results in growth arrest.

\subsubsection{Model 2: Slow-Growth model}
In the slow-growth model, we imagine two processes: (i) checkpoint process X and (ii) other processes. The cell will divide after whichever process finishes last. Other processes will finish after time predicted by the metabolic load, identical to the threshold model defined above. However, we model checkpoint process X as the completion of a fixed amount of activity in an irreversible process. We will therefore assume it will take a time inversely proportional to the amount of enzyme X ($N_p$). The amount of activity is set by effective threshold $n_p$: 
\begin{equation}
T = \tau_0\max\{ \textstyle N_0+N_p,\frac{2n_pN_0}{N_p} \}
\end{equation}
such that $n_p$ defines the level of protein required to make the growth rate half the metabolic limit. 

Unlike the arrest model, cell growth slows but does not stop for $N_p < n_p$. This model will test whether the results of the RLTO model are an artifact of the assumed arrest-based slow growth.

\subsubsection{Model 3: Symmetric model}
For the symmmetric model, we choose the model parameters such that the single-cell optimum was close to the other models: $n_0 = 8.5$, $\sigma_n=5$. The cell-cycle duration is:
\begin{equation}
T = \tau_0 N_0 \exp( \textstyle \frac{(N_p-n_0)^2}{2\sigma_n^2} ),
\end{equation}
such that the noise-free growth rate will be Gaussian is $N_p$.

\subsection{Results: Overabundance is a robust prediction}

The growth rates as a function of the mean expression level $\mu_p$ are shown in Fig.~\ref{fig:threemod}.  The symmetric model has a population optimum in close proximity to its single-cell optimum, as we intuitively expect. However, both the arrest (RLTO) model and the slow-growth model have optima far above the threshold number $n_p$. We therefore conclude that it is fitness asymmetry rather than growth arrest that is responsible for the overabundance phenomenon.

Why doesn't growth arrest of a sub-population lead to a stronger effect than the same sub-population growing slowly?
In Ref.~\cite{Huang:2022fu}, we showed that the population doubling time $\overline{T}$ can be understood as the exponential mean of the stochastic cell-cycle duration:
 \begin{equation}
 \overline{T} \equiv f^{-1} [\mathbb{E}_T f(T)], \label{eqnexpmean}
 \end{equation}
 where $\mathbb{E}_T$ is the expectation over the stochastic duration $T$ and $f(t) \equiv \exp(-kt)$, where $\inliner{k=\overline{T}^{-1}\ln 2}$ is the population growth rate. Due to the functional form of $f(t)$, any long cell cycles are exponentially suppressed in their contribution to the exponential mean. Therefore, low-probability extremely-long-duration cell cycles only contribute to the growth rate by reducing the fraction of growing cells.

\begin{table*}
\resizebox{\textwidth}{!}{\begin{tabularx}{1.15 \textwidth}{ X | X | X  X X | X  X  X | X X }
                         &            &                   &                             &                               &  \multicolumn{3}{c|}{Total number of }       &    \multicolumn{2}{c}{Average}      \cr
\centering{Model organism}                &  \centering{Growth condition} & \centering{Doubling time}:  & \centering{Message lifetime}: &  \centering{Message recycling ratio:} & \centering{messages /cell:}      &   \centering{messages /cell-cycle:}          &    \centering{proteins:}     &  \centering{translation efficiency:}  &  \centering{translation rate:} \cr
                         &                               & \centering{$T$}    &  \centering{$\tau_m=\gamma_m^{-1}$} & \centering{ $m=T/\tau_m$ }             &   \centering{$N^{\rm tot}_{m/c}$}  &   \centering{$N^{\rm tot}_{m}$}    &  \centering{$N^{\rm tot}_{p}$}   & \centering{$\varepsilon$}  &  \centering{$\beta_p$ (h$^{-1}$)}  \cr
\hline
\hline
 \textit{Escherichia coli}  & \raggedleft   LB & \raggedleft  30 min \cite{Bernstein:2002rp}  & \raggedleft $2.5$ min \cite{Chen:2015wt} & \raggedleft 12  & \raggedleft $7.8 \times 10^3$ \cite{Bartholomaus:2016df}   & \raggedleft $9.4 \times 10^4$ & \raggedleft $3 \times 10^6$ \cite{Milo:2013yn} & \raggedleft  $22$  & \raggedleft  $530$   \cr 
 (\textit{E.~coli})         & \raggedleft   M9 & \raggedleft  90 min \cite{Bernstein:2002rp} & \raggedleft $2.5$ min  \cite{Chen:2015wt} & 36 \raggedleft & \raggedleft $2.4 \times 10^3$ \cite{Bartholomaus:2016df} & \raggedleft $8.6 \times 10^4$ & \raggedleft $3 \times 10^6$ \cite{Milo:2013yn}  & \raggedleft $24$ & \raggedleft  $580$  \cr 
                     \hline
 \textit{Sacchromyces cerevisiae (Yeast--haploid)}     & \raggedleft  YEPD    & \raggedleft  90 min \cite{Albe2002book} & \raggedleft  22 min \cite{Chia:1979jm} & \raggedleft 4 & \raggedleft $2.9 \times 10^4$ \cite{Pelechano:2010gz} & \raggedleft $1 \times 10^5$ & \raggedleft  $5 \times 10^7$ \cite{Futcher:1999sb} & \raggedleft $4 \times 10^2$   & \raggedleft $410$  \cr 
                                          \hline
\textit{Mus musculus (Mammalian-mouse)}    & \raggedleft   Tissue        & \raggedleft  27.5 h \cite{Schwanhausser:2011ec} & \raggedleft  15 h \cite{Schwanhausser:2011ec} & \raggedleft 1.8  & \raggedleft $1.7 \times 10^5$ \cite{Schwanhausser:2011ec} & \raggedleft $3 \times 10^5$ \cite{Schwanhausser:2011ec} & \raggedleft  $3 \times 10^9$ \cite{Schwanhausser:2011ec} & \raggedleft $1 \times 10^4$   & \raggedleft $660$  \cr                                                                              
                                          \hline
 \textit{Homo sapiens (Human)}         & \raggedleft   Tissue   & \raggedleft  24 h  \cite{Cooper:2000af}  & \raggedleft  14 h  \cite{Yang:2003mf} & \raggedleft  1.7 & \raggedleft   $3.6 \times 10^5$ \cite{Albe2002book} & \raggedleft $5 \times 10^5 $  & \raggedleft $2 \times 10^9$ \cite{Milo:2013yn} & \raggedleft  $4 \times 10^3$ & \raggedleft  120  \cr 
                     \hline
\end{tabularx}}

\caption{ \textbf{Central dogma parameters for three model organisms with detailed references.}  Columns three through seven hold representative values for  measured central-dogma parameters for the model organisms described in the paper. Each value is followed by a reference for its source. \label{tab1refs} 
}
\end{table*}

\section{Quantitation of central dogma parameters for one-message-rule}

The RLTO model predicts the \textit{one-message-rule} for the lower threshold  on transcription for essential genes. In this section, we use transcriptome data from the literature to test this prediction. We first describe the sources of the data (Sec.~\ref{sec:desriptonTab1}), how the estimates are computed (Sec.~\ref{sec:methodsCDPs}),  the results (Sec.~\ref{sec:omrress}) and discussion (Sec.~\ref{sec:discussrule}).

\label{sec:desripton}

\subsection{Methods: Selection of  central dogma parameter estimates}
\label{sec:desriptonTab1}

The estimates for central dogma model parameters come from two types of data: (i) quantitative measurement of cellular-scale parameters for each organism (total number of messages in the cell, cell cycle duration, \textit{etc}) and (ii) genome-wide studies quantitative of mRNA and protein abundance.

For the cellular-scale central dogma parameters, we relied heavily on an online compilation of biological numbers: BioNumbers \cite{Milo:2010nw}. This resource provides a collection of curated quantitative estimates for biological numbers, as well as their original source. In the interest of conciseness, we have cited only the original source in the Tab.~\ref{tab1refs}, although we are extremely grateful and supportive of the creators of the BioNumbers website for helping us very efficiently identify consensus estimates for the parameters of the central dogma parameters.

For the selection of genome-wide studies on abundance, we used many of the same resources cited in BioNumbers as well as studies selected by a previous study of a quantitative analysis of the central dogma: Hausser \textit{et al.} \cite{Hausser:2019fi}.

\subsubsection{\textit{E.~coli} data}

\idea{Message lifetimes:} The message lifetimes (and median lifetime) were taken from a recent transcriptome-wide study by Chen \textit{et al.}  \cite{Chen:2015wt}. These investigators measured the lifetime in both rapid (LB) and slow growth (M9). 

\idea{Noise:} Taniguchi \textit{et al.} 
have performed a beautiful  simultaneous study of the proteome and transcriptome with single-molecule sensitivity \cite{Taniguchi2010}. Although we use the noise analysis data from this study for our supplemental analysis of \textit{E.~coli} noise, it is not the source for our \textit{E.~coli} transcriptome data due to the extremely slow growth of the cells in this study (150 minute doubling time), which is not consistent with the growth conditions for the other sources of data. 

\idea{mRNA abundance:} Instead, we used data from the more recent Bartholomaus \textit{et al.} study \cite{Bartholomaus:2016df}, which characterizes the transcriptome in both rapid (LB) and slow growth (M9). 

\idea{Total cellular message number.} This study was chosen since it was the source of the BioNumbers estimates of cellular message number in \textit{E.~coli} (BNID 112795 \cite{Milo:2010nw}). 

\idea{Doubling time:} The source of the doubling times for rapid (LB) and slow (M9) growth of \textit{E.~coli} comes from Bernstein \cite{Bernstein:2002rp}. 

\idea{Essential gene classification.} The classification of essential genes in \textcolor{red}{\textit{E.~coli}} comes from the construction of the Keio knockout collection from Baba \textit{et al.} \cite{Baba:2008vn}.

\idea{Protein number.} The total protein number in \textit{E.~coli} came from Milo's recent review of this subject \cite{Milo:2013yn}.

\subsubsection{Yeast data}

\label{sec:yeastdatasource}

\idea{Message lifetimes:} The message lifetimes (and median lifetime) were taken from Chia \textit{et al.} \cite{Chia:1979jm}.

\idea{Noise:} The noise data was taken from the Newman \textit{et al.} study, which used flow cytometry of a library of fluorescent fusions to characterize protein abundance with single-cell resolution \cite{Newman:2006nl}.

\idea{mRNA abundance:} The transcriptome data comes from the very recent Blevins \textit{et al.} study \cite{Blevins:2019yj}.

\idea{Total cellular message number.} There are a wide-range of estimates for the total cellular message number in yeast: $1.5\times 10^4$ \cite{Hereford:1977au} (BNID 104312 \cite{Milo:2010nw}), $1.2\times10^4$ \cite{Haar:2008eg} (BNID 102988 \cite{Milo:2010nw}), $6.0\times 10^4$ \cite{Zenklusen:2008fo} (BNID 103023 \cite{Milo:2010nw}), $2.6\times 10^4$ \cite{Pelechano:2010gz} (BNID 106763 \cite{Milo:2010nw}) and $3.0\times 10^4$ \cite{Miura:2008zn}. We used the compromise value of $2.9\times 10^4$.

\idea{Doubling time:}  The doubling time was taken from \cite{Albe2002book}.

\idea{Protein number.} The total protein number in yeast comes from Futcher \textit{et al.} \cite{Futcher:1999sb}.

\idea{Essential gene classification.} The classification of essential genes in yeast comes from van Leeuwen \textit{et al.} \cite{Leeuwen:2020eh}.

\idea{Proteome abundance data:} The proteome abundance data came from two sources: flow cytometry of fluorescent fusions from Newman \textit{et al.} \cite{Newman:2006nl} as well as mass-spec data from de Godoy \textit{et al.} \cite{Godoy:2008eb}.

\subsubsection{Human data}

\label{sec:humandatasource}

\idea{Message lifetimes:} The message lifetimes (and median lifetime) were taken from Yang \textit{et al.} \cite{Yang:2003mf} who reported a median half life of 10 h which corresponds to a lifetime of 14 h.

\idea{mRNA abundance:} The transcriptome data comes from the data compiled by the Human Protein Atlas \cite{Uhlen:2015jm}, which we averaged over tissue types.

\idea{Total cellular message number.} The total cellular message number in human comes from Velculescu \textit{et al.} \cite{Velculescu:1999hf} (BNID 104330 \cite{Milo:2010nw}).

\idea{Doubling time:}  The doubling time was taken from \cite{Cooper:2000af}.

\idea{Protein number.} The total protein number in human came from Milo's recent review of this subject \cite{Milo:2013yn}.

\idea{Essential gene classification.} The classification of essential genes in human comes from Wang \textit{et al.} \cite{Wang:2015lb}.

\begin{figure*}
  \centering
    \includegraphics[width=0.95\textwidth]{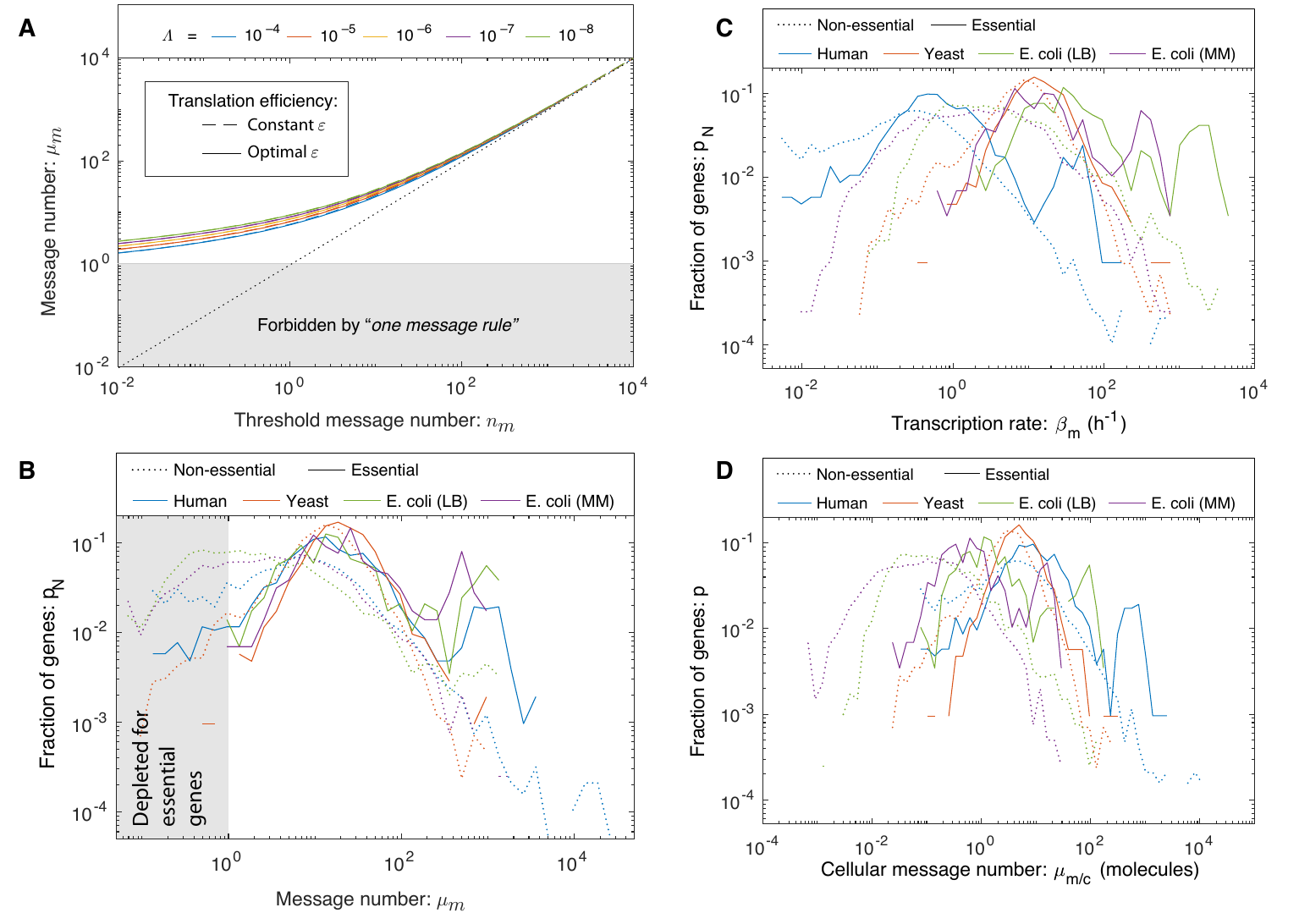}
      \caption{ {\color{red} \textbf{The one message rule. Panel A: One-message-rule for essential genes.}} For highly transcription genes (high $\mu_m$), little compensation for noise is required and the optimal message number tracks with the threshold message number $n_m$. However, as the threshold message number approaches one ($n_m\rightarrow 1$), the noise is comparable to the mean, and the optimal message number $\mu_m$  increases to compensate for the noise. As a result, a lower threshold of roughly one message per cell-cycle is required for essential genes.  This threshold is predicted for both fixed (dashed) and optimized translation efficiency (solid). The threshold is weakly dependent on relative load $\Lambda$. \textbf{Panel B: A one message threshold is observed in three evolutionarily-divergent organisms.} As predicted by the RLTO model, essential, but not nonessential genes, are observed to be expressed above a one message per cell-cycle threshold. All organisms have roughly similar distributions of message number for essential genes, which are not observed for message numbers below a couple per cell cycle. \textbf{Panel C: The distribution of gene transcription rate.} The typical transcription rate varies by two orders-of-magnitude between organisms.  \textbf{Panel D: The distribution of gene cellular message number.} There is also a two-order-of-magnitude variation between typical cellular message numbers.  No consistent lower threshold is observed for either statistic.
            \label{fig:wallAB}
      \label{SIfig:onemessagerule}}
\end{figure*}

\subsection{Methods: Quantitative estimates of central dogma parameters}

\label{sec:methodsCDPs}

\subsubsection{Estimating the \textit{cellular message number}: $\mu_{m/c}$}

For each model organism (and condition), we found a consensus estimate from the literature for the total number of mRNA messages per cell $\inliner{N_{m/c}^{\rm tot}}$. This number and its source are provided in Tab.~\ref{tab1refs}. To estimate the number of messages corresponding to gene $i$, we re-scaled the un-normalized abundance level $r_i$:
\begin{equation}
    N_{m/c, i} = N_{m/c}^{\rm tot}  \textstyle\frac{r_i}{\sum_j r_j}, \label{eqn:rescalcmn}
\end{equation}
where the sum over gene index $j$ runs over all genes.

\subsubsection{Estimating the \textit{transcription rate}: $\beta_{m}$}
\label{sec:esttransrate}
To estimate the transcription rate for gene $i$, we start from the estimated cellular message number $N_{m/c,i}$ and use the \textcolor{orange}{canonical steady-state noise model}  prediction for the cellular message number:
\begin{equation}
    N_{m/c,i} = \beta_{m,i}/\gamma_{m,i},
\end{equation}
where $\gamma_{m,i}$ is the message decay rate. Since gene-to-gene variation in message number is dominated by the transcription rate (\textit{e.g}~\cite{Chen:2015wt}), we estimate the decay rate as the inverse gene-median message lifetime:
\begin{equation}
\gamma_{m,i} = \tau_m^{-1}, 
\end{equation}
for which a consensus value was found from the literature. This number and its source are provided in Tab.~\ref{tab1refs}. We then estimate the gene-specific transcription rate:
\begin{equation}
    \beta_{m,i}=N_{m/c,i}/\tau_m.
\end{equation}

\subsubsection{Estimating the \textit{message number}:  $\mu_{m}$}
\label{sec:CCMN}

To estimate the message number of gene $i$, we use the predicted value from the \textcolor{red}{canonical steady-state noise} model (see \ref{sectelegraph}):
\begin{equation}
    N_{m,i} = T \beta_{m,i} = \textstyle \frac{T}{\tau_m} N_{m/c,i},
\end{equation}
where $T$ is the doubling time and $N_{m/c,i}$ is the cellular message number (Eq.~\ref{eqn:rescalcmn}).


\subsection{Results: Histograms of central dogma transcriptional statistics}

\label{sec:omrress} 
We generated histograms for each of the three transcriptional statistics: transcription rate $\beta_m$, cellular message number $\mu_{m/c}$, and message number $\mu_m$. The histograms for  transcription rate and cellular message number do not show a consistent lower limit (as predicted) and are shown in Fig.~\ref{fig:wallAB}; however, the histogram for message number does show a consistent lower bound for the three model organisms and is shown in Fig.~\ref{SIfig:onemessagerule}B.

\begin{table*}
\resizebox{\textwidth}{!}{\begin{tabularx}{1.15\textwidth}{ c | c | X  |c c c| }
Gene  & Message   & Annotated function & Essential (E)/   \\
name &  number:       &  from Ecocyc                  &   Nonessential (N)      \\ 
& $\mu_m$&  & Ref.~\cite{Baba:2008vn}, \cite{Gerdes:2003ys}, \cite{Goodall:2018rm} \\ \hline   
\textit{alsK} &        0.3  & The alsK gene encodes a D-allose kinase. Its role in the degradation of D-allose is unclear; AlsK is not required for utilization of a D-allose carbon source; this effect may be due to the presence of other ambiguous sugar kinases within \textit{E.~coli} K-12. &  E, N, N \\
& & & \\
    \textit{bcsB} &    0.4 & BcsB is encoded in a predicted operon together with \textit{bcsA}, \textit{bcsZ} and \textit{bcsC}. In other organisms, these genes are involved in cellulose biosynthesis, a characteristic of the rdar (red, dry and rough) morphotype. However, the K-12 laboratory strain of \textit{E.~coli} does not show a rdar morphotype and does not produce cellulose. &  E, N, N \\ 
    & & & \\
    \textit{entD} &    0.4 & AcpS is the founding member of a 4'-phosphopantetheinyl (P-pant) transferase protein family that includes \textit{E.~coli} EntD, \textit{E.~coli} o195 protein, and \textit{Bacillus subtilis} Sfp; family members share two conserved motifs but relatively low sequence identity overall. & E, N, N  \\
    & & & \\
    \textit{yafF} &    0.4 & No information about this protein was found by a literature search conducted on April 19, 2017. & E,-, N
 \\
    & & & \\
    \textit{yagG} &    0.6 & \textit{yagGH} is predicted to be a member of the XylR regulon; its products may mediate transport (YagG) and hydrolysis (YagH) of xylooligosaccharides; putative XylR and CRP binding sites are identified upstream of \textit{yagGH}. &  E,-, N \\
    & & & \\
    \textit{yceQ} &    0.2 & No information about this protein was found by a literature search conducted on July 12, 2017. & E, E, N \\
    & & & \\
    \textit{ydiL} &    0.2 & No information about this protein was found by a literature search conducted on April 7, 2017. & E, N, N \\ 
    & & & \\
    \textit{yhhQ} &   0.4 & YhhQ is an inner membrane protein implicated in the uptake of queuosine (Q) precursors - 7-cyano-7-deazaguanine (preQ0) and 7-aminomethyl-7-deazaguanine (\textit{preQ1}) - for Q salvage. Q-modified tRNA is absent in $\Delta$\textit{queD} and $\Delta$\textit{queD} $\Delta$\textit{yhhQ} strains grown in minimal media with glycerol; Q-modified tRNA is detected when a $\Delta$\textit{queD} strain is grown in minimal media plus 10 nM \textit{preQ0} or \textit{preQ1} but is absent when a $\Delta$\textit{queD} $\Delta$\textit{yhhQ} strain is grown under these conditions. \textit{yhhQ} expressed from a plasmid restores the presence of Q-modified tRNA in a $\Delta$\textit{queD} $\Delta$\textit{yhhQ} strain. &  E,-, N  \\
    & & & \\
    \textit{yibJ} &    0.3 & No information about this protein was found by a literature search conducted on July 9, 2018. & E, N, N \\
    & & & \\
    \textit{ymfK} &    0.4 & YmfK is a component of the relic lambdoid prophage e14 and is likely the SOS-sensitive repressor. It is similar to the P34 gene of the \textit{Shigella flexneri} bacteriophage SfV and belongs to the LexA group of SOS-response transcriptional repressors. & E, E, E 
 \end{tabularx}}

\caption{ \textbf{Below-threshold essential genes identified in \textit{E.~coli}.} This table describes the message numbers and annotations for essential genes that we estimated to have expression below the threshold of one message per cell cycle. However, in the final column, we show classifications from three different studies. Only one of the identified genes, \textit{ymfK}, was consistently defined as essential. } \label{ess_tab}

\end{table*}

\subsection{Discussion:  \textit{E.~coli} essential genes below the one-message-rule threshold}

\label{sec:discussrule}

Since our own preferred model system is \textit{E.~coli}, we focus here. Our essential gene classification was based on the construction of the Keio knockout library \cite{Baba:2008vn}. By this classification, 10 essential genes were below threshold. (See Tab.~\ref{ess_tab}.) Our first step was to determine what fraction of these genes were also classified as essential using transposon-based mutagenesis \cite{Gerdes:2003ys,Goodall:2018rm}. Of the 10 initial candidates, only one gene, \textit{ymfK}, was consistently classified as an essential gene in all three studies, and we estimate that its message number is just below the threshold ($\mu_m = 0.4$). \textit{ymfK} is located in the lambdoid prophage element e14 and is annotated as a CI-like  repressor which regulates lysis-lysogeny decision \cite{Mehta:2004sy}. In $\lambda$ phase, the CI repressor represses lytic genes to maintain the lysogenic state. A conserved function for \textit{ymfK} is consistent with it being classified as essential, since its regulation would prevent cell lysis.
However, since \textit{ymfK} is a prophage gene, not a host gene, it is not clear that its expression should optimize host fitness, potentially at the expense of phage fitness. 
In summary, closer inspection of  below-threshold essential genes supports the threshold hypothesis.

\section{Analysis of gene-expression noise}

\label{sec:aonS}

This section provides a detailed development of gene expression noise. \textcolor{red}{We continue the discussion of the model from Sec.~\ref{sec:noise_intro} that provided a self-contained development of the noise models developed by others which are the input to the RLTO model.} Secs.~\ref{sec:non-canon}-\ref{sec:end_noise_s} describe the RLTO prediction of non-canonical noise scaling and the test of this model.

\label{sec:statmodelnoise}

\subsection{Results: RLTO model predicts non-canonical noise scaling}

\label{sec:non-canon}

The predicted scaling of the optimal translation efficiency with message number has many important implications, including on the global characteristics of noise. Based both on theoretical and experimental evidence, it is widely claimed that gene-expression noise should be inversely proportional to protein abundance  \cite{Bar-Even:2006rv,Taniguchi2010}:
\begin{equation}
{\rm CV}_p^2 \propto \mu_p^{-1}, \label{eqn:canonical}
\end{equation}
for low-expression proteins, as observed in \textit{E.~coli} \cite{Taniguchi2010}; however, the more fundamental prediction is that the noise is inversely proportional to the message number:  
\begin{eqnarray}
\mu_p &=& \mu_m \varepsilon, \label{eqn:meanprot}\\
{\rm CV}^2_p &=& \textstyle \frac{\ln 2}{\mu_m}. \label{eqn:CVP2}
\end{eqnarray}
In \textit{E.~coli}, the translation efficiency is roughly constant (\textit{i.e.}~$\inliner{\hat{\mu}_p \propto \hat{\mu}_m}$, Fig.~\ref{fig:proteom}D) and therefore Eq.~\ref{eqn:CVP2} is consistent with the canonical noise model (Eq.~\ref{eqn:canonical}). However, in eukaryotes, the translation efficiency grows with message number (\textit{i.e.}~$\inliner{\hat{\mu}_p \propto \hat{\mu}_m^{2}}$, Fig.~\ref{fig:proteom}BC). If we substitute this proportionality into Eq.~\ref{eqn:CVP2}, we predict the non-canonical noise scaling:
\begin{equation}
{\rm CV}_p^2 \propto \mu_p^{-1/2}, \label{eqn:noncanonical}
\end{equation}
for eukaryotic cells.

\subsection{Methods: Analysis of gene expression noise}
\label{sec:SMnoise} 

The quantitative model for gene expression noise includes multiple contributions: 
\begin{equation}
    {\rm CV}^2_p \approx \textstyle\frac{1}{\mu_p} + \textstyle\frac{\ln 2}{\mu_m} + c_0, 
\end{equation}
where the first term can be understood to represent the Poisson noise from translation, the second term the Poisson noise from transcription, and the last term, $c_0$, is called the \textit{noise floor} and is believed to be caused by the cell-to-cell variation in metabolites, ribosomes, and polymerases \textit{etc.} \cite{Elowitz:2002tb,Swain:2002te}.

\begin{figure}
    \centering
    \includegraphics[width=0.48\textwidth]{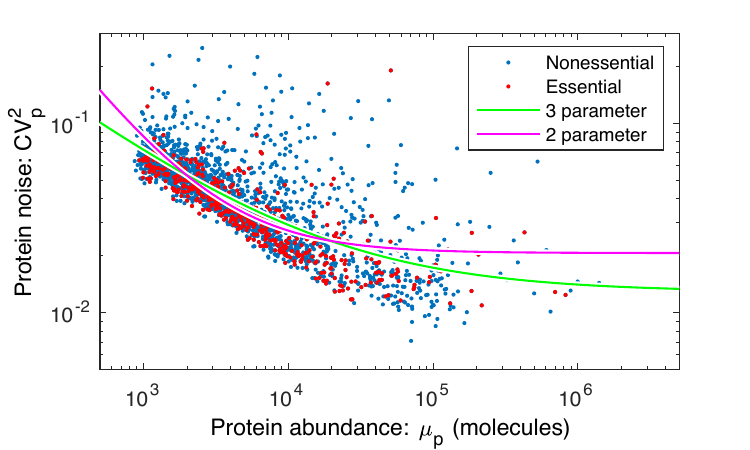}
    \caption{\textbf{ Yeast noise fit against canonical noise model, with a noise floor. } Yeast noise data fit with the 2- (null hypothesis with $\mu_p^{-1}$ dependence) and 3- parameter ($\mu_p^a$) models. The two-parameter model corresponds to the canonical noise model (Eq.~\ref{eqn:canonical}) and fails to quantitatively fit the data.}
    \label{fig:yeastnoise}
\end{figure}


 In the main text of the paper, we have ignored the role of the noise floor in the analysis of noise in yeast. Unlike \textit{E.~coli}, where the noise floor is high (${\rm CV}^2_p=0.1$) and is determinative of the noise associated with almost all essential genes \cite{Taniguchi2010,Elowitz:2002tb,Swain:2002te}, in yeast the noise floor is much lower (${\rm CV}^2_p=0.01$) and therefore affects only genes with the highest expression.  
 
 In this section, we will consider models that include the noise floor, since its presence can make the noise scaling more difficult to interpret. To determine if the scaling of the noise is consistent with the canonical assumption that the noise is proportional to $\mu_p^{-1}$ for low expression, we  will consider two competing empirical models for the noise (Fig. \ref{fig:yeastnoise}). In the null hypothesis, we will consider a model:
 \begin{equation}
 \eta_0(\mu_p; b,c ) = \textstyle\frac{b}{\mu_p}+c,  \label{eqn:newnullS}  
 \end{equation}
and an alternative hypothesis with an extra exponent parameter $a$:
 \begin{equation}
 \eta_1(\mu_p; a,b,c ) = \textstyle\frac{b}{\mu_p^a}+c. \label{eqn:althyp}  
 \end{equation}
We will assume that ${\rm CV}^2_p$ is normally distributed about $\eta$ with unknown variance $\sigma^2_\eta$.

In this context, a maximum likelihood analysis is equivalent to least-squares analysis. Let the sum of the squares be defined:
\begin{equation}
 S_I({\bm \theta}) \equiv \sum_i [{\rm CV}_{p,i}^2- \eta_I(\mu_{p,i}; {\bm \theta} )]^2, 
\end{equation}
for model $I$ where $\bm \theta$ represents the parameter vector. The maximum likelihood parameters are 
\begin{equation}
\hat{\bm \theta} = \arg \max_{\bm \theta} S_I({\bm \theta}),
\end{equation}
with residual norm:
\begin{equation}
\hat{S}_I = S_I(\hat{\bm \theta}).
\end{equation}
To test the null hypothesis, we will use the canonical likelihood ratio test with the test statistic:
\begin{equation}
Z \equiv 2\ln \frac{q_1}{q_0}, 
\end{equation}
where $q_0$ and $q_1$ are the likelihoods of the null and alternative hypotheses, respectively. Wilks' theorem states that $Z$ has a chi-squared distribution of dimension equal to the difference of the dimension of the alternative and null hypotheses ($3-2=1$). 

\subsubsection{Hypothesis test I}

In our first analysis, we will estimate the variance directly. We computed the mean-squared difference for successive ${\rm CV}^2_p$ values, sorted by mean protein number $\mu_p$. The variance estimator is 
\begin{equation}
\hat{\sigma}^2_\eta = \textstyle\frac{1}{2} \left<({\rm CV}^2_{p,i}-{\rm CV}^2_{p,i+1})^2 \right>_i=6.3\times 10^{-4},
\end{equation}
where the brackets represent a standard empirical average over gene $i$ for the $\mu_p$-ordered gene ${\rm CV}_p^2$ values. The test statistic can now be expressed in terms of the residual norms:
\begin{eqnarray}
Z &=& (\hat{S}_1-\hat{S}_2)/\hat{\sigma}^2_\eta, \\
&=& 3.3\times 10^4,
\end{eqnarray}
which corresponds to a p-value far below machine precision. We can therefore reject the null hypothesis.

\subsubsection{Hypothesis test II}

In a more conservative approach, we can use maximum likelihood estimation to estimate the variance of each model independently as a model parameter. In this case, the test statistic can again be expressed in terms of the residual norms:
\begin{eqnarray}
Z &=& N \ln \textstyle\frac{\hat{S}_1}{\hat{S}_2}, \\
&=& 1.6\times 10^2,
\end{eqnarray}
where $N$ is the number of data points. (Details of derivation are in Sec.~\ref{sec:statdetails}.)  In this case, the p-value can be computed assuming the Wilks' theorem (\textit{i.e.}~the chi-squared test):
\begin{equation}
p = 6\times10^{-36},    
\end{equation}
again, strongly rejecting the null hypothesis.

\subsubsection{Maximum likelihood estimates of the parameters}

\label{sec:yeastnoiseempiricalmodel}

In the alternative hypothesis, the maximum likelihood estimate (MLE) of the empirical noise model (Eq. \ref{eqn:althyp}) parameters are (Fig. \ref{fig:yeastnoise}):
\begin{eqnarray}
a &=& 0.57 \pm 0.02,\\
b &=& 3.0 \pm 0.5,\\        
c &=& 0.013  \pm 0.001,    
\end{eqnarray}
where the parameter uncertainty has been estimated using the Fisher Information in the usual way using the MLE estimate of the variance \cite{Rao1945,Cramer1946}.

%

%

\subsubsection{Details: Statistical details MLE estimate of the variance}

\label{sec:statdetails}

The minus-log-likelihood for the normal model $I$ is:
\begin{equation}
h_I(\hat{\bm \theta},\sigma^2) = \textstyle\frac{N}{2} \ln 2\pi \sigma^2 + \textstyle\frac{1}{2\sigma^2}\hat{S}_I, 
\end{equation}
where $\hat{S}_I$ is the least-square residual. We then minimize $h_I$ with respect to the variance $\sigma^2$:
\begin{equation}
\partial_{\sigma^2} h|_{\hat{\sigma}^2} = 0,
\end{equation}
to solve for the MLE  $\hat{\sigma}^2$:
\begin{equation}
\hat{\sigma}^2 = \textstyle \frac{1}{N} \hat{S}_I.
\end{equation}
Next we evaluate $h$ at the variance estimator:
\begin{equation}
h_I(\hat{\bm \theta},\hat{\sigma}^2) = \textstyle\frac{N}{2} \left[\ln 2\pi \frac{\hat{S}_I}{N} + 1\right].
\end{equation}
The test statistics can be written in terms of the $h$'s:
\begin{eqnarray}
  Z &=& 2h_0(\hat{\bm \theta},\hat{\sigma}^2)-2h_1(\hat{\bm \theta},\hat{\sigma}^2),\\
  &=& N \ln \textstyle \frac{\hat{S}_0}{\hat{S}_1},
\end{eqnarray}
which can be evaluated directly in terms of the residual norms for the null and alternative hypotheses.

\subsection{Results: Non-canonical noise scaling is observed in yeast.}

To test the RLTO model predictions for noise scaling, we reanalyze the dataset collected by Newman \textit{et al.}, who performed a single-cell proteomic analysis of yeast by measuring the abundance of fluorescent fusions by flow cytometry \cite{Newman:2006nl}. Since the competing models (Eqs.~\ref{eqn:canonical} and \ref{eqn:noncanonical}) make different scaling predictions, we first apply a statistical test to determine whether the observed scaling is consistent with the canonical model (Eq.~\ref{eqn:canonical}).
We consider the null hypothesis of the canonical model (Eq.~\ref{eqn:newnullS}) and the alternative hypothesis with an unknown scaling exponent (Eq.~\ref{eqn:althyp}).  To test the models, we perform a null hypothesis test. (A detailed description of the statistical analysis, which includes the contribution of the noise floor, is given in the Sec.~\ref{sec:statmodelnoise}.)
We reject the null hypothesis with a p-value of $\inliner{p = 6\times 10^{-36}}$. 
The observed scaling exponent is $\hat{a} = -0.57 \pm 0.02$, which is close to our predicted estimated exponent from the RLTO model ($-\frac{1}{2}$).


\subsection{Methods: Parameter-free prediction of noise from protein-message relation.}

By combining the noise model (Eq.~\ref{eqn:CVP2}) with a protein-message abundance relation, the relation between protein abundance and noise can be predicted without additional fitting parameters. (We call this prediction \textit{parameter-free} since, although a parameter is fit when determining the protein-message abundance, once this relation has been established, no new parameters are fit in order to predict the noise.)
To test this prediction, we will compare  three competing models: (i) the RLTO model, (ii) an empirical protein-message abundance model, and (iii) the constant-translation-efficiency model. 

\subsubsection{Estimating \textit{protein number} ($\mu_p$) for the noise analysis }
\label{sec:MS}
\label{sec:Fl}


The protein abundance data for yeast grown in YEPD media and measured with flow cytometry fluorescence \cite{Newman:2006nl} were given in arbitrary units (AU). In order to convert from AU to protein number, the fluorescence values were rescaled by comparing with mass-spectrometry protein abundance data for yeast grown in YNB media \cite{Godoy:2008eb}. Since the protein abundance from mass-spectrometry was given in terms of intensity, the intensity values were first rescaled by the total number of proteins in yeast, $5 \times 10^7$. (See Sec.~\ref{sec:yeastdatasource}.) The mass-spectrometry protein data was thresholded at 10 proteins, based on the assumption that the noise of the data for 10 and fewer proteins makes the data unreliable. Next, the log of the fluorescence protein abundance in AU as a function of the log of thresholded mass-spectrometry protein abundance was fit as a linear function with an assumed slope of 1 to find the offset, 3.9, (Fig.\ \ref{fig:AUconv}) which corresponds to a multiplicative scaling factor. We then used that offset value to rescale the fluorescence data from AU to protein number. We also compared to yeast grown in SD media \cite{Newman:2006nl} and found a similar offset result.


\begin{figure}
    \centering
    \includegraphics[width=0.48\textwidth]{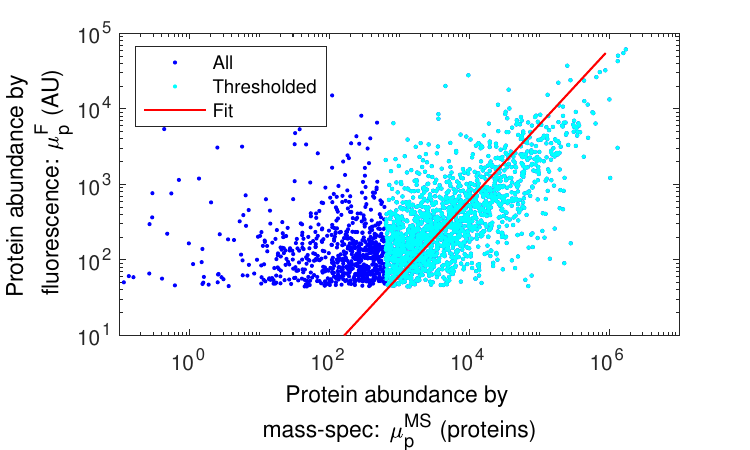}
    \caption{\textbf{ Fit to rescale fluorescence intensity to protein number. } Protein abundance from flow cytometry fluorescence \cite{Newman:2006nl} as a function of mass-spectrometry scaled abundance \cite{Godoy:2008eb}. The mass-spectrometry data was thresholded at 10 proteins, and then a linear fit was performed to find the multiplicative offset of 3.9, which was used to convert protein fluorescence AU to number.}
    \label{fig:AUconv}
\end{figure}

\subsubsection{Empirical models for yeast gene expression}

To generate the empirical model for protein number as a function of message number, we used protein abundance data from Newman \textit{et al.} \cite{Newman:2006nl}, re-scaled to estimate protein number (Sec.~\ref{sec:Fl}) and transcriptome data from Lahtvee \textit{et al.} \cite{Lahtvee:2017al}, re-scaled to estimate message number (Sec.~\ref{sec:CCMN}).

\subsubsection{Empirical model for protein number}

\label{sec:empmodel0}
We initially fit the empirical model for protein number,
\begin{equation}
    \mu_p = C_0 \mu_m^{\alpha_0}, \label{eqn:pn}
\end{equation}
to the data using a standard least-squares approach; however, the algorithm led to a very poor fit since it does not account for uncertainty in both independent and dependent variables. We therefore used an alternative approach \cite{Hellton2014}, which assumes comparable error in both variables. The model parameters are: 
\begin{eqnarray}
\alpha_0 &=& 2.1\pm 0.04,\\
C_0 &=& 8.0 \pm 1.0,
\end{eqnarray}
where the uncertainties are the estimated standard errors. The result of the empirical model fit is shown in Fig.~\ref{fig:transEff}A, along with the constant-translation-efficiency model, and the RLTO model.

\subsubsection{Empirical model for message number}

For the prediction of the coefficient of variation, it is useful to invert Eq.~\ref{eqn:pn} to generate a model for message number as a function of protein number:
\begin{eqnarray}
    \mu_m &=& C_{0}^{-1/\alpha_{0}} \mu_p^{1/\alpha{0}}, \\
          &=& C_1 \mu_p^{\alpha_1}, \label{eqn:ccmn}
\end{eqnarray}
where the last line defines two new parameters: a coefficient $C_1$ and an exponent $\alpha_1$. The resulting parameters and uncertainties are: 
\begin{eqnarray}
\alpha_1 &\equiv& 1/\alpha_0,\\
&=& 0.48\pm 0.01,\\
C_1 &\equiv& C_0^{-1/\alpha_0},\\
&=& 0.37\pm 0.02,
\end{eqnarray}
where the uncertainties are the estimated standard errors.

\subsubsection{Empirical model for translation efficiency}

\label{sec:empmodel2}

To generate an empirical model for translation efficiency, we started from the empirical model for protein number (Eq.~\ref{eqn:pn}),  and then use Eq.~\ref{eqn:meanprot} to relate protein number, message number, and translation efficiency: 
\begin{eqnarray}
        \varepsilon &=& \textstyle \frac{\mu_p}{\mu_m},\\ &=& C_0 \mu_m^{\alpha_0-1}, \\
        &=& C_2 \mu_m^{\alpha_2},
\end{eqnarray}
where the last line defines two new parameters: a coefficient $C_2$ and an exponent $\alpha_2$. The resulting parameters and uncertainties are: 
\begin{eqnarray}
\alpha_2 &=& \alpha_0-1,\\
&=& 1.07 \pm 0.04,\\
C_2 &=& C_0,\\
&=& 8.0\pm 1.0,
\end{eqnarray}
where the uncertainties are the estimated standard errors.

\begin{figure}
  \centering
   \includegraphics[width=0.48\textwidth]{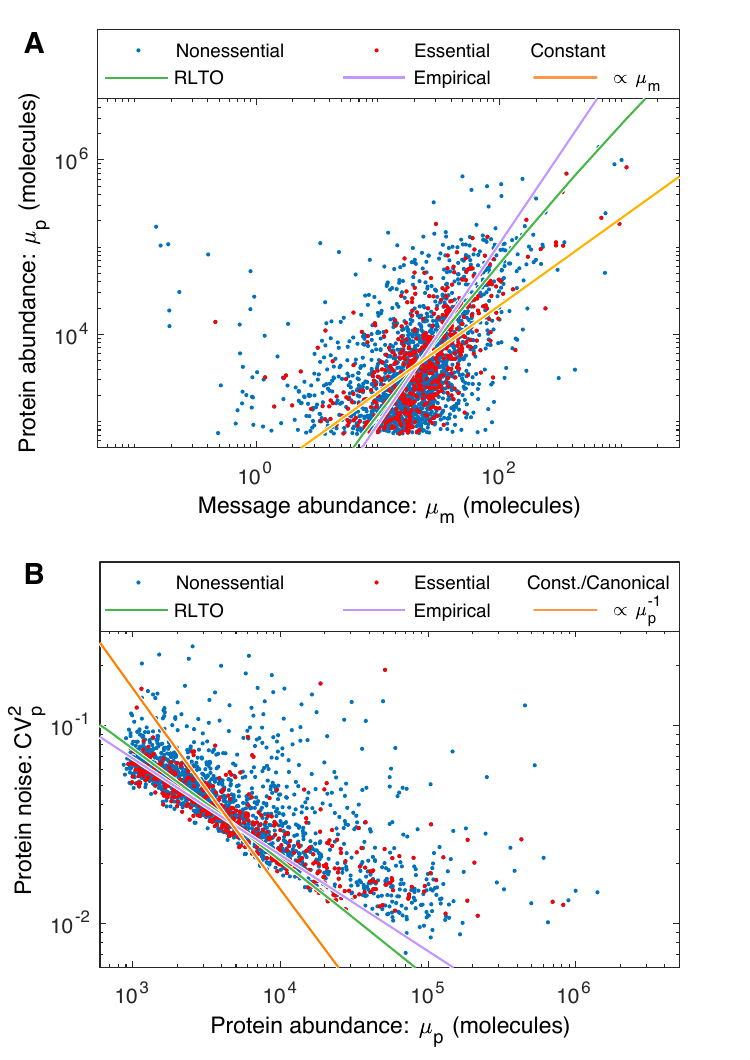}
      \caption{ {\color{red} \textbf{Load balancing predicts the scaling of noise. Panel A: Three competing models for protein abundance in yeast.}} The empirical model (purple) fits the slope and the y offset. The RLTO (green) and constant-translation-efficiency (orange) models fit a parameter corresponding to the y offset only. As discussed in the analysis of the proteome fraction, the RLTO model qualitatively captures the scaling of the protein abundance with message number better than the constant translation efficiency model; however, the predicted fit does not correspond to the optimal power law, which is represented by the empirical model.
      The protein abundance has a cutoff near $10^1$ due to  autofluorescence  \cite{Newman:2006nl}.
      \label{fig:transEff}
  \textbf{Panel B: Predictions of the noise-protein abundance relation.} Using each competing protein abundance model, the noise-protein abundance relation can be predicted using Eq.~\ref{eqn:CVP2}. The canonical noise model (Eq.~\ref{eqn:canonical}) fails to capture even the scaling of the noise. In contrast, both the RLTO and empirical models quantitatively predict both the scaling and magnitude of the noise. The empirical model has the highest performance, presumably due to its two-parameter fit to the protein abundance in Panel A.  A fit accounting for the noise floor is shown in  Fig.~\ref{fig:yeastnoise}.
      \label{fig:noisefigS} }
\end{figure}

\subsubsection{Empirical model for the coefficient of variation}

\label{sec:empmodel3}

To generate an empirical model for the coefficient of variation, we started from the empirical model for message number (Eq.~\ref{eqn:ccmn}),  and then substitute this into the statistical model prediction for CV$_p^2$ (Eq.~\ref{eqn:CVP2}):
\begin{eqnarray}
{\rm CV}_p^2 &=& \textstyle \frac{\ln 2}{\mu_m},\\
 &=&  C_0^{1/\alpha_0}\ln 2 \cdot \mu_p^{-1/\alpha_0},\\
 &=& C_3 \mu_p^{\alpha_3}, 
\end{eqnarray}
where the last line defines two new parameters: a coefficient $C_3$ and an exponent $\alpha_3$. The resulting parameters and uncertainties are: 
\begin{eqnarray}
\alpha_3 &\equiv& -1/\alpha_0,\\
&=& -0.48\pm0.01, \\
C_3 &\equiv& C^{1/\alpha_0}_0\ln 2,\\
&=& 1.9\pm 0.1,
\end{eqnarray}
where the uncertainties are the estimated standard errors.

\subsection{Results: Parameter-free prediction of noise-abundance in yeast}

The fit of the competing protein-message abundance models are shown in Fig.~\ref{fig:noisefig}A.  
Using each model, we can now predict the relation between protein abundance and noise without additional fitting parameters. The predictions of the three competing models are compared to the experimental data in Fig.~\ref{fig:noisefig}B.

In both its ability to capture the protein abundance and predict the noise, the RLTO model vastly outperforms the constant-translation-efficiency model. The purely empirical model that best captures the protein abundance data, due to directly fitting both the y-offset and slope, also performs best in predicting the noise.
It is important to emphasize that the prediction of the noise in all models is non-trivial since there are no free parameters fit, once the protein abundance relation is determined. We therefore conclude that the noise model (Eq.~\ref{eqn:CVP2})  quantitatively predicts the observed noise from the message number and that eukaryotic noise has non-canonical scaling due to load balancing.

\subsection{Discussion: Implications of noise}

\label{sec:end_noise_s}

What are the biological implications of gene expression noise? Many important proposals have been made, including bet-hedging strategies, the necessity of feedback in gene regulatory networks, \textit{etc.}\ \cite{Raser:2005we}.  Our model suggests that robustness to noise fundamentally shapes the central dogma regulatory program. With respect to message number, the one-message-rule sets a lower bound on the transcription rate of essential genes. (See Fig.~\ref{fig:principles}B.) With respect to protein expression, robustness to noise has two important implications: Protein overabundance significantly increases protein levels above what would be required in the absence of noise and therefore reshapes the metabolic budget.
(See Fig.~\ref{fig:principles}A.)  Robustness to noise also gives rise to load balancing, the proportionality of the optimal transcription and translation rates.  (See Fig.~\ref{fig:principles}C.) Not only does robustness to noise affect central dogma regulation, but there is an important reciprocal effect: Load balancing changes the global scaling relation between  noise and protein abundance.  (See Fig.~\ref{fig:noisefigS}B.) 

{\color{red}
\section{Data Tables}

\subsection{Datasets for one-message-rule analysis.}

\idea{Data S1}: Organism: \textit{S.~cerevisiae} (yeast). Original source: \cite{Lahtvee:2017al},  \cite{Blevins:2019yj}, \cite{Godoy:2008eb} \& \cite{Newman:2006nl}. Essential gene classification: \cite{Leeuwen:2020eh}.
Processing: We merged these datasets. We added message number (messages per cell-cycle) as described in Sec.~\ref{sec:desripton}.

\idea{Data S2}: Organism: \textit{Homo sapiens} (human). Original sources: mRNA abundances: Human Protein Atlas \cite{Uhlen:2015jm}. Essential gene classification: \cite{Yang:2003mf}. Processing: We merged these datasets. We added message number (messages per cell-cycle) as described in Sec.~\ref{sec:desripton}.

\idea{Data S3}: Organism:  \textit{E.~coli} grown in rich media (LB). Original sources: mRNA abundance: \cite{Bartholomaus:2016df}. Essential gene classification: \cite{Baba:2008vn}. Processing: We merged these datasets. We added message number (messages per cell-cycle) as described in Sec.~\ref{sec:desripton}.

\idea{Data S4}: Organism:  \textit{E.~coli} grown in minimal media (MM), Original sources: mRNA abundance: \cite{Bartholomaus:2016df}. Essential gene classification: \cite{Baba:2008vn}. Processing: We merged these datasets. We added message number (messages per cell-cycle) as described in Sec.~\ref{sec:desripton}.

\subsection{Datasets for load-balancing analysis.}

\idea{Data S5}: Load-balancing data for yeast cells. 
Organism: \textit{S.~cerevisiae} (yeast). Original sources: Protein abundance \cite{Ghaemmaghami:2003jj}. mRNA abundance: \cite{Lahtvee:2017al}. Processing: We merged these datasets. We added protein number and message number (messages per cell-cycle) as described in Sec.~\ref{sec:desripton}.

\idea{Data S6}: Load-balancing data for mammalian cells. 
Organism: \textit{Mus musculus} (mammalian).
Original source: \cite{Schwanhausser:2011ec}.  Processing: We added protein number and message number (messages per cell-cycle) as described in Sec.~\ref{sec:desripton}.

\idea{Data S7}: Load-balancing data for \EC. 
Organism:  \textit{E.~coli}. Original source: \cite{Balakrishnan:2022ai}.  Processing: We added protein number and message number (messages per cell-cycle) as described in Sec.~\ref{sec:desripton}.

\subsection{Datasets for noise analysis.}

\idea{Data S8}: Noise data for yeast.
Organism: \textit{S.~cerevisiae} (yeast). Original sources: Noise measurements: \cite{Newman:2006nl}. Essential gene classifications: \cite{Leeuwen:2020eh}. Protein fluorescence data, rescaled by mass-spec data as described in Sec. \ref{sec:MS}: \cite{Newman:2006nl}. Processing: We merged these datasets. We added protein number as described in Sec.~\ref{sec:desripton}.

\subsection{Datasets for one-message-rule exceptions.}

\color{orange}{
\idea{Data S9}:
Exceptions to the one-message-rule for yeast cells.
Organism: \textit{S.~cerevisiae} (yeast). Original sources: Message abundances: \cite{Lahtvee:2017al}. Essential gene classifications: \cite{Leeuwen:2020eh}. 

\idea{Data S10}:
Exceptions to the one-message-rule for human cells.
Organism: \textit{H.~sapiens} (human). Original sources: Message abundances: \cite{Uhlen:2015jm}. Essential gene classifications: \cite{Yang:2003mf}. 

\idea{Data S11}:
Exceptions to the one-message-rule for \EC grown in rich media (LB).
Organism: \EC. Original sources: Message abundances: \cite{Bartholomaus:2016df}. Essential gene classifications: \cite{Baba:2008vn}. 
}

}

\end{document}